\documentstyle[preprint,12pt,bbold,eqsecnum,aps,amssymb]{revtex}

\tightenlines

\newcommand{\vecr}{\mbox{\bf r}}
\newcommand{\vecrs}{\mbox{\bf \scriptsize r}}
\newcommand{\vep}{\vecr_\perp}
\newcommand{\veps}{\vecrs_\perp}
\newcommand{\dvep}{d^2r_\perp}
\newcommand{\dveps}{d^2r'_\perp}
\newcommand{\ver}{\mbox{\bf R}}
\newcommand{\tvec}{\tilde{\mbox{\bf r}}}
\newcommand{\tr}{\tilde r}
\newcommand{\eh}{\frac{1}{2}}
\newcommand{\oml}{\omega(l)}
\newcommand{\omsl}{\Omega_s(l)}
\newcommand{\be}{\begin{eqnarray}}
\newcommand{\ee}{\end{eqnarray}}
\newcommand{\nn}{\nonumber}
\newcommand{\ew}{\epsilon_{wf}}
\newcommand{\sw}{\sigma_{wf}}
\newcommand{\ewnull}{\epsilon_{wf}^{(0)}}
\newcommand{\eweins}{\epsilon_{wf}^{(1)}}
\newcommand{\swnull}{\sigma_{wf}^{(0)}}
\newcommand{\sweins}{\sigma_{wf}^{(1)}}
\newcommand{\siglp}{\sigma_{l,g}^{(p)}}
\newcommand{\siglc}{\sigma_{l,g}^{(c)}}
\newcommand{\sigls}{\sigma_{l,g}^{(s)}}
\newcommand{\sils}{\sigma_{l,s}}
\newcommand{\drho}{\Delta\rho}
\newcommand{\dmu}{\Delta\mu}
\newcommand{\dom}{\Delta\Omega}

\newcommand{\dt}{\delta T}
\newcommand{\dmupre}{\Delta\mu_{pre}}

\newcommand{\mupre}{\mu_{pre}}
\newcommand{\mupret}{\mu_{pre}(T)}
\newcommand{\rl}{\rho_l}
\newcommand{\rg}{\rho_g}
\newcommand{\rw}{\rho_w}
\newcommand{\resi}{\rw\ew\sw}
\newcommand{\void}[1]{}
\newcommand{\gsim}{{\scriptstyle{\stackrel{>}{\sim}}}}
\newcommand{\lsim}{{\scriptstyle{\stackrel{<}{\sim}}}}

\begin{document}

\title{WETTING OF CURVED SURFACES}
\author{T.~Bieker and S.~Dietrich}
\address{Fachbereich Physik, Bergische Universit\"at Wuppertal,
D-42097 Wuppertal, \\ Federal Republic of Germany}
\maketitle

{\small
As a first step towards a microscopic understanding of the effective
interaction between colloidal particles suspended in a solvent we
study the wetting behavior of one-component fluids at spheres and
fibers. We describe these phenomena within density functional theory
which keeps track of the microscopic interaction potentials governing
these systems. In particular we properly take into account the
power-law decay of both the fluid-fluid interaction potentials and the
substrate potentials. The thicknesses of the wetting films as a function
of temperature and chemical potential as well as the wetting phase
diagrams are determined by minimizing an effective interface potential
which we obtain by applying a sharp-kink approximation to the density
functional. We compare our results with previous approaches to this
problem.}


\newpage

\centerline{\large \bf I. Introduction}
\vskip 1cm

\addtocounter{section}{1}

If colloidal particles are dissolved in a solvent consisting of a
binary liquid mixture, necessarily one of these two components is
preferentially adsorbed on the spherical surfaces of the colloidal
particles. Near a first-order phase separation of the bulk solvent
this adsorption may lead to the coating of the colloids by wetting
films which snap into a bridgelike structure if two colloids come
close to each other. This coagulation can become so pronounced that it
results in flocculation \cite{beysens,gallagher,kiraly,jaya,woermann}. 
Such experiments have inspired several
theoretical investigations
\cite{pomeau,boucher,dobbs,osborn,carnie,loewen,degennes,henderson,saez,lee,ipsen}
which have shed light on various features
of these complex phenomena.

Although  there are indications that for higher concentrations of
colloidal particles their effective mutual interactions are beyond a
superposition of effective pair potentials \cite{loewen}, 
for low concentrations the
knowledge of the effective interaction between two isolated colloidal
particles at a fixed distance immersed into the solvent is
important. Moreover, this latter configuration is interesting in its
own, because the bridge formation between individual pairs is not
solely accessible indirectly via flocculation but can be studied also
directly with atomic or surface force apparatuses \cite{christenson}. 
Besides colloidal
systems these problems are in addition relevant for technical
applications such as the contact between lubricant films on the disk
surfaces in magnetic recording with the recording head 
\cite{mate}, sintering \cite{basa}, latex film formation \cite{lin},
catalyst wetting efficiency in trickle-bed reactors \cite{aldahhan}, or for
the bonding mechanism of emulsified adhesives in granular and fibrous
substrates \cite{debisschop}. 

As a prerequisite for understanding the effective interaction between
such spherical particles immersed into a solvent one must know the
structural and thermal properties of these systems in the case that
the distance between two colloidal particles is macroscopically
large. Near a first-order phase transition of the solvent this amounts
to studying the wetting behavior of the solvent at the surface of a
spherical substrate. The paradigmatic case is the formation of a
liquidlike film at the curved substrate-vapor interface upon
approaching the liquid-vapor coexistence curve in the bulk of a
one-component fluid from the vapor side. This problem has been studied
theoretically by several authors
\cite{philip,holyst,lipowsky,indekeu,swift,gil,had} 
based on various versions of phenomenological
models. They confirm the general expectation (see Subsect. X.B in
Ref.~\cite{dietrich}) that the positive curvature of the surface of a substrate
prevents the build-up of macroscopically thick wetting films because, in
contrast to a planar geometry, the area of the emerging liquid-vapor
interface increases with the thickness of the wetting film. This
increasing cost of the free energy of the film has effectively the
same consequence as if the bulk fluid is kept off liquid-vapor
coexistence. Therefore continuous wetting transitions are eliminated
and first-order wetting transitions are reduced to quasi--first-order
transitions between small and large, but finite, film thicknesses,
smeared out due to the finite size of the substrate area.

However, beyond these general aspects it is important to know to which
extent the film thicknesses of these wetting films are limited and how
these limitations depend on the radius of the colloidal particles, on
the character and the form of the substrate potential and the
interaction potential between the solvent molecules, and on the size
of the solvent molecules. The aforementioned phenomenological mo\-dels
do not allow one to answer these questions because they do not keep track
of the microscopic details of the system. In particular the experience
with wetting phenomena on flat substrates tells that it is essential
to take into account equally the power-law decay of the substrate
potential and of the interaction potential between the fluid
particles \cite{dietrich,schick}. Inter alia, the long range of 
the forces between the fluid
particles causes the breakdown of a gradient expansion for treating
the deviations of the emerging liquid-vapor interface from its flat
configuration and leads to a {\it nonlocal} Hamiltonian \cite{napi}. 
For simple
geometric shapes of the interfaces involved this long-ranged character
of the dispersion forces acting in and on fluids is taken into account
by the Dzyaloshinskii-Lifshitz-Pitaevskii (DLP) theory \cite{dzy} which has been
applied to spheres and cylinders \cite{parsegian}. For a given system this allows one
to compute the cost in free energy $\Omega_s(l)$ to maintain a thick
liquid film of prescribed thickness $l$ adsorbed on the substrate in
terms of the frequency dependences of the permittivities of the
substrate, the bulk liquid, and the bulk vapor. Although this approach
has the advantage to take into account many-body forces and
retardation, it suffers also from some shortcomings. (i) The
DLP theory gives access only to the leading asymptotic behavior of
$\Omega_s(l\to\infty)$ and thus does not allow one to describe
critical wetting transitions which result from the competition between
the leading and next-to-leading order term in $\Omega_s(l)$ for large
$l$ \cite{dietrich,schick}. (ii) Details of the substrate potential, 
which are important for
first- and second-order wetting transitions \cite{dietrich,schick}, 
are not captured by the DLP theory. 
(iii) $\Omega_s^{(DLP)}(l)$ exhibits an unphysical divergence for
$l\to 0$ because this approach ignores the repulsive part of both the
substrate potential and the fluid-fluid interaction potential and thus
the structure of the emerging substrate-liquid interface. (iv) The
broadening of the emerging liquid-vapor interface upon raising the
temperature cannot be accounted for. (v) The dependence of
$\Omega_s^{(DLP)}(l)$ on temperature and chemical pressure, in
particular close to phase transitions in the bulk fluid, is not
transparent. (vi) In the present context, as expounded above, the most
severe drawback of the DLP approach is, that the DLP results for
wetting of a single sphere or cylinder \cite{parsegian}
cannot be used as a building block
for investigating later on the effective interactions between two such
objects because the shape of the bridging wetting film between them
is not known in advance which precludes the practical application of 
the DLP theory to the problem described in the beginning.

Therefore in this paper we compute (Sect.~II) and discuss (Sect.~III) the
effective interface potential $\Omega_s(l)$ for the wetting of a single
sphere and of a single cylinder on the basis of density functional
theory \cite{evans}. This microscopic approach allows one to address the
aforementioned points (i)-(vi) and overcomes the shortcomings of the
phenomenological theories. The price to be paid is that the results do
not account for the effects of many-body forces.

In the beginning of this introduction we have focused on the physical
interest in spheres exposed to a fluid. It turns out that there is
also substantial interest in the cylindrical geometry. In view of its
importance for technical processes such as the lubrication of textile
fibers, optical fiber processing, and the formation of fiber-reenforced
resins \cite{connor} there are many studies of the statics and dynamics of
adsorption on cylinders \cite{osborn,indekeu,levinson,quere,%
rooks,wagner,yarin,carroll,chiao,darbellay,groot,mchale,bruyn}. 
In addition the adsorption on cylinders is
important for the vibrating-wire microbalance as a standard technique
to study wetting phenomena \cite{bartosch}. However, one has to keep
in mind that the
convoluted surface of graphite fibers impedes the interpretation of
these experiments on the basis of theoretical models which assume a
smooth cylindrical surface \cite{wilen}. (Such surface inhomogeneities may even
play a role
for spherical colloidal particles \cite{czarnecki}.) 
For the same reasons given above
for the spherical geometry we apply a microscopic density functional
theory also to the problem of wetting of wires in order to obtain a
detailed picture of their behavior in systems with long-ranged
interactions.

We conclude the introduction with two remarks. First, the present
study is dealing with volatile liquids, i.~e., it is important that
the liquid phase and the surrounding vapor are in thermal
equilibrium. Second, we do not address the problem of {\em interfacial}
wetting at curved surfaces, i.~e., the substrate is passive. Therefore
our results do not apply directly to such phenomena as surface melting
of small particles and wires.

\vskip 1cm

\centerline{\large \bf II. Effective interface potential from density}
\centerline{\large \bf functional theory}

\subsection{Density functional}

\addtocounter{section}{1}

As for the description of wetting phenomena in planar geometries we
describe the one-component fluid by a simple grand canonical density
functional \cite{dietrich,evans}: 

\be
&&\Omega\left[\{\rho (\vecr)\};T,\mu\right]=\int_V d^3r
f_h\left[\{\rho(\vecr)\},T\right] \nn\\
&+& \eh \int_V d^3r\int_V d^3r'\tilde w\left(|\vecr -\vecr'|\right)
\rho(\vecr) \rho(\vecr ')
+\int_V d^3r \rho(\vecr)\left(\rw v(\vecr)-\mu\right).\nn\\
\label{ansatz}
\ee

\noindent $V$ is a macroscopic but finite volume into which the substrate ${\cal S}$ is
embedded but is not a part of it.
$\Omega\left[\{\rho (\vecr)\};T,\mu\right]$ is the grand canonical
free energy for a given density configuration $\rho(\vecr)$ as a
function of the temperature $T$ and the chemical potential $\mu$.
We assume a spherically symmetric pair interaction potential $w(r)$
for the fluid
particles which according to standard procedures \cite{evans} is divided
into a short-ranged repulsive and a
long-ranged attractive part $\tilde w(r)$ with
$\tilde w(r\to\infty)\sim r^{-6}$. Here we do not consider retardation
effects which would lead to $\tilde w(r\to\infty)\sim r^{-7}$.
The divergent repulsive part of the interaction potential gives rise to a
reference free energy which is mapped onto $f_h(\rho,T)$ as the free
energy density of a homogeneous system of hard spheres with an
effective diameter $d(T)$ \cite{barker} and number density $\rho$.
An appropriate expression for $f_h$ is given by the Carnahan-Starling
formula \cite{carnahan}
\be
f_h(\rho,T)=k_BT\rho\left(\ln(\eta)-1+\frac{4\eta-3\eta^2}
{(1-\eta)^2} \right)
\ee
where $\eta=\frac{\pi}{6}\rho d(T)^3$ is the dimensionless packing
fraction.
The second term in Eq. (2.1) accounts for the long-ranged
attractive interaction
which we treat according to the Barker-Henderson scheme \cite{barker},
\be
\tilde w(r)=\Theta(r-\sigma)w(r),
\ee
where $\Theta$ denotes the
Heaviside function. For many purposes it may be useful to adopt the
Lennard-Jones potential $\phi_{LJ}(r)$ as the actual full interaction
potential $w(r)$:
\be
w(r)=\phi_{LJ}(r)=4\epsilon\left(\left(\frac{\sigma}{r}\right)^{12}
-\left(\frac{\sigma}{r}\right)^6\right) 
\ee
with depth $\epsilon$ and dia\-me\-ter $\sigma$ of the fluid
par\-ti\-cles so that $d(T)=\int_0^\sigma dr \left(1-e^{-\beta\phi_{LJ}(r)}\right)$.
For this interaction potential Eq.~(2.1) leads to a phase diagram of the bulk
fluid with a critical point at $k_BT_c/\epsilon=1.09858$,
$\rho_c\sigma^3=0.27201$, and $\mu_c/\epsilon=-3.77713$ \cite{werner}.
The fluid is exposed to a substrate which in a first approximation one
may consider to follow from a superposition of Lennard-Jones
interaction potentials $\phi_{LJ}^{wf}$ between wall and fluid
particles,
\be
\phi_{LJ}^{wf}=4\epsilon_{wf}\left(\left(\frac{\sw}{r}\right)^{12}
-\left(\frac{\sw}{r}\right)^6\right),
\ee
i.~e., 
$v(\vecr)=\int_S d^3r'\phi^{wf}_{LJ}(|\vecr-\vecr'|)$
where $S$ is the substrate volume. Ignoring the exponentially decaying
corrugation effects of an atomically structured substrate,
$v(\vecr)$ 
depends only on the radial distance to the substrate surface, 
$v(\vecr)=v(r)$. In general, however, $v(r)$ is more complicated than
a linear superposition of Eq.~(2.5) and enters the problem as a free
parameter function.
The equilibrium number density, which minimizes Eq.~(2.1), depends
also only on $r$ and yields the corresponding grand canonical
potential of the system.

For the above interaction potentials this minimum can only be
determined numerically. Here, however, we construct an effective
interface potential $\Omega_s(l)$ which is the cost in free energy to
maintain a liquidlike film of a prescribed thickness $l$ near the
substrate although the bulk fluid is in the vapor phase. The
equilibrium thickness $l_0$ minimizes this effective interface
potential. This approach yields a transparent insight into the various
wetting transitions which can occur \cite{dietrich2}. It has turned 
out that the
so-called sharp-kink approximation, in which the trial-functions
$\rho(\vecr)$ compatible with the thickness $l$ are approximated by 
piecewise constant density profiles, yields a surprisingly accurate
expression for the actual effective interface potential
$\Omega_s(l)$ \cite{napi2}. 
(In the present context there are also approaches beyond the
sharp-kink approximation \cite{patrykiejew}. These numerical studies
yield valuable information about specific systems but no overall
picture for the wetting phenomena near curved surfaces.)
For spherical and cylindrical substrates the sharp-kink approximation
amounts to using the following
trial function (see Fig.~1(a)):
\be
\hat\rho(r)=\left[\rl\Theta(h-r)
+\rg\Theta(r-h)\right]\Theta(r-r_1).
\label{ska}
\ee
$h$ and $r_0$ denote the radius of the emerging gas-liquid interface
and of the substrate, respectively, so that $l=h-r_0\geq d_w$. The
repulsive part of the substrate potential leads to an excluded volume
close to the substrate. This is taken into account by the radius
$r_1=r_0+d_w$; approximately $d_w=\eh(\sigma+\sw)$. Actually $d_w$ is
given by the zeroth moment of the wall-liquid interface profile
$\rho_{wl}(z)$, $d_w=\int_0^\infty dz (1-\rho_{wl}(z)/\rl)$, at
gas-liquid coexistence $\mu=\mu_0(T)$ [55(c)]. $\rl=\rl(T,\mu=\mu_0(T))$
denotes the equilibrium liquid density at two-phase coexistence,
whereas $\rg=\rg(T,\mu)$ is the actual vapor density.
From Eq.~(2.6) one obtains the 
sharp-kink profile for the flat geometry
if one sets $r_0=0$ and replaces $r$ by the distance $z$ perpendicular
to a half-space filled with substrate atoms (see Fig.~1(b)).

If Eq.~(2.6) is inserted into Eq.~(2.1) $\Omega$ can be decomposed
into a bulk contribution proportional to $V^{(i)}$, $i=s,c$, and into
subdominant terms. The superscripts $s$ and $c$ correspond to a sphere
and a cylinder, respectively, so that $V^{(s)}=\frac{4\pi}{3}(L^3-r_0^3)$ and
$V^{(c)}=\pi M(L^2-r_0^2)$; $M$ is the length of the cylinder. In the
cylindrical case we consider the thermodynamic limit $M\to\infty$
which leaves one with a wire. In the planar geometry, $i=p$,
$V^{(p)}=AL$ where $A$ is the substrate area. In all cases $L$ denotes
a macroscopic system size (see Fig.~1).
The systematic application of the sharp-kink approximation to
Eq.~(2.1) leads for all geometries $i=s,c,p$ and independent of the
form of $w(r)$ and $v(\vecr)$ to (sometimes we omit the superscripts)
\be
\Omega[\hat\rho(r);T,\mu]=V\Omega_b(\rg,T,\mu)+
\Sigma_{g,vac}+\bar\omsl.
\ee
$\Omega_b$ is the bulk free energy density of the vapor phase.
The finite volume $V^{(i)}$ introduces an artificial gas-vacuum
interface which gives rise to the surface contribution
$\Sigma_{g,vac}$. This and other surface contributions
$\Sigma_{\alpha,\beta}(r)$ are proportional to the surface area $A(r)$
of the corresponding interface with radius $r$ so that
$\lim_{r\to\infty}\Sigma_{\alpha,\beta}(r)/A(r)=\sigma_{\alpha,\beta}$  
is finite and is the surface tension between phases (substrate, liquid,
or vapor) $\alpha$ and
$\beta$ in planar geometry:
\be
\Sigma_{\alpha,\beta}(r)=A(r)\sigma_{\alpha,\beta}(r).
\ee
We denote the $r$-dependent surface tension as $\sigma_{\alpha,\beta}(r)$.
The third contribution in Eq.~(2.7) is the
effective interface potential $\bar\omsl$ which contains
the dependence on the film thickness $l$:
\be
\bar\omsl=V_{liq}\Delta\Omega_b
+\Sigma_{l,g}+\Sigma_{l,s}+\bar\oml\label{efipo}
\ee
In Eq.~(\ref{efipo})
$V_{liq}\Delta\Omega_b=V_{liq}(\rl-\rg)(\mu_0-\mu)\equiv
V_{liq}\drho\Delta\mu$ is 
the cost in free energy to fill the volume
$V^{(s)}_{liq}=\frac{4\pi}{3}(h^3-r_0^3)$ and $V^{(c)}_{liq}=\pi
M(h^2-r_0^2)$, respectively, with liquid although $\mu$ and $T$ 
favor the vapor phase. Upon approaching
liquid-vapor coexistence $\mu_0(T)$ from below,
i.~e.~$\mu < \mu_0$, $\Delta\Omega_b$ is positive, so that
$V_{liq}\Delta\Omega_b$ increases with the radius of the liquid
volume. Thus for all geometries the formation of an macroscopically thick 
layer is prohibited as long as one is off coexistence.

If the equilibrium value $l_0(T,\mu)$, which minimizes
$\bar\Omega_s(l)$, diverges for $\mu\to\mu_0(T)$ one has a so-called
complete wetting transition \cite{dietrich,schick}. However, for
curved geometries a macroscopically thick film cannot emerge even when
coexistence is reached because the free energy contribution due to the
liquid-gas interface,
$\Sigma_{l,g}=A(h)\sigma_{l,g}(h)$, increases with the area $A(h)$
of the interface, i.~e.~with the film thickness $l$.
Therefore the equilibrium film thickness $l_0$ on spherical and cylindrical substrates 
will {\it always} be limited to a finite value \cite{dietrich}. 
The free energy of the liquid-solid
interface does, by contrast, not depend on $l$, so that 
$\Sigma_{l,s}=A(r_0)\sigma_{l,s}(r_0)$ is a constant 
with respect to the minimization of $\bar\Omega_s(l)$.

As long as the liquid-gas and the solid-liquid interface are not
macroscopically far apart from each other, there will be a nonzero 
interaction between them. This is taken into account by the last term 
in the effective interface potential, $\bar\oml$, which by
construction vanishes for $l\to\infty$. 
In the case of a planar substrate, interfacial phase transitions
along the coexistence curve $\mu_0(T)$ are determined by $\bar\oml$
only \cite{dietrich,schick,dietrich2}. 
For spheres and cylinders, however, the wetting behavior at 
$\Delta\mu\equiv\mu_0-\mu=0$ is governed by $\bar\oml$ as well as $\Sigma_{lg}$.

It is convenient to normalize the effective
interface potential $\bar\omsl$ by the area of the
substrate surface, $A(r_0)$, so that (see Fig.~1)
\be
\omsl&=&\frac{1}{A(r_0)}\bar\omsl\nn\\
&=&\frac{V_{liq}}{A(r_0)}\Delta\Omega_b+\frac{A(r_0+l)}{A(r_0)}\sigma_{l,g}(r_0+l)
+\sigma_{l,s}(r_0)+\oml
\ee
where
\be
\oml=\frac{1}{A(r_0)}\bar\oml.
\ee
The geometrical prefactors $V_{liq}/A(r_0)$ and $A(r_0+l)/A(r_0)$ have
the following explicit forms:
\be
\frac{V_{liq}}{A(r_0)}=\frac{r_0}{\tau+1}\left(
\left(1+\frac{l}{r_0}\right)^{\tau+1}-1\right)
\ee
and
\be
\frac{A(r_0+l)}{A(r_0)}=\left(1+\frac{l}{r_0}\right)^\tau
\ee
with $\tau=2$ for a sphere and $\tau=1$ for a cylinder.
Before determining the surface phase diagrams from the minimization of
$\Omega_s(l)$ we first
discuss the two $l$-dependent terms $\Sigma_{l,g}$ and $\oml$,
which are interesting in their own right and allow for a comparison
with results obtained by other approaches.

\subsection{Liquid-vapor and solid-liquid surface tension}

From $\Sigma_{l,g}$ we obtain an expression for the
curvature dependence of the liquid-vapor surface tension of a spherical
drop and of a liquid thread. Whereas apart from Ref.~\cite{lovett}  
we are not aware of studies of liquid fibers, 
the surface tension of
liquid drops has been subject to partially controversial discussions
\cite{buff}.
Here we calculate the surface tension within our approach for both cases explicitly in
order to work out the influence of the curvature.

Within the sharp-kink approximation of Eq.~(2.1) the planar
liquid-vapor surface tension is given by \cite{dietrich}
\be
\siglp=\frac{1}{A}\Sigma_{l,g}^{(p)}&=&-\eh (\drho)^2\int_0^\infty dz\ t^{(p)}(z),
\ee
where $t^{(p)}(z)$ is the interaction potential of a fluid particle at
distance $z>0$ of a half-space filled with
fluid particles, too, interacting with a pair potential $\tilde w$:
\be
t^{(p)}(z)=\int_z^\infty\ dz'\int_{R^2}\ d^2r_\parallel\ \tilde w
\left(\sqrt{\vecr_\parallel^2+z'^2}\right),
\ee
$\vecr^2_\parallel=x^2+y^2$.
If one inserts the 
intermolecular potential $\tilde w$ chosen above, one obtains for the surface
tension of the planar interface
$\siglp=\frac{3}{4}\pi\epsilon\sigma^4(\drho)^2$.
Close to $T_c$ this expression for $\sigma_{l,g}$ vanishes $\sim\tilde
t$ where $\tilde t=(T_c-T)/T_c\to 0$ instead of the expected
mean-field behavior $\tilde t^{3/2}$ because the sharp-kink
approximation fails to take into account the broadening of the
interface on the scale of the bulk correlation length $\xi\sim\tilde
t^{-1/2}$. Thus Eq.~(2.14) takes into account the long-ranged
character of the forces but should not be used close to $T_c$.
The corresponding expressions for a cylinder and a sphere of radius
$h$ have a similar form:
\be
\siglc(h)=\frac{1}{2M\pi h}\Sigma_{l,g}^{(c)}
=-\eh (\drho)^2\int_h^\infty dr\ \frac{r}{h}\ t^{(c)}(r;h)
\ee
and
\be
\sigls(h)=\frac{1}{4\pi h^2}\Sigma_{l,g}^{(s)}
=-\eh (\drho)^2\int_h^\infty dr\ \left(\frac{r}{h}\right)^2\ t^{(s)}(r;h).
\ee
$t^{(c)}(r;h)$ is the interaction potential at distance $r>h$ from
the axis of an infinite liquid cylinder of radius $h$:
\be
t^{(c)}(r;h)=\int_{-\infty}^{+\infty}\ dz\int_0^{2\pi} d\phi \int_0^h r'\ dr'
\tilde w\left(\sqrt{r^2 +r'^2-2rr'\cos\phi+z^2}\right).
\ee
Correspondingly, $t^{(s)}(r;h)$ is the interaction potential at distance
$r>h$ from the
center of a liquid drop of radius $h$:
\be
t^{(s)}(r;h)=\frac{2\pi}{r}\int_0^h\ dr'\ r'\int_{r-r'}^{r+r'}ds\ s\tilde
w(s).
\ee

If one inserts Eqs.~(2.3) and (2.4) into Eq.~(2.19), one obtains for $r\gg h$
\be
t^{(s)}(r;h)\stackrel{r\gg h}{\longrightarrow}
-\frac{4\pi}{3}h^3\frac{4\epsilon}{(r/\sigma)^6}
+{\cal O}\left(\frac{1}{r^8}\right),
\ee
i.~e., viewed from a large distance the liquid drop appears as a
single molecule with an effective radius $h$, exerting a
van-der-Waals interaction on any other fluid particle.
Equations (2.16) and (2.17) are valid for a general form of the
interaction potential $w(r)$. For the specific choice given by
Eqs.~(2.3) and (2.4) the expansion in powers of the curvature $1/h$
yields the following expressions for the surface tensions (see
Eqs. (A7) and (B35) in Appendices A and B, respectively):
\be
\frac{1}{A}\Sigma^{(p)}_{l,g}&=&\siglp,\\
\siglc(h)
&=&\siglp\left(1-\frac{1}{6}\frac
{\ln (h/\sigma)}{(h/\sigma)^2} -\frac{1}{(h/\sigma)^2}\left(\frac{5}{3}\ln 2+\frac{1}{144}\right)
-{\cal O}\left(\frac{1}{(h/\sigma)^4}\right)\right),
\ee
and
\be
\sigls(h)=\siglp\left(1-\frac{2}{9}
\frac{\ln (h/\sigma)}{(h/\sigma)^2}-\frac{1}{(h/\sigma)^2}\left(\frac{2}{9}\ln
2+\frac{4}{27}\right)
-{\cal O}\left(\frac{1}{(h/\sigma)^8}\right)\right).\label{surftension}
\ee
If one compares these results
with the usual ansatz for the surface tension of a drop \cite{ergaenzung},
\be
\sigls(h)=\siglp\left(1-\frac{2\delta}{h}+\frac{\alpha}{h^2}+\ldots\right),
\ee
first, one notices that our approach predicts that the surface tensions of the
curved interfaces are smaller than the planar one and that both attain the
latter one asymptotically,
$\lim_{h\to\infty}\sigma_{l,g}^{(s,c)}(h)=\sigma_{l,g}^{(p)}$. 
Second, since our density profile (Eq.(2.6)) is antisymmetric around
$r=h$ and does not contain a symmetric component, within our approach
the Tolman length $\delta$ must be zero \cite{buff}. Third, the long-ranged forces
give rise to logarithmic corrections which dominate the usual
subdominant term $\sim \alpha h^{-2}$. 
Within our approach
$\sigma_{l,g}^{(i)}$ is invariant with respect to interchanging liquid
with vapor and therefore they do not depend on the sign of curvature
of the interfaces. The actual behavior of $\sigls(h)$ and $\siglc(h)$
as obtained from Eqs.~(2.16) and (2.17) is shown in Fig.~2.

The solid-liquid surface tension entering Eq.~(2.10) depends only on
$r_0$ and thus it is constant with respect to the minimization of $\Omega_s(l)$. 
Within the sharp-kink approximation (Eq.~(2.6)) it has the
following form ($\tau=2$ for a sphere, $\tau=1$ for a cylinder):
\be
\sigma_{l,s}(r_0)=&-&\eh\rl^2\int_{r_1}^\infty
dr\,\left(\frac{r}{r_0}\right)^\tau 
t^{(\tau)}(r,r_1)+\rl\rw\int_{r_1}^\infty dr\,
\left(\frac{r}{r_0}\right)^\tau v^{(\tau)}(r,r_0)\nn\\
&-&\Omega_b(\rl)V^{(\tau)}_{exc}/A(r_0)
\ee
where $v^{(\tau)}(r;r_0)$ is the substrate potential at a distance $r$ of a
sphere and a cylinder with radius $r_0$, respectively, and with
\be
V^{(\tau)}_{exc}/A(r_0)=\frac{r_0}{\tau+1}
\left(\left(1+d_w/r_0\right)^{\tau+1}-1\right),
\ee
where $V_{exc}$ is the excluded volume due to the repulsive parts of
the interaction potentials (see Fig.~1). (For additional information
see Ref.~\cite{stecki}.)

\subsection{Interaction between the interfaces}

For the flat substrate, within the sharp-kink approximation, the correction term
$\omega^{(p)}(l)=\frac{1}{A}\bar\omega^{(p)}(l)$ 
is given by 
\be
\omega^{(p)}(l)&=&\drho\left(\rl\int_{l-d_w}^\infty\ dz\ t^{(p)}(z) 
-\rw\int_l^\infty\ dz\ v^{(p)}(z)\right),\quad l\geq d_w\\
&=& \frac{a}{l^2}+\frac{b}{l^3}+\ldots\quad,\quad l\gg d_w.\nn
\ee
Approximately the substrate potential may be expressed in terms of
$\phi_{LJ}^{wf}$ (Eq.~(2.5)):
\be
v^{(p)}(z)=\int_z^\infty\ dz'\int_{R^2}\ d^2r_\parallel\,\phi_{LJ}^{wf}\left(
\sqrt{\vecr_\parallel^2+z'^2}\right).
\ee
In contrast to $t^{(p)}(z)$ the function $v^{(p)}(z)$ diverges for $z\to 0$. However,
according to Eq.~(2.6), in Eq.~(2.27) one has $l\geq d_w$.

In the case of a planar substrate and for the above parametrization of
the interaction in terms of Lennard-Jones potentials
the order of the wetting transition and the wetting temperature $T_w$ 
are determined by $\epsilon$, $\sigma$, $\ew/\epsilon$, 
$\sw/\sigma$, and $\rw\sw^3$.
In case of a first-order wetting transition $\oml$ has two
minima separated by a free energy barrier. One of these minima is
localized at a finite value $l_0$ and represents the global
minimum for temperatures $T<T_w$, while the second minimum is given by
$l=\infty$ where $\oml=0$. At $T=T_w$, the minimum at $l=\infty$
changes from a local to a global minimum, so that the film thickness on
the planar substrate jumps from a microscopic value to infinity. 
In case of a second order wetting transition, $\oml$ has
only one minimum, which is shifted continuously from a small value 
$l_1$ to $l=\infty$ as the temperature is raised from $T<T_w$ towards
$T=T_w$ \cite{dietrich,schick,dietrich2}.

For the curved geometries, the systematic analysis of the sharp-kink
approximation to the density functional in Eq.~(2.1) leads to the
result that $\oml$ has essentially the same
structure as for the planar system ($l\geq d_w$):
\be
\omega^{(\tau)}(l)=\drho\left(\rl\int_{r_0+l}^\infty dr\ 
\left(\frac{r}{r_0}\right)^\tau\ t(r;r_1)
-\rw\int_{r_0+l}^\infty dr\  \left(\frac{r}{r_0}\right)^\tau
\ v(r;r_0)\right)
\ee
with $\tau=1$ for the cylindrical and $\tau=2$ for
the spherical system.
$t(r;r_1)$ is defined in terms of $\tilde w(r)$, and thus $w(r)$,
according to Eq.~(2.18) and Eq.~(2.19), respectively; $v(r;r_0)$ is the
substrate potential of a sphere and a cylinder of radius $r_0$,
respectively. 
It should be emphasized that Eq.~(2.29) is valid for any
functional form of $w(r)$ and $v(r)$, provided only that they decay rapidly
enough in order to guarantee the convergence of the integrals;
Eq.~(2.29) also does not depend on the explicit form of Eq.~(2.2).

However, due to the unlimited increase of $A(r_0+l)/A(r_0)$ (see
Eq.~(2.13)) in Eq.~(2.10) even at coexistence, i.~e.,
$\Delta\Omega_b=0$, 
the film thickness $l$ cannot take on a macroscopic value
on cylinders or spheres, so that first order wetting transitions 
are reduced to a jump from a small to a large but finite value of $l$.
In order to be able to discuss these transitions quantitatively the
general expression for the effective interface potential (Eq.~(2.29))
must be evaluated for specific interaction potentials $w(r)$ and
$v(r)$. According to Appendices A and B the 
model potentials in Eqs.~(2.3)-(2.5) lead to (see Eq.~(A13))
\be
\omega^{(\tau)}(l)=\omega^{(\tau)}_{attr}(l)
+\omega^{(\tau)}_{rep}(l)
\ee
where
\be
\omega^{(s)}_{rep}(l)&=&\frac{1}{r_0^2}\frac{\pi}{540}
\drho(\rl\epsilon\sigma^{12}-\resi^{12})
\left(\frac{h^2+8hr_0+r_0^2}{(h+r_0)^8}-\frac{h^2-8hr_0+r_0^2}{(h-r_0)}\right)\nn\\
& &\quad +\frac{\pi}{90}\drho\rl\epsilon\sigma^{12}
\left(\frac{9h-r_0}{(h-r_0)^9}-\frac{9h+r_0}{(h+r_0)^9}\right)\frac{d_w}{r_0}
+{\cal O}\left(\left(\frac{d_w}{r_0}\right)^2\right),
\ee
\be
\omega^{(s)}_{attr}(l)&=&-\frac{1}{r_0^2}\frac{\pi}{3}
\drho(\rl\epsilon\sigma^6-\resi^6)
\left(2hr_0\frac{h^2+r_0^2}{(h^2-r_0^2)^2}-\ln\frac{h+r_0}{h-r_0}\right)\nn\\
& &\quad-\frac{16\pi}{3}\drho\rl\epsilon\sigma^6
\frac{h^3r_0}{(h^2-r_0^2)^3}\frac{d_w}{r_0}
+{\cal O} \left(\left(\frac{d_w}{r_0}\right)^2\right),
\ee
(see Eqs.~(B2), (B20), and (B22))
\be\lefteqn{
\omega^{(c)}_{rep}(l)=\frac{7\pi^2}{64}
\drho(\rl\epsilon\sigma^{12}-\resi^{12})\frac{r_0}{h^9}\,
{}_2F_1\left(\frac{11}{2},\frac{9}{2};2;\left(\frac{r_0}{h}\right)^2\right)}\nn\\
& &+\frac{7\pi^2}{64}\drho\rl\epsilon\sigma^{12}\frac{r_0}{h^9}
\left(2\,{}_2F_1\left(\frac{11}{2},\frac{9}{2};2;\left(\frac{r_0}{h}\right)^2\right)
+\frac{99}{4}\frac{r_0^2}{h^2}\,
{}_2F_1\left(\frac{13}{2},\frac{11}{2};3;\left(\frac{r_0}{h}\right)^2\right)\right)
\frac{d_w}{r_0}\nn\\
& &+{\cal O}\left(\left(\frac{d_w}{r_0}\right)^2\right),
\ee
and
\be\lefteqn{
\omega^{(c)}_{attr}(l)=-\frac{\pi^2}{2}
\drho(\rl\epsilon\sigma^6-\resi^6)\frac{r_0}{h^3}\,
{}_2F_1\left(\frac{5}{2},\frac{3}{2};2;\left(\frac{r_0}{h}\right)^2\right)}\nn\\
& &-\frac{\pi^2}{2}\drho\rl\epsilon\sigma^6\frac{r_0}{h^3}
\left(2\,{}_2F_1\left(\frac{5}{2},\frac{3}{2};2;\left(\frac{r_0}{h}\right)^2\right)
+\frac{15}{4}\frac{r_0^2}{h^2}\,
{}_2F_1\left(\frac{7}{2},\frac{5}{2};3;\left(\frac{r_0}{h}\right)^2\right)\right)
\frac{d_w}{r_0}\nn\\
& &+{\cal O}\left(\left(\frac{d_w}{r_0}\right)^2\right).
\ee
Here we have assumed that the radius of the substrate is large
compared with the diameter of the fluid particles. 
Equations (2.32) and (2.34) can be compared with other
approaches \cite{philip,parsegian}, in particular with the dispersion theory developed in
Ref.~[34(c)] which considers a system composed of three concentrically
arranged media characterized by different dielectric constants
$\epsilon_i$. The summation of the electromagnetic surface normal modes
yields the dispersion function. Its expansion in powers of the dielectric
reflection coefficient
$\Delta_{ij}=\frac{\epsilon_i-\epsilon_j}{\epsilon_i+\epsilon_j}$
results in a factorization of the dispersion energies into a
frequency-dependent and a geometrical factor. The geometrical factors
(Eqs.~(6.5) and (6.9) in Ref.~[34(c)]) are in full agreement with our
results for the attractive contributions (Eqs.~(2.32) and (2.34)). 
It is reassuring to see that despite of the differences between these
two approaches as discussed in the Introduction both yield the same
functional dependence on the film thickness for large $l$. Furthermore one
should keep in mind that the density functional approach yields a
well-defined expression of $\omega(l)$ even for small values of $l$,
whereas the dispersion theory leads to diverging energies in this
limit. In addition the density functional can capture subdominant
contributions to $\omega(l)$ which arise both from the repulsive parts
of the interaction potentials and from detailed structures of the substrate
determining the wetting behavior at coexistence. This will be
discussed in the following section.

For the model defined by Eqs.~(2.4) and (2.5)
the effective interface potential of the curved substrates,
$\omega^{(\tau)}(l,r_0)=\omega^{(\tau)}_{attr}(l,r_0)
+\omega^{(\tau)}_{rep}(l,r_0)$,
can be related to that of the planar substrate as follows:
\be
\omega^{(\tau)}(l,r_0)=\omega^{(p)}_{attr}(l)
{\cal S}^{(\tau)}_{attr}\left(\frac{r_0}{l}\right)
+\omega^{(p)}_{rep}(l)
{\cal S}^{(\tau)}_{rep}\left(\frac{r_0}{l}\right)
\ee
where the scaling functions
\be
{\cal S}^{(\tau)}_{attr}\left(\frac{r_0}{l}\right)
=\frac{\omega^{(\tau)}_{attr}(l,r_0)}{\omega^{(p)}_{attr}(l)},
\quad
{\cal S}^{(\tau)}_{rep}\left(\frac{r_0}{l}\right)
=\frac{\omega^{(\tau)}_{rep}(l,r_0)}{\omega^{(p)}_{rep}(l)}
\ee
turn out to depend indeed only on the ratio $x=r_0/l$ so that 
\be
\omega^{(\tau)}(l,r_0)=\omega^{(p)}(l)\frac{{\cal S}^{(\tau)}_{attr}
(r_0/l)
+\frac{\omega^{(p)}_{rep}(l)}{\omega^{(p)}_{attr}(l)}
{\cal S}^{(\tau)}_{rep}(r_0/l)}
{1+\frac{\omega^{(p)}_{rep}(l)}{\omega^{(p)}_{attr}(l)}}.
\ee
Since 
$\omega^{(p)}_{attr}(l)=a/l^2$ and 
$\omega^{(p)}_{rep}(l)=c/l^8$
with the Hamaker constant
$a=\pi\drho(\resi^6-\rl\epsilon\sigma^6)/3$ 
and with $c=\pi\drho(\rl\epsilon\sigma^{12}-\resi^{12})/90$,
one has
$\omega^{(p)}_{rep}(l)/\omega^{(p)}_{attr}(l)=c/(al^6)$ 
and thus due to $\omega^{(p)}(l)=\omega^{(p)}_{attr}(l)+\omega^{(p)}_{rep}(l)$
\be
\omega^{(\tau)}(l,r_0)=\omega^{(p)}(l)\left({\cal S}^{(\tau)}_{attr}
+\frac{c}{a}l^{-6}\left({\cal S}^{(\tau)}_{rep}\left(\frac{r_0}{l}\right)
-{\cal S}^{(\tau)}_{attr}\left(\frac{r_0}{l}\right)\right)\right)
+{\cal O}(l^{-12}).
\ee
Therefore for $l\gg\sigma$ 
the effective interface potential attains the scaling form
\be
\omega^{(\tau)}(l,r_0)=\omega^{(p)}(l)
{\cal S}_\tau\left(\frac{r_0}{l}\right),\quad l\gg\sigma,
\ee
where 
${\cal S}_\tau(x)
\equiv{\cal S}^{(\tau)}_{attr}(r_0/l)$.
These scaling functions exhibit the following limiting behavior:
\be
{\cal S}_\tau(x\to\infty)=1+\frac{\tau}{2x}+
{\cal O}\left(x^{-2}\right)
\ee
and
\be
{\cal S}_\tau(x\to 0)=\alpha_\tau x+\beta_\tau x^2 +
{\cal O}\left(x^{3}\right)
\ee
where
$\alpha_c=\frac{3\pi}{2}$, $\beta_c=-\frac{9\pi}{2}$,
$\alpha_s=\frac{16}{3}$, and $\beta_s=-16$.
The full scaling function ${\cal S}_\tau (x)$ is shown in Fig.~3(a).
It exhibits a non-monotonic behavior.

\vskip 1cm

\centerline{\large \bf III. Wetting behavior}

\addtocounter{section}{1}
\addtocounter{subsection}{-3}
\setcounter{equation}{0}

\subsection{Film thickness at coexistence and above the wetting
transition temperature}

We consider a curved substrate-fluid system at a temperature
$T>T_w^{(\tau)}(r_0)$ above its (quasi--)wetting transition 
temperature $T_w^{(\tau)}(r_0)$
and at liquid-vapor coexistence. (Typically one has
$T_w^{(\tau)}(r_0)>T_w^{(\tau)}(\infty)=T_w^{(p)}$, see below).
The curvature prevents the build-up of a macroscopically
thick wetting film as it would  form on the corresponding planar
substrate. If this film thickness is sufficiently large it can be
determined from the leading terms in the effective interface
potential ($a>0$):
\be
\Omega_s(l)\simeq\frac{a}{l^2}\, {\cal S}_\tau\left(\frac{r_0}{l}\right)+
\left(1+\frac{l}{r_0}\right)^\tau\siglp+\sigma_{l,s}(r_0).
\ee
Due to $T>T_w^{(\tau)}$ and $\mu-\mu_0(T)=0^-$ the equilibrium
thickness corresponding to Eq.~(3.1) is the maximum thickness the
film can attain and it is given by 
\be
l_{max}^{(\tau)}=\frac{1}{x_\tau^*(\kappa)}r_0
\ee
where $x^*_\tau(\kappa)$ is the solution of the implicit equation 
\be
x=\kappa\left(\frac{\tau}{2}\right)^{1/3}s_\tau(x)
\ee
with (see Eq.~(2.36) and Fig.~3(b))
\be
s_\tau(x)=(1+x^{-1})^{(\tau-1)/3}\left({\cal S}_\tau(x)+\frac{x}{2}
{\cal S}_\tau'(x)\right)^{-1/3}.
\ee
This shows that the limiting thickness of the wetting film above $T_w$
is governed by the dimensionless parameter
\be
\kappa =\left(\frac{r_0^2\siglp}{a}\right)^{1/3}.
\ee
Due to
$s_\tau(x\to\infty)=1+\left(\frac{\tau}{4}-\frac{1}{3}\right)x^{-1}
+\left(\frac{2}{9}-\frac{\tau}{4}+\frac{\tau^2}{24}\right)x^{-2}
+{\cal O}(x^{-3})$ one finds 
\be
x_\tau^*(\kappa\to\infty)=\left(\frac{\tau}{2}\right)^{1/3}\kappa+
\frac{\tau}{4}-\frac{1}{3}+{\cal O}\left(\kappa^{-1}\right).
\ee
In the limit $x\to 0$ one has 
$s_1(x\to 0)=\left(\frac{4}{9\pi}\right)^{1/3}x^{-1/3}
\left(1+\frac{4}{3}x-\frac{59}{72}x^2
+{\cal O}(x^3)\right)$ and
$s_2(x\to 0)=\eh x^{-2/3}
\left(1+\frac{5}{3}x-\frac{4}{9}x^2
+{\cal O}(x^3)\right)$
so that
\be
x_{\tau=1}^*(\kappa\to 0)=
\left(\frac{2\kappa^3}{9\pi}\right)^{1/4}
+\left(\frac{2\kappa^3}{9\pi}\right)^{1/2}
+{\cal O}(\kappa^{9/4})
\ee
and
\be
x_{\tau=2}^*(\kappa\to 0)=
\left(\frac{\kappa}{2}\right)^{3/5}
+\left(\frac{\kappa}{2}\right)^{6/5}+{\cal O}(\kappa^{9/5}).
\ee
This implies that for a given system, i.~e., $\siglp$ and $a$
fixed and for large $r_0$ the maximum thickness of the wetting film is
given by
\be
l_{max}^{(\tau)}(r_0\to\infty)
=\left(\frac{2ar_0}{\tau\siglp}\right)^{1/3}
=\left(\frac{2a}{\drho\frac{\tau\siglp}{\drho r_0}}\right)^{1/3},
\quad\kappa\gg 1,
\ee
(see also, c.~f., the first paragraph of Subsect.~III.~B)
whereas for a small liquid-vapor surface tension $\siglp$ and a large
Hamaker constant $a$ the maximum thicknesses are (see Eq.~(3.7))
\be
l_{max}^{(c)}\simeq \left(\frac{9\pi ar_0^2}{2\siglp}\right)^{1/4},
\quad\kappa\ll 1,
\ee
and
\be
l_{max}^{(s)}\simeq \left(\frac{8ar_0^3}{\siglp}\right)^{1/5},
\quad\kappa\ll 1.
\ee

For large $\kappa$ $l_{max}$ is larger for a
cylinder than for a sphere by a constant factor $2^{1/3}\simeq 1.26$ 
(Eq.~(3.6)) whereas for small $\kappa$ this ratio diverges weakly as 
$(9\pi)^{1/4}2^{-17/20}\kappa^{-3/20}$ (Eqs.~(3.7) and (3.8)). 
The full dependence of $l_{max}^{(\tau)}/r_0=1/x_\tau^*$ 
on $\kappa$ is shown in Fig.~4.

In principle $\kappa$ can be any positive number. Typically, at low
temperatures, $\siglp$ ranges between $10^{-3}J/m^2$ and
$10^{-2}J/m^2$ \cite{israel}
and $a$ between $10^{-20}J$ and $10^{-19}J$ \cite{israel} so that
$\kappa=\kappa_0\cdot(r_0/\mbox{\AA})^{2/3}$ with $\kappa_0=0.05\ldots 0.20$.
With $\kappa_0=0.05$ according to Fig.~4 this implies
that for $r_0=100\mbox{\AA}$ $l_{max}^{(c)}\simeq 130\mbox{\AA}$ and
$l_{max}^{(s)}\simeq 90\mbox{\AA}$ whereas for $r_0=1\mu m$
$l_{max}^{(c)}\simeq 550\mbox{\AA}$ and $l_{max}^{(s)}\simeq
430\mbox{\AA}$. 
Due to finite
resolutions for pressure and temperature equilibrium film thicknesses
on a planar substrate can be followed up to at most a few hundred
$\mbox{\AA}$, say $500\mbox{\AA}$. In this case and for the above quoted values of
$\siglp$ and $a$ $r_0$ must be smaller than $50\mu m$ for a cylinder
and $100\mu m$ for a sphere in order to cause a deviation from the
corresponding planar film thickness of more than $5\%$.

If for a given system, i.~e. $r_0$ fixed, the temperature is raised
towards $T_c$, $\siglp$ vanishes as $\sigma_0\tilde t^{2\nu}$,
$\nu\simeq 0.632$, where $\tilde t=(T_c-T)/T_c\to 0$, and $a\sim\drho$
as $a_0\tilde t^\beta$, $\beta\simeq0.328$, so that $\kappa$ goes to
zero as $\left(\frac{r_0^2\sigma_0}{a_0}\right)^{1/3} \tilde
t^{(2\nu-\beta)/3} \sim\tilde t^{0.312}$ and Eqs.~(3.7) and (3.8) 
apply. From
these one infers that the maximum film thicknesses diverge as 
$\l_{max}^{(c)}
\simeq(9\pi r_0^2a_0/2\sigma_0)^{1/4}\tilde t^{-(2\nu-\beta)/4}
\sim\tilde t^{-0.234}$ and 
$l_{max}^{(s)}
\simeq(8r_0^3a_0/\sigma_0)^{1/5}\tilde t^{-(2\nu-\beta)/5}
\sim\tilde t^{-0.187}$, 
respectively.
At $T_c$ the critical fluctuations in the bulk lead to a power-law
decay of the density profile towards the bulk density. As function of
the temperature $T$ this mesoscopic interfacial structure known as
critical adsorption emerges from a film whose thickness diverges like
the bulk correlation length 
$\xi=\xi_0^-\tilde t^{-\nu}\simeq\xi_0^-\tilde t^{-0.632}$ 
where $\xi_0^-$ is typically of the order of 1.5\AA. This picture for
a planar substrate \cite{gero} holds also for spherical and
cylindrical substrates provided $\xi$ is large compared with their
radii of curvature \cite{indekeu}. On this basis we expect that ultimately
for $\tilde t\to 0$ both $l_{max}^{(c)}$ and $l_{max}^{(s)}$ should
diverge as $\tilde t^{-\nu}$. Our above results obtained from the
effective interface potential signal such a divergence but the
corresponding exponent is smaller. This difference must be attributed
to our present sharp-kink approximation of a mean-field
theory. Nonetheless one should keep in mind that at least within the
mean-field theory for a planar geometry the leading behavior of the
effective interface potential is captured correctly by the sharp-kink
approximation in spite of the actual broadening of the emerging
liquid-vapor interface close to $T_c$ \cite{napi2}; it is reasonable to
expect that this quality of the sharp-kink approximation carries over
to the case of curved surfaces. However, as the discussion of
Eqs.~(2.46) and (2.47) in Ref.~[55(c)] shows even this quality of the
sharp-kink approximation does not suffice to describe properly the
crossover between complete wetting and critical adsorption. The
crossover condition $l\simeq\xi$ must be inserted by hand. Our above
findings for $l^{(\tau)}_{max}$ must be judged and augmented similarly.

The above analysis relies on a positive value of $a$ and, because we
have considered only the leading terms in $\omega(l)$, on the fact
that $l^{(\tau)}_{max}$ is large. This is fulfilled for $T>T_w^{(\tau)}(r_0)$ where
$T_w^{(\tau)}(r_0)$ is the thin-thick transition temperature in the case of a
(quasi--) first-order wetting transition. If the wetting transition on the
planar substrate is continuous, there is no transition at coexistence
and we consider simply $T>T_w^{(p)}$, i.~e., $a>0$.
The solid-vapor surface tension is the minimum value of $\Omega_s(l)$
so that at coexistence
\be
\sigma_{s,g}=\min_l\Omega_s(l)=\Omega_s(l_0)\leq\frac{A(r_0+l_0)}{A(r_0)}
\sigma_{l,g}(r_0+l_0)+\sils(r_0)
\ee
because $\omega(\infty)=0$ so that $\omega(l_0)\leq 0$ (see
Eq.~(2.10)). One should note that in Eq.~(3.12) the inequality never
turns into an equality if $r_0<\infty$ because $l_0(r_0<\infty)<\infty$ so
that $\omega(l_0(r_0<\infty))<0$. Only in the limit $r_0\to\infty$ one
recovers the standard relation that
$\sigma_{s,g}^{(p)}\leq\siglp+\sigma_{l,s}^{(p)}$ turns into an
equality at $T=T_w^{(p)}$. Since in Eq.~(3.12) the right hand side
$\sigma_{l,g}$ is enhanced by the factor $A(r_0+l_0)/A(r_0)$ as compared
with the planar case one can surmise that $T_w^{(\tau)}(r_0)$ is larger than
$T_w^{(p)}$. However, this argument must be checked more carefully,
because at $T_w^{(\tau)}(r_0)$ the right hand side of Eq.~(3.12) is not
attained. This will be done in Subsec. III.~C.

\subsection{Complete wetting off coexistence and above the wetting
transition temperature}

According to Eq.~(3.9) the thickness $l^{(\tau)}_{max}$ of a wetting film on a
curved substrate at liquid-vapor coexistence $\Delta\mu=0$ is the same as if the
identical system in planar geometry is taken off from liquid-vapor
coexistence with a $\Delta\mu_{eff}(\Delta\mu=0)=\frac{\tau\siglp}{r_0\drho}$. In
this subsection we want to discuss how a wetting film on a curved
substrate attains this maximum thickness $l^{(\tau)}_{max}$ when one approaches
liquid-vapor coexistence, i.~e.~$\dmu\to 0$, above $T_w^{(\tau)}(r_0)$. In this
case we have (see Eqs.~(2.10), (2.12), (2.13), and (3.1)), 
with $a>0$,
\be
\Omega_s(l\gg d_w)&\simeq& \frac{a}{l^2}S_\tau\,\left(\frac{r_0}{l}\right)
+\left(1+\frac{l}{r_0}\right)^\tau\siglp\nn\\
&+&\frac{1}{\tau+1}\left(
\left(1+\frac{l}{r_0}\right)^{\tau+1}-1\right)r_0\drho\dmu
+\sigma_{l,s}(r_0)
\ee
so that
\be
l_0=\frac{1}{y_\tau^*(\kappa,\lambda)}r_0
\ee
where $y_\tau^*(\kappa,\lambda)$ is the solution of the 
implicit equation
\be 
y=\kappa\left(\frac{\tau}{2}\right)^{1/3}u_\tau(y,\lambda)
\ee
with 
\be
u_\tau(y,\lambda)=\left(1+\frac{\lambda}{\tau}
\left(1+\frac{1}{y}\right)\right)^{1/3}s_\tau(y).
\ee
The dimensionless parameter $\lambda$ measures the distance from
liquid-vapor coexistence,
\be
\lambda=\frac{\drho r_0}{\siglp}\dmu,
\ee
so that $y_\tau^*(\kappa,\lambda=0)=x_\tau^*(\kappa)=r_0/l_{max}(\kappa)$.
With 
$u_\tau(y\to\infty,\lambda)=\left(1+\frac{\lambda}{\tau}\right)^{1/3}
+y^{-1}(\tau+\lambda-\frac{4}{3})/4(1+\lambda/\tau)^{2/3}+{\cal O}(y^{-2})$
one has 
\be
y_\tau^*(\kappa\to\infty,\lambda)=
\left(\frac{\tau}{2}\right)^{1/3}\kappa
\left(1+\frac{\lambda}{\tau}\right)^{1/3}+
\frac{\tau+\lambda-\frac{4}{3}}{1+\frac{\lambda}{\tau}}
+{\cal O}\left(\kappa^{-1}\right).
\ee
In the limit $\kappa\to 0$ one finds 
$u_1(y\to 0)=\left(\frac{4}{9\pi}\right)^{1/3}
\left(\frac{\lambda}{y^2}\right)^{1/3}
\left(1+\frac{1+5\lambda}{3}\frac{y}{\lambda}
+{\cal O}\left(\left(\frac{y}{\lambda}\right)^2\right)\right)$ and
$u_2(y\to 0)=\left(\frac{\lambda}{16}\right)^{1/3}\frac{1}{y}
\left(1+\frac{2+6\lambda}{3}\frac{y}{\lambda}
+{\cal O}\left(\left(\frac{y}{\lambda}\right)^2\right)\right)$ so that
\be
y_1^*(\kappa\to 0,\lambda)
=\left(\frac{2\kappa^3\lambda}{9\pi}\right)^{1/5}
+\frac{1+5\lambda}{5\lambda}
\left(\frac{2\kappa^3\lambda}{9\pi}\right)^{2/5}
+{\cal O}(\kappa^{9/5})
\ee
and
\be
y_2^*(\kappa\to 0,\lambda)
=\left(\frac{\kappa^3\lambda}{16}\right)^{1/6}
+\frac{1+3\lambda}{3\lambda}
\left(\frac{\kappa^3\lambda}{16}\right)^{1/3}
+{\cal O}(\kappa^{3/2}).
\ee
In the case $\kappa\gg 1$ Eq.(3.16) implies
\be
l_0(r_0,\dmu)=\frac{l_{max}}{\left(1+\frac{\lambda}{\tau}\right)^{1/3}},
\quad \kappa\gg 1,
\ee
whereas for $\kappa\ll 1$ one has 
\be
l_0^{(c)}(r_0,\dmu)=\left(\frac{9\pi ar_0^2}{2\drho\dmu}\right)^{1/5},
\quad \kappa\ll 1,
\ee
and
\be
l_0^{(s)}(r_0,\dmu)=\left(\frac{16ar_0^3}{\drho\dmu}\right)^{1/6},\quad
\kappa\ll 1.
\ee
Equations (3.16), (3.22), and (3.23) show that the limits 
$\lambda\to 0$ and $\kappa\to 0$ cannot be interchanged.

Since for a small undersaturation $\dmu$ 
the thickness of the complete wetting film 
on the corresponding planar substrate is ($T>T_w^{(p)}$)
\be
l_0(r_0=\infty,\Delta\mu\to 0)=l_0^{(p)}(\Delta\mu\to 0)=
\left(\frac{2a}{\drho\Delta\mu}\right)^{1/3},
\ee
the film thickness on the corresponding curved
substrate can be expressed in terms of an effective
$\dmu_{eff}(\Delta\mu;r_0,\tau)$:
$$
l_0(r_0,\dmu\to 0)=l_0^{(p)}(\dmu_{eff}),\eqno(3.25a)
$$
$$
\dmu_{eff}=\dmu+\delta\mu(\kappa,\lambda).\eqno(3.25b)
$$
\addtocounter{equation}{1}
In the following we determine $\dmu_{eff}$ on the basis of Eq.~(3.13)
which captures the {\em leading} behavior of $\Omega_s(l)$ for large $l$. 
(One should keep in mind that prewetting phenomena depend also on the
behavior of $\Omega_s(l\,\gsim\, d_w)$ so that one cannot expect that the
(quasi--) prewetting line on curved substrates is simply the
prewetting line of the planar substrate, $\dmu_{pw}^{(p)}(T)$, shifted
upwards by $\dmu_{eff}$. Thus Eq.~(3.25a) holds only above the
prewetting line where $l_0$ is sufficiently large. Therefore we concur
with Upton et al.~(Sect.~XI in Ref.~[26(b)] in that the effects of
curvature cannot be fully subsumed into an effective
$\Delta\mu_{eff}$.) 
With Eqs.~(3.14) and (3.17) one has
\be
\lambda_{eff}(\kappa,\lambda)=\frac{r_0\drho\dmu_{eff}}{\siglp}
=2\left(\frac{y^*(\kappa,\lambda)}{\kappa}\right)^3.
\ee
In the limit $\kappa\gg 1$ Eq.~(3.18) applies so that
$\dmu_{eff}$ is given by
\be
\dmu_{eff}(\Delta\mu;r_0,\tau)
=\dmu+\frac{\tau\siglp}{\drho r_0},\quad\kappa\gg 1. 
\ee
Thus in the limit $\kappa\gg 1$ the curved substrates resemble the
same  film thickness as the planar substrate if the liquid-vapor
coexistence curve $\mu_0(T)$ is shifted upwards by
\be
\delta\mu_{\infty}^{(\tau)}=\frac{\tau\siglp}{\drho r_0}=
\frac{\tau\left(\siglp\right)^{3/2}}{\drho a^{1/2}}\kappa^{-3/2},\quad\kappa\gg 1.
\ee
This is in accordance with our previous finding 
$\dmu_{eff}(\dmu=0)=\delta\mu^{(\tau)}_\infty$ (see Eq.~(3.9)).

From Eq.~(3.27) one finds that for spheres ($\tau=2$) the effect of the
curvature is twice as large as for cylinders ($\tau=1$). It should be
emphasized that the shift $\delta\mu_{\infty}^{(\tau)}$
is independent of the Hamaker constant $a$
and depends on surface properties only via the planar liquid-vapor surface
tension.

Equation (3.27) holds only in the limit $\kappa=(\siglp r_0^2/a)^{1/3}\gg 1$,
i.~e., for $r_0\to\infty$ and $T$ fixed. In the opposite case, i.~e.,
for $T\to T_c$ and $r_0$ fixed one has $\kappa\sim\tilde
t^{(2\nu-\beta)/3}\to 0$ so that the limit $\kappa\ll 1$ must be
considered (see Eqs.~(3.19) and (3.20)). 
These latter equations are valid provided that (i) $y^*\ll 1$ and (ii)
$\left(1+1/y^*\right)\lambda/\tau\gg 1$ (compare Eq.~(3.16)). Condition (i)
is fulfilled if $\kappa\ll (9\pi/2\lambda)^{1/3}$ for a cylinder and if
$\kappa\ll 2^{4/3}/\lambda^{1/3}$ for a sphere, respectively. Condition (ii) is
always fulfilled for $\lambda>\tau$; for $\lambda<\tau$ it
requires $y^*\ll \lambda/(\tau-\lambda)$, i.~e., 
$\kappa\ll (9\pi/2)^{1/3}\lambda^{4/3}/(1-\lambda)^{5/3}$ for a
cylinder and $\kappa\ll 2^{4/3}\lambda^{5/3}/(2-\lambda)^2$ for a
sphere, respectively. In this regime, denoted as I (see Figs.~6(a) and (b)),
$\lambda_{eff}$ is given by
\be
\lambda_{eff}^{(c)}\simeq
2\left(\frac{2\lambda}{9\pi}\right)^{3/5}\kappa^{-6/5},
\quad\mbox{regime I},
\ee
and
\be
\lambda_{eff}^{(s)}\simeq
\left(\frac{\lambda}{4}\right)^{1/2}\kappa^{-3/2},
\quad\mbox{regime I}.
\ee
On the other hand according to Eq.~(3.18) 
the limit $y^*\gg 1$ applies if $\kappa\gg
2^{1/3}/(\tau+\lambda)^{1/3}$ (regime II); in this case Eq.~(3.26)
yields
\be
\lambda_{eff}^{(\tau)}\simeq\tau+\lambda,
\quad\mbox{regime II},
\ee
which is equivalent to Eq.~(3.27).
In the intermediate regime III, i.~e., $1\gg y^*\gg  \lambda/(\tau-\lambda)$,
the asymptotic formulae in Eqs.~(3.19) and
(3.20) do not hold. Instead of expanding $y^*$ in powers of $\kappa$,
in this case
one has to evaluate Eqs.~(3.15) and (3.16) for small values of
$\lambda$ first and then to consider the limit $\kappa\to 0$:
\be
y^*_\tau\left(\kappa,\lambda\to 0\right)\simeq
\kappa\left(\frac{\tau}{2}\right)^{1/3}
\left(1+\frac{\lambda}{3\tau}\left(1+\frac{1}{y}\right)\right)s_\tau(y).
\ee
With $s_\tau(y)$ taken in the limit $y\to 0$ one finds in this regime III
\be
y_1^*\simeq
\left(\frac{2\kappa^3}{9\pi}\right)^{1/4}
\left(1+\frac{\lambda}{4}\left(\left(\frac{2\kappa^3}{9\pi}\right)^{-1/4}
+2+{\cal O}\left(\kappa^{3/4}\right)\right)
+{\cal O}\left(\lambda^2\right)\right)
\ee
and 
\be
y_2^*\simeq
\left(\frac{\kappa}{2}\right)^{3/5}
\left(1+\frac{\lambda}{10}\left(\left(\frac{\kappa}{2}\right)^{-3/5}
+2+{\cal O}\left(\kappa^{3/5}\right)\right)
+{\cal O}\left(\lambda^2\right)\right),
\ee
respectively, so that
\be
\lambda_{eff}^{(c)}&\simeq&
2\left(\frac{2}{9\pi\kappa}\right)^{3/4}
\left(1+\frac{3}{4}\lambda\left(\left(\frac{2\kappa^3}{9\pi}\right)^{-1/4}
+4+{\cal O}\left(\kappa^{3/4}\right)\right)+{\cal
O}\left(\lambda^2\right)\right),\nn\\
&&\mbox{regime III},
\ee
and
\be
\lambda_{eff}^{(s)}&\simeq&
\left(\frac{1}{16\kappa^6}\right)^{1/5}
\left(1+\frac{3}{10}\lambda\left(\left(\frac{\kappa}{2}\right)^{-3/5}+4
+{\cal O}\left(\kappa^{3/5}\right)\right)+{\cal
O}\left(\lambda^2\right)\right),\nn\\
&&\mbox{regime III}.
\ee
Figures 7(a) and (b) visualize this crossover behavior of $\lambda_{eff}$
as function of $\kappa$ for $\lambda$  fixed in the case of a cylinder
and a sphere, respectively.

\subsection{First-order wetting transitions}

The results of the previous subsection allow us to discuss the effect
of curvature on a wetting transition which is first order on a  planar
substrate. 
For suitably chosen interaction potentials 
Fig.~8 shows such a phase diagram 
and the temperature dependence of the corresponding film thickness at
coexistence for  a cylinder and a sphere, respectively, as
compared to those of the corresponding  planar substrate.
In accordance with 
Refs.~\cite{hauge,night} we find for the planar substrate that
the prewetting line $\mu_{pw}^{(p)}(T)$ joins the gas-liquid coexistence
curve $\mu_0(T)$ tangentially,
\be
\mu^{(p)}_{pw}\left(T\to T_w^{(p)}\right)-\mu_0(T)
=k_\mu|T-T_w^{(p)}|^{\gamma(\zeta)},
\quad\gamma(\zeta)=\frac{\zeta}{\zeta -1},
\ee
if for large distances the pair interaction potentials in the system decay like
$r^{-(3+\zeta)}$ \cite{hauge}. 
Upon crossing the prewetting line the wetting film thickness undergoes
a thin-thick transition where the corresponding jump 
$\Delta l(T)$ diverges at the first-order transition temperature
$T_w^{(p)}$ as
\be
\Delta l\left(T\to T_w^{(p)}\right)=k_l|T-T_w|^{-\gamma(\zeta)/\zeta}.
\ee
In Subsects.~III.A and III.B we focused on the region $T>T_w^{(p)}$ of
the phase diagram where the behavior of the film thickness is
independent of the order of the wetting transition.
Since, however, in the region close to $T_w^{(p)}$ the order of the
transition plays an essential role, the present and the following
subsection are dedicated separately to 
first- and second-order transitions, respectively. 
A first-order wetting transition is accompanied by a
prewetting line which implies the occurrence of discontinuities in the
film thickness also off coexistence. In the region between the
prewetting line and the coexistence curve, i.~e., $T>T_w^{(p)}$ and
$\dmu\leq \dmu_{pw}^{(p)}(T)$ (see Subsect.~III.B), the film thickness
on the curved substrates and on the corresponding planar substrate are
related to each other via
$\dmu_{eff}$. In the limit of large radii, $\kappa\gg 1$,
this relationship is given by the shift $\dmu_{eff}=\dmu+\delta\mu$.
Thus one is inclined to surmise that for $\kappa\gg 1$ also the prewetting
line of the curved substrates can be obtained by shifting the
prewetting line of the corresponding planar substrate upwards in the
phase diagram by the amount $\delta\mu$. In this case the
knowledge of the location of the prewetting lines relative to each
other yields information about the shift of the transition temperature,
$T_w^{(\tau)}(r_0)-T_w^{(p)}$, and the jump in the film thickness
$\Delta l^{(\tau)}$ on the curved substrates.
Since Eq.~(3.37) holds
especially for $T=T_w^{(\tau)}(\kappa\to\infty)$, one has with
$\mu_{pw}^{(p)}(T_w^{(\tau)}(\kappa\to\infty)\to T_w^{(p)})-
\mu_0(T_w^{(\tau)}(\kappa\to\infty))
=\delta\mu_\infty^{(\tau)}$,
\be
T_w^{(\tau)}(\kappa\to\infty)-T_w^{(p)}=\left(\frac{1}{k_\mu}
\delta\mu_\infty^{(\tau)}\right)^{1/\gamma(\zeta)} 
\ee
so that for the jump in thickness at coexistence on the
curved substrate one has
\be
\Delta l^{(\tau)}\left(T_w^{(\tau)}(\kappa\to\infty)\right)
=k_l\left(\frac{1}{k_\mu}
\delta\mu_\infty^{(\tau)}\right)^{-1/\zeta}.
\ee
Thus $\zeta=3$ implies
$T_w^{(\tau)}(\kappa\to\infty)-T_w^{(p)}=(3\pi\tau/(4k_\mu))^{3/2}
(\drho/r_0)^{3/2}$ and 
$\Delta l^{(\tau)}(T_w^{(\tau)}(\kappa\to\infty)) = 
k_l(4 k_\mu/(3\pi\tau))^{1/3} (r_0/\drho)^{1/3}$.

However, the results of Subsect.~III.B hold only for large $l$ whereas
the study of the prewetting line involves also the behavior of
$\Omega_s(l)$ for $l\,\gsim\, d_w$. This requires to test the above
considerations numerically which takes into account the full function
$\Omega_s(l)$. 
Therefore we have calculated the prewetting lines for the planar,
spherical, and cylindrical geometries
for fixed fluid-fluid and fluid-substrate 
interaction potential parameters and for
different radii. The analysis of the dependence of the jump in the
film thickness and of the shift of the transition temperature yields 
$(T_w^{(\tau)}(r_0\to\infty)-T_w^{(p)})/T_w^{(p)}\sim r_0^{-2/3}$
and $\Delta l^{(\tau)}(r_0\to\infty)\sim r_0^{1/3}$ which confirms the
above heuristic considerations.
However, the inspection of the prewetting lines in Fig.~8(a)
demonstrates that the phase diagrams of the curved systems do not
correspond to a simple shift of the phase diagram of the planar
system. 

Since 
$\kappa=\left(9(\frac{r_0}{\sigma})^2\drho/(4(\rw(\frac{\ew}{\epsilon})
(\frac{\sw}{\sigma})^6-\rl))\right)^{1/3}$,
the limit $\kappa\ll 1$ requires (i) a small radius
of the substrate, (ii) large substrate potential parameters
$\rw\ew\sw^6$ in comparison to $\rl$, or (iii) $T$ close to
$T_c$. Simultaneously the parameter space of $\rw$, $\ew$, and $\sw$
is restricted by the requirement to have a first-order wetting
transition in the temperature interval between 
the triple point temperature $T_3$ and $T_c$. 
Within the framework of our numerical analysis a set of parameters which
fulfills these conditions has not been found. Thus a numerical
analysis of the case $\kappa\ll 1$ would require to enlarge the
parameter space considered here which we have not pursued.

\subsection{Critical wetting transitions at coexistence}

Up to now we have considered the case that the Hamaker constant $a$ of
the planar system (Eq.~(2.27)) is positive. 
This is always valid for $T>T_w^{(p)}$.
If the wetting transition is first order this holds even for
$T=T_w^{(p)}$. In the latter case the description of the system for $T<T_w^{(p)}$
requires one to consider the full effective interface potential and
not only its leading behavior for large film thicknesses (Eq.~(2.27)).

If, however, the substrate potential is sufficiently weak, a favorable
choice of interaction potentials allows for the occurrence of a
continuous, so-called critical wetting transition such that $a$ 
changes sign at $T_w^{(p)}$ and
the coefficient $b$ of the next-to-leading order term in $\omega(l)$ is
positive at that temperature \cite{dietrich} (Eq.~(2.27)):
\be
a(T\lessgtr T_w)\lessgtr 0,\quad a(T=T_w)=0,\quad b(T_w)>0
\quad\mbox{(critical wetting)}.
\ee
For critical wetting there is no prewetting
line. 
Recently such a continuous wetting transition at coexistence has been
observed experimentally \cite{bonn}.
This implies that on a curved substrate any wetting
transitions, which are continuous in the planar geometry, are wiped
out. Due to the effective upward shift of the bulk
phase diagram caused by curvature (see Subsect.~III.B), 
the wetting behavior on a curved
substrate upon raising the temperature along the liquid-vapor
coexistence curve $\mu=\mu_0(T)-0$ is that of the corresponding
planar substrate if in the latter case one passes the critical 
wetting transition at $T_w^{(p)}$
with a constant undersaturation $\delta\mu^{(\tau)}_\infty$ 
parallel and below
$\mu_0(T)$ (Eq.~(3.28)). This results in a steep but smooth increase of the
thickness of the wetting film on a curved substrate upon passing $T_w^{(p)}$
at $\mu_0(T)$ and a levelling off at a finite value for $T>T_w^{(p)}$ 
(see Fig.~9) for
which the previous considerations (Subsects.~III.B and C) hold. Since,
however, these previous results have been derived for the case
$a>0$, this analysis must be refined in
order to cover the case that $a$ changes sign at $T_w$.

To this end we first consider the circumstances under which 
critical wetting transitions can occur. As mentioned 
above, the substrate potential must be sufficiently weak so that
$a(T)<0$ for $T$ close to the triple point. However, for pure
Lennard-Jones interactions (see Eqs.~(2.4) and (2.5)), it is
well known that the excluded volume effect near the substrate 
(i.~e., $d_w>0$) leads to $b(T_w)<0$ and thus induces a first-order 
wetting transition \cite{dietrich,napi2}. One can fulfill 
the necessary condition $b(T_w)>0$ (Eq.~(3.45)) by coating the
substrate with an appropriate thin film consisting of a different 
material which leaves the {\em leading}
asymptotic behavior of the substrate potential and thus $a$ unchanged but
introduces a suitable subdominant contribution which can
overcompensate  the
effect of the excluded volume within the effective interface
potential.

Therefore we consider a 
planar substrate filling the half-space $z\leq 0$ 
and consisting of two types of particles which are located in the
region $z\leq -\delta$ and in the film $-\delta<z\leq0$ and which
interact with the fluid particles according
to Eq.~(2.5) with interaction potential parameters
$\{\swnull, \ewnull\}$ and $\{\sweins,
\eweins\}$, respectively.
With the simple assumption that the two species exhibit the same
number density $\rw$ this leads to a substrate potential
(Eq.~(2.28)) $V^{(p)}(z)=\rw v^{(p)}(z)$ with 
\be
v^{(p)}(z\to\infty)=-\left[\frac{u_3}{z^3}+\frac{u_4}{z^4}+\ldots\right]
\ee
where 
\be
u_3=\frac{2\pi}{3}\ewnull\left(\swnull\right)^6,\quad u_4=2\pi\delta
\left(\eweins(\sweins)^6-\ewnull(\swnull)^6\right).
\ee
(This expression for $u_4$ does not yet include contributions due to a
discrete arrangement of the substrate atoms on a lattice; in this
respect see Ref.~\cite{dietrich}.) Thus with (see Eqs.~(2.4) and (2.15))
\be
t^{(p)}(z\to\infty)=-\left[\frac{t_3}{z^3}+{\cal O}(z^{-5})\right]
\ee
and
\be
t_3=\frac{2\pi}{3}\epsilon\sigma^6,\quad t_4=0,
\ee
one has
\be
a&=&\eh\drho\left(u_3\rw-t_3\rl\right)\\
b&=&\frac{1}{3}\drho\left(u_4\rw-3d_wt_3\rl\right).
\ee
Thus $b>0$ is realized for a sufficient large value of $u_4$ which is,
e.~g., the case if the coating of the
substrate consists of particles whose interaction with the fluid
particles is stronger than that of the particles in the main part of
the substrate beneath.

Analogously, for curved substrates we consider coated spheres and
cylinders which consist of particles characterized by interaction
potential parameters $\{\ewnull,\swnull\}$ and $\{\eweins,\sweins\}$
for $0\leq r\leq r_0-\delta$ and $r_0-\delta<r\leq r_0$,
respectively. Based on the results of Appendix C Eq.~(2.35) can be
generalized in order to be able to describe continuous wetting transitions:
\be 
\omega^{(\tau)}(l,r_0)=\frac{a}{l^2}{\cal
S}_\tau\left(\frac{r_0}{l}\right) +\frac{b}{l^3}{\cal
R}_\tau\left(\frac{r_0}{l}\right) +{\cal O}\left(l^{-4}\right)
\ee
where the scaling function ${\cal R}_\tau(x)$ (see Fig.~10 and
Eqs.~(C4)-(C7)) is given by ($x=r_0/l$) 
\be
{\cal R}_\tau(x)=\lim_{l\to\infty}\left[\frac{l^3}{b}
\left(\omega^{(\tau)}(l,r_0)-\frac{a}{l^2}{\cal S}_\tau(x)\right)\right].
\ee
For large $x$,  ${\cal R}_\tau(x)$ takes on the form
\be
{\cal R}_\tau(x\to\infty)=1+\frac{3\tau}{4x}+{\cal O}(x^{-2})
\ee
whereas for $x\to 0$ one has
\be
{\cal R}_c(x\to 0)=\frac{3\pi}{2}\left(1-3x+{\cal O}(x^2)\right)
\ee
and
\be
{\cal R}_s(x\to 0)=8\left(1-3x+{\cal O}(x^2)\right).
\ee 
The minimization of the corresponding effective interface potential
\be
\Omega_s(l)=\frac{a}{l^2}{\cal S}_\tau\left(\frac{r_0}{l}\right)
+\frac{b}{l^3}{\cal
R}_\tau\left(\frac{r_0}{l}\right)+\siglp\left(1+\frac{l}{r_0}\right)^\tau
+\sils
\ee
yields for the equilibrium film thickness the implicit equation (note
that here $\kappa\leq 0$ is possible)
\be
x=\kappa\left(\frac{\tau}{2}\right)^{1/3}r_\tau(x,\gamma)
\ee
with
\be
r_\tau(x,\gamma)=
\left(1+\gamma x q_\tau(x)\right)^{-1/3}
s_\tau(x)
\ee
and
\be
q_\tau(x)=
\frac{{\cal R}_\tau(x)+\frac{x}{3}{\cal R}'_\tau(x)}
{{\cal S}_\tau(x)+\frac{x}{2}{\cal S}'_\tau(x)}.
\ee
The functions $s_\tau(x)$ and $q_\tau(x)$ depend only on the
geometrical properties of the substrate whereas the fluid-fluid and
fluid-substrate interaction potential parameters enter in form of 
the dimensionless numbers $\kappa=\left(r_0^2\siglp/a\right)^{1/3}$
and $\gamma=3b/\left(2ar_0\right)$. For the explicit forms of the
functions $q_\tau(x)$ see Eqs.~(C6) and (C7). The knowledge of the
behavior of $q_\tau(x)$ and $s_\tau(x)$ in the limits $x\to 0$
and $x\to\infty$ allows one to solve Eq.~(3.54) approximately for
specific regions of the ($\kappa$, $\gamma$) parameter space
(see Fig.~11).
Since, however, one has $x=r_0/l\geq 0$, only the regions ($\kappa<0$,
$\gamma<0$) (for $T\geq T_w^{(p)}$) and ($\kappa\geq 0$, $\gamma\geq 0$)
(for $T\geq T_w^{(p)}$) are of interest. 

With
$q_\tau(x\to\infty)=1+\frac{\tau}{4}x^{-1}+\frac{\tau^2}{16}x^{-2}
+{\cal O}\left(x^{-3}\right)$ and
$s_\tau(x\to\infty)=1+\left(\frac{\tau}{4}-\frac{1}{3}\right)x^{-1}
+\left(\frac{2}{9}-\frac{\tau}{4}+\frac{\tau^2}{24}\right)x^{-2}
+{\cal O}\left(x^{-3}\right)$
one has for $|\gamma|\gg\frac{1}{x}$:
\be
r_\tau(x\to\infty,\left|\gamma\right|\gg \frac{1}{x})=\frac{1}{(\gamma
x)^{1/3}}\left(1+\frac{1}{\gamma
x}\frac{\gamma\tau-2\gamma-2}{6}+{\cal O}\left(\left(\gamma x^2\right)^{-1}\right)
\right)
\ee
so that the solution $x^*_\tau(\kappa,\gamma)$ of Eq.~(3.54) reads
(note that $\kappa^3/\gamma$ is always positive)
\be
x^*_\tau(|\kappa|\to\infty,|\gamma|\to\infty)
=\left(\frac{\tau}{2}\frac{\kappa^3}{\gamma}\right)^{1/4}-
\frac{1}{4\gamma}\left(1+\frac{\gamma}{2}(2-\tau)\right)
+{\cal O}\left(\left(\kappa\gamma\right)^{-1}\right).
\ee
Thus in the limit of fixed but large $r_0$ and $a\to 0$
one finds for the film thickness $l_0^{(\tau)}=r_0/x^*_\tau$ 
($b$ must be positive for critical wetting to occur)
\be
l_0^{(\tau)}=\left(\frac{3}{\tau}\frac{br_0}{\siglp}\right)^{1/4}
=\left(\frac{3b}{\drho\delta\mu_\infty^{(\tau)}}\right)^{1/4},
\quad T=T_w^{(p)},\quad r_0\gg\left(\frac{3b}{2\siglp}\right)^{1/3},
\ee
where $\delta\mu_\infty^{(\tau)}$ is defined by Eq.~(3.28).
For the corresponding planar substrate the film thickness along a
complete wetting path $\dmu\to 0$ approaching the transition
temperature $T=T_w^{(p)}$ of the second order wetting
transition is given by 
$l_0(r_0=\infty,\dmu,T=T_w^{(p)})=(\frac{3b}{\dmu\drho})^{1/4}
=l_0^{(p)}(T=T_w^{(p)},\dmu)$ due to $a(T=T_w^{(p)})=0$ (see Eq.~(2.27)). 
Thus in the limit considered in Eq.~(3.58) the film thickness at
coexistence on a curved substrate equals the film thickness of the
corresponding planar system which is off coexistence at
$\dmu=\delta\mu_\infty^{(\tau)}$. This implies that the picture, which
has emerged in the previous subsection for $a>0$, carries over to the
case $a=0$. 

For $x\to\infty$, $|\gamma|\ll\frac{1}{x}$ the sign of $\gamma$ has to
be considered. If $\gamma$ is positive, i.~e., $T>T_w^{(p)}$, one can
expand $r_\tau(x)$ in powers of $\gamma x$ as before, which leads to 
\be
r_\tau(x\to\infty,\gamma x\ll 1)&=&
1+\frac{3\tau-4}{12}x^{-1}
+\frac{\gamma x}{3}\left(1+\frac{3\tau-2}{6}x^{-1}
+{\cal O}\left(x^{-2}\right)\right)\nn\\
&+&{\cal O}\left(x^{-2},\gamma^2x^2\right)
\ee
so that one finds for the solution of Eq.~(3.54)
\be
&&x^*_\tau(\kappa\to+\infty,\gamma\to 0^+)
=\kappa\left(\frac{\tau}{2}\right)^{1/3}
+\frac{3\tau-4}{12}\nn\\
&&-\frac{\gamma\kappa}{3}\left(\frac{\tau}{2}\right)^{1/3}
\left(\kappa\left(\frac{\tau}{2}\right)^{1/3}
+\frac{3\tau-2}{6}+{\cal O}\left(\kappa^{-1}\right)\right)
+{\cal O}\left(\kappa^{-1},\gamma^2\kappa^3\right),\quad\gamma\kappa\ll 1.
\ee
Hence for large $r_0$ and $b/a\to 0$ 
the film thickness attains its asymptotic value given by 
Eq.~(3.9) as
\be
l_0^{(\tau)}(T)&=&\frac{r_0}{\kappa}\left(\frac{2}{\tau}\right)^{1/3}
\left[1+\frac{\gamma}{3}\left(\left(\frac{\tau}{2}\right)^{1/3}\kappa
+\frac{1}{3}\right)+{\cal O}
\left(\kappa^{-1},\gamma^2\right)\right]\nn\\
&&T>T_w^{(p)},\quad r_0\gg\frac{3b}{2a}>0.
\ee
For $\gamma\to 0^-$, $\kappa\to-\infty$, i.~e., for $T<T_w^{(p)}$,
$r_0\to\infty$, and $a$ fixed, the film thickness is determined by the
equation
\be
x^3+\gamma x^4=\frac{\tau}{2}\kappa^3.
\ee
Since $x^3+\gamma x^4$ has zeros at $x=0$ and $x=-1/\gamma$ and a
maximum at $x=-3/(4\gamma)$, Eq.~(3.63) implies $x^*>-1/\gamma\equiv
r_0/l_0^{(p)}$ and thus $l_0^{(p)}>l_0^{(\tau)}$. With the ansatz
$x^*=-1/\gamma +\delta x$ Eq.~(3.63) leads to 
\be
x^*=-\frac{1}{\gamma}\left(1+\frac{\tau}{2}\gamma^3\kappa^3
+{\cal O}\left(\left(\gamma\kappa\right)^6\right)\right),
\quad\gamma\kappa\ll 1,
\ee
so that one has for the film thickness
\be
l_0^{(\tau)}(T)&=& l_0^{(p)}\left(T,\dmu=0\right)
\left(1-\frac{27\tau}{16}\frac{b^3\siglp}{r_0 a^4}
\right),\nn\\
&&T<T_w^{(p)},\quad r_0\gg\left(\frac{3b}{2\siglp}\right)^{1/3}.
\ee

So far we have discussed the limit $x\to\infty$. For $x\to 0$ one has
$s_c(x\to 0)=\left(\frac{4}{9\pi}\right)\frac{1}{x^{1/3}}
\left(1+\frac{4}{3}x-\frac{59}{72}x^2+{\cal O}(x^3)\right)$,
$s_s(x\to 0)=\frac{1}{2x^{2/3}}\left(1+\frac{5}{3}x-\frac{4}{9}x^2
+{\cal O}(x^3)\right)$ and 
$q_c(x\to 0)=\frac{2}{3x}\left(1+\frac{25}{8}x^2
+{\cal O}(x^3)\right)$,
$q_s(x\to 0)=\frac{1}{x}\left(1+2x^2+{\cal O}(x^3)\right)$.
Thus $r_\tau(x\to 0,\gamma)$ is given by
$r_c(x\to 0,\gamma)=
\left(\frac{4}{9\pi}\right)^{1/3}\left(1+\frac{2}{3}\gamma\right)^{-1/3}
\frac{1}{x^{1/3}}\left(1+\frac{4}{3}x\right)$
and $r_s(x\to 0,\gamma)=\eh\left(1+\gamma\right)^{-1/3}
\frac{1}{x^{2/3}}\left(1+\frac{5}{3}x\right)$
which leads to
\be
x^*_c(|\kappa|\ll 1,\gamma)=\left(\frac{2}{9\pi}
\frac{\kappa^3}{1+\frac{2}{3}\gamma}\right)^{1/4}
+\left(\frac{2}{9\pi}\frac{\kappa^3}
{1+\frac{2}{3}\gamma}\right)^{1/2}
+{\cal O}\left(\left(
\frac{\kappa^3}{1+\frac{2}{3}\gamma}\right)^{3/4}\right)
\ee
for a cylinder ($\tau=1$) and
\be
x^*_s(|\kappa|\ll 1,\gamma)=\left(\frac{1}{8}
\frac{\kappa^3}{1+\gamma}\right)^{1/5}
+\left(\frac{1}{8}
\frac{\kappa^3}{1+\gamma}\right)^{2/5}+
{\cal O}\left(\left(
\frac{\kappa^3}{1+\gamma}\right)^{3/5}\right) 
\ee
for a sphere ($\tau=2$), respectively. 
Obviously for $a<0$ there exists a physically meaningful solution only
if $\gamma<-\frac{4-\tau}{2}$.
The film thicknesses are thus given by
\be
l_{0}^{(c)}=\left(\frac{9\pi}{2}
\frac{a r_0^2}{\siglp}
\left(1+\frac{b}{ar_0}\right)\right)^{1/4},\quad 
r_0\ll\left(\frac{|a|}{\siglp}\right)^{1/2}
\ee
and 
\be
l_{0}^{(s)}=\left(8\frac{ar_0^3}{\siglp}
\left(1+\frac{3b}{2ar_0}\right)\right)^{1/5},\quad
r_0\ll\left(\frac{|a|}{\siglp}\right)^{1/2}.
\ee
Upon crossing $T_w^{(p)}$ the parameter $\gamma$ jumps from $-\infty$
at $T_w^{(p)}-0$ to $+\infty$ at $T_w^{(p)}+0$ because at $T=T_w^{(p)}$
the Hamaker constant $a$ changes sign.
At $T_w^{(p)}$ $x^*_\tau$ depends only on the parameter
$\kappa^3/\gamma=2r_0^3\siglp/(3b)$ which is positive both above and
below $T_w^{(p)}$: 
\be
&&x_c^*(|\kappa|\ll 1,|\gamma|\to\infty)
=\left(\frac{\kappa^3}{3\pi\gamma}\right)^{1/4}
+\left(\frac{\kappa^3}{3\pi\gamma}\right)^{1/2}
+{\cal O}\left(\left(\frac{\kappa^3}{\gamma}\right)^{3/4}\right),\nn\\    
&&\frac{\kappa^3}{\gamma}\ll 1,
\ee
and 
\be
&&x_s^*(|\kappa|\ll 1,|\gamma|\to\infty)
=\left(\frac{\kappa^3}{8\gamma}\right)^{1/5}
+\left(\frac{\kappa^3}{8\gamma}\right)^{2/5}
+{\cal O}\left(\left(\frac{\kappa^3}{\gamma}\right)^{3/5}\right),\nn\\
&&\quad\frac{\kappa^3}{\gamma}\ll 1,
\ee
so that one obtains for the film thicknesses
at $T_w^{(p)}$ the results
\be
l_{0}^{(c)}=\left(\frac{9\pi b r_0}{2\siglp}\right)^{1/4},
\quad T=T_w^{(p)},\quad r_0\ll\left(\frac{3b}{2\siglp}\right)^{1/3},
\ee
and 
\be
l_{0}^{(s)}=\left(\frac{12 r_0 b}{\siglp}\right)^{1/5},
\quad T=T_w^{(p)},\quad r_0\ll\left(\frac{3b}{2\siglp}\right)^{1/3}.
\ee
In the limit $\gamma\to 0^+$, i.~e., $r_0$ fixed but small and 
$b/a\to 0$, one recovers the first-order results
given by Eqs.~(3.10) and (3.11), respectively:
\be
&&l_0^{(c)}(|\kappa|\ll 1,\gamma\to 0^+)=
r_0\left(\frac{9\pi}{2\kappa^3}\right)^{1/4}
\left[\left(1+\frac{1}{6}\gamma+{\cal O}\left(\gamma^2\right)\right)
+{\cal O}\left(\kappa^{3/4}\right)\right]\nn\\
&&T>T_w^{(p)},\quad 0<\frac{b}{a}\ll\frac{2r_0}{3}
\ee
and
\be
&&l_0^{(s)}(|\kappa|\ll 1,\gamma\to 0^+)=
r_0\left(\frac{2}{\kappa}\right)^{3/5}
\left[\left(1+\frac{1}{5}\gamma+{\cal O}\left(\gamma^2\right)\right)
+{\cal O}\left(\kappa^{3/5}\right)\right]\nn\\
&&T>T_w^{(p)},\quad 0<\frac{b}{a}\ll\frac{2r_0}{3}.
\ee
For those regions of the $(\kappa,\gamma)$ parameter space which are not
covered by the above considerations of the limits $x\to 0$ and
$x\to\infty$ the solution of Eq.~(3.54) requires a numerical
treatment of Eqs.~(3.54) - (3.65) using the full expressions for the
functions $q_\tau(x)$ and $s_\tau(x)$.

\vskip 1cm

\centerline{\large \bf IV. Finite-size effects}
\vskip 1cm

\addtocounter{section}{1}
\setcounter{equation}{0}

In Sect.~III  we have shown that the positive curvature of a substrate
prevents the formation of infinitely thick wetting films. At
coexistence a critical wetting transition is smeared out whereas a
first-order wetting transition is reduced to a discontinuos transition
between a thin and a thick, but finite, film. However, the latter
nonanalytic behavior as well as the discontinuities associated with
the corresponding prewetting line are artefacts of the mean-field
theory we have used. Since the surface area of spheres is
quasi-zerodimensional and that of cylinders quasi-onedimensional,
fluctuations erase any phase transition and lead to a rounding of the
above mentioned discontinuities. With increasing radii of the spheres
and cylinders this rounding sharpens and a first-order transition
emerges gradually. In this sense our mean-field results become
increasingly more accurate as the size of the spherical and
cylindrical substrates becomes larger. 

Based on the results of Privman and Fisher \cite{privman} for the
influence of finite-size effects  on first-order transitions we are
able to estimate the rounding of the first-order wetting
transitions described above (see also Refs.~\cite{lipowsky} and
\cite{indekeu}). 
In Ref.~\cite{privman} the scaling functions for the singular part of the
free energy and the magnetization in an external field of a finite
Ising spin lattice, 
i.~e., a stripe and a block geometry, are derived. The shape of these
functions allows one to determine the region over which the
phase transition is smeared out. 
In order to translate the results of Ref.~\cite{privman} 
onto the case of the prewetting transition 
we analyze the 
effective interface potential $\Omega_s(l)$ given by Eq.~(2.10). In
the vicinity of the prewetting line $\mupret$ $\Omega_s(l)$ exhibits two
local minima, $l_<(T,\mu)$ and $l_>(T,\mu)$ with the corresponding values 
$\Omega_s\left(l_<(T,\mu)\right)=:\Omega_s^<(T,\mu)$ and 
$\Omega_s\left(l_>(T,\mu)\right)=:\Omega_s^>(T,\mu)$ of the effective
interface potential. One of the two thicknesses $l_<(T,\mu)$ 
and $l_>(T,\mu)$ corresponds to the
global minimum which represents the equilibrium film thickness
$l^{(0)}(T,\mu)$ at the point $(T,\mu)$ of the phase diagram. On the
prewetting line one has
$\Omega_s^<(T,\mupret)=\Omega_s^>(T,\mupret))$ so that the thin film
($l_<^{(0)}(T,\mupret)$) and the thick film ($l_>^{(0)}(T,\mupret)$)
coexist. We now consider the difference
$\Delta\Omega_s(T,\mu)=\Omega_s^>(T,\mu)-\Omega_s^<(T,\mu)$ between
the two minimum values for a given temperature $T$ as a function of
the chemical potential $\mu$. According to Ref.~\cite{privman}   
the width $\delta\mu(T)$ of the rounding region for a 
spherical substrate is given implicitly by the inequality 
\be
\left|\Delta\Omega_s\left(T,\mu=\mupret+\delta\mu(T)\right)\right|
r_0^2\quad\lsim\quad
 k_BT.
\ee
On a cylinder along its axis the liquidlike film is expected to break
up into domains 
with small ($l_<^{(0)}$) and large ($l_>^{(0)}$) film thickness,
respectively. The corresponding kinks at the domain boundaries are
accompanied by an 
additional free energy per unit length, i.~e., a line tension
$\Sigma_l$. Following Ref.~\cite{indekeu} we estimate $\Sigma_l$ by
the product 
$\Sigma_l\simeq\Delta l\siglp\equiv
\left(l_>-l_<\right)\siglp\simeq l_>\siglp$. Reference \cite{widom}
provides a more detailed picture of the line tension associated with
the coexistence of thin and thick films on a planar substrate along
the prewetting line. We note that from these analyses one expects that
in systems governed by nonretarded van der Waals forces $\Sigma_l$
diverges logarithmically
upon approaching coexistence at $T_w$. Since for large 
$r_0$ one has $\Delta l\sim|T-T_w|^{-\eh}$ (Eq.~(3.38)), our
ansatz for $\Sigma_l$ overestimates
the actual value. Since we aim only for a rough estimate we refrain
from a more detailed analysis along these lines. 
(For a discussion of finite-size effects on the line tension see the
work by Indekeu and Dobbs in Ref.~\cite{widom}.)
The rounding region of the prewetting phase transition on a
cylindrical substrate is then given by \cite{privman}
\be
\left|\Delta\Omega_s\left(T,\mu=\mupret+\delta\mu(T)\right)\right|
r_0\xi\exp\left(r_0\Sigma_l(T,\mu=\mupret)/k_BT\right)
\quad\lsim\quad
k_BT,
\ee
where $\xi$ is the bulk correlation length.
The width $\delta\mu(T)$ as determined by Eqs.~(4.1) and (4.2) can be
translated into a temperature interval $\delta T(T)$
by taking into account
the slope $m_{pre}^{(\tau)}(T)=\partial\left(\mupre^{(\tau)}(T)/\epsilon\right)
/\partial \left(k_BT/\epsilon\right)$ of the prewetting line:
\be
\left|k_B\delta T\right|\simeq \left|m_{pre}^{(\tau)}(T)\right|^{-1}
\left|\delta\mu(T)\right|. 
\ee
Now consider $\dom_s$ at a point $(T,\mupret+\delta\mu(T))$ on the
boundary of the rounding region given by Eqs.~(4.1) and (4.2),
respectively. The calculation of $\dom_s(T,\mupret+\delta\mu(T))$
requires 
the evaluation of the effective interface potential for the two minima
at $l_>$ and $l_<$. Whereas $\Omega_s$ in the
asymptotic form (Eq.~(3.13)) of Eq.~(2.10) 
yields sufficient information about the
second minimum $l_>$, this is less transparent for 
$l_<(T,\mu)$ and $\Omega_s^<(T,\mu)$. However, because
$\Omega_s^<(T,\mu)$ varies only weakly upon a deviation
$\delta\mu$ from the prewetting line
$\mu = \mupret$ \cite{lipowsky}, one has 
$\Omega_s^<\left(T,\mupret+\delta\mu\right)\simeq
\Omega_s^<\left(T,\mupret\right)=\Omega_s^>(T,\mupret)$. 
Thus one obtains 
\be
\left|\Delta\Omega_s\left(T,\mupret+\delta\mu\right)\right|
\simeq
|\delta\mu|\left|\frac{\partial\Omega_s^>(T,\mu)}
{\partial\mu}\right|_{\mu=\mupret}.
\ee
With Eqs.~(4.1) - (4.3) this implies
\be
\left|k_B\delta T\right|\quad\lsim\quad\frac{k_BT}{r_0^2}
\left|m_{pre}^{(s)}\right|^{-1}
\left|\frac{\partial\Omega_s^>(T,\mu)}{\partial\mu}\right|^{-1}_{\mu=\mupret}
\ee
for a sphere and 
\be
\left|k_B\delta T\right|\quad\lsim\quad\frac{k_BT}{r_0\xi}
\left|m_{pre}^{(c)}\right|^{-1}
\left|\frac{\partial\Omega_s^>(T,\mu)}{\partial\mu}\right|^{-1}_{\mu=\mupret}
\exp\left(-r_0\Sigma_s/k_BT\right)
\ee
for a cylinder, respectively. 
The factor $\partial\Omega_s^>/\partial\mu$ consists of four contributions:
\be
\frac{\partial\Omega_s^>}{\partial\mu}
=\left(\frac{\partial\Omega_s^>}{\partial\mu}\right)_{l_>,\rl,\rg}
&+&\left(\frac{\partial\Omega_s^>}{\partial l_>}\right)_{\mu,\rl,\rg}
\frac{\partial l_>}{\partial\mu}+\nn\\
&+&\left(\frac{\partial\Omega_s^>}{\partial \rl}\right)_{\mu,l_>,\rg}
\frac{\partial\rl}{\partial\mu}
+\left(\frac{\partial\Omega_s^>}{\partial \rg}\right)_{\mu,l_>,\rl}
\frac{\partial\rg}{\partial\mu}.
\ee
Since the variation of $(\rg(\mu=\mu_0)-\rg(\mu))/\rg(\mu=\mu_0)$ and
$(\rl(\mu=\mu_0)-\rl(\mu))/\rl(\mu=\mu_0)$ are both of the order $10^{-3}$ 
along the whole prewetting line, the last two terms of Eq.~(4.7)
can be neglected. With Eq.~(3.13) for the effective interface
potential $\Omega_s(l)$ one obtains 
\be
\frac{\partial\Omega_s^>}{\partial\mu}&=&\frac{r_0\drho}{\tau+1}
\left(\left(1+\frac{1}{y}\right)^{\tau+1}-1\right)\nn\\
&-&\frac{\tau r_0\drho}{y^2}\left(1+\frac{1}{y}\right)^{\tau-1}
\left(1+\frac{\lambda}{\tau}\left(1+\frac{1}{y}\right)
-\frac{2}{\tau}\frac{y^3}{\kappa^3}s_\tau^{-3}(y)\right)
\frac{\partial y}{\partial\lambda}
\ee
with $y=r_0/l_>$.
The second minimum $l_>$ is given as the solution of the implicit
equation (3.15). Depending on the value of the dimensionless
parameters $\kappa=\left(r_0^2\siglp/a\right)$ and
$\lambda=\left(3b/(2ar_0)\right)$ there are three different regimes in
each of which the implicit equation (3.15) provides an approximate
solution $l_>$ (see Sect.~III.B and
Fig.~6). Accordingly, this leads to three different limiting cases of
the estimates (4.5) and (4.6) for the rounded regions. 
\begin{enumerate}
\item
In region I one has $\kappa\to 0$, $\lambda$ fixed, and $y$ given by
Eqs.~(3.19) and (3.20) which leads to 
$\left|\frac{\partial\Omega_s^>}{\partial\mu}\right|^{-1}
\simeq\frac{3}{2}\left(\frac{r_0\dmu}{a\drho}\right)^{\eh}$ for a
sphere and
$\left|\frac{\partial\Omega_s^>}{\partial\mu}\right|^{-1}\simeq
5\left(\left(\frac{2\dmu}{9\pi a}\right)^2\frac{r_0}{\drho^3}\right)^{1/5}$ 
for a cylinder, respectively. Thus  with $\dmupre=\mu_0-\mupre$ one
finds for the width of the 
rounding region $|k_B\delta T|$:
\be
|k_B\delta T|\quad\lsim\quad
\frac{3}{2}\frac{k_BT}{|m_{pre}^{(s)}|}
\left(\frac{\dmupre}{r_0^3a\drho}\right)^{1/2}
\ee    
for a sphere and
\be
|k_B\delta T|\quad&\lsim&\quad
5\frac{k_BT}{\xi|m_{pre}^{(c)}|}
\left(\frac{2}{9\pi}\frac{\dmupre}{a}
    \right)^{2/5}\left(\frac{1}{r_0^4\drho^3}\right)^{1/5}\times\nn\\
&\times&\exp\left(-\frac{\siglp}{k_BT}
\left(\frac{9\pi}{2}\frac{ar_0^7}{\drho\dmupre}\right)^{1/5}\right)
\ee
for a cylinder.
\item
For film thicknesses $l_>$ small compared with the radius $r_0$ (region
II) one has $\kappa\to\infty$, $\lambda$ fixed, and $y$ is given by
Eq.~(3.32). With
$\left|\frac{\partial\Omega_s^>}{\partial\mu}\right|^{-1}
=\frac{3}{r_0\drho} 
\left(\frac{2}{\tau}\frac{\siglp}{r_0a}\right)^{1/3}
\left(1+\frac{r_0\drho\dmu}{\tau\siglp}\right)^{1/3}$
one finds
\be
|k_B\delta T|\quad\lsim\quad\frac{3k_BT}{\drho|m_{pre}^{(s)}|r_0^2}
\left(\frac{\frac{2\siglp}{r_0}+\drho\dmupre}{2a}\right)^{1/3}
\ee for a sphere and
\be
|k_B\delta T|\quad&\lsim&\quad\frac{3k_BT}{\drho\xi|m_{pre}^{(c)}|r_0}
\left(\frac{\frac{\siglp}{r_0}+\drho\dmupre}{2a}\right)^{1/3}\times\nn\\
&\times&\exp\left(-\frac{r_0\siglp}{k_BT}\left(\frac{2a}{\frac{\siglp}{r_0}
+\drho\dmupre}\right)^{1/3}\right)
\ee
for a cylinder, respectively.
\item
In region III one has $\lambda/\kappa\to 0$ and $y$ given by
Eqs.~(3.33) and (3.34), so that one finds 
\be
|k_B\delta T|\quad\lsim\quad3\frac{k_BT}{|m_{pre}^{(s)}|}\frac{1}{\drho} 
\left(\frac{\siglp}{8ar_0^3}\right)^{3/5}
\left(1+\frac{3}{10}\left(\frac{8ar_0^3}{{\siglp}^6}\right)^{1/5}
\drho\dmupre\right)
\ee 
for a sphere and
\be
&|k_B\delta T|&\quad\lsim\quad2\frac{k_BT}{\xi|m_{pre}^{(c)}|}\frac{1}{\drho}
\left(\frac{2}{9\pi}\frac{\siglp}{r_0^2a}\right)^{1/2}
\left(1+\left(\frac{9\pi}{32}\frac{r_0^2a}{{\siglp}^5}\right)^{1/4}
\drho\dmupre\right)\times\nn\\
&\times&\exp\left(-\frac{1}{k_BT}\left(\frac{9\pi}{2}r_0^6a{\siglp}^3
\right)^{1/4}
\left(1+\left(\frac{9\pi}{512}\frac{r_0^2a}{{\siglp}^5}\right)^{1/4}
\drho\dmupre\right)^{-1}\right)
\ee
for a cylinder, respectively.
\end{enumerate}

From Eqs.~(4.9)-(4.14) one infers that for a sphere the width of the 
rounded region shrinks for $r_0\to\infty$ always algebraically
but exponentially for a cylinder.
In the
case of large radii and small film thickness (region II) our result
for $\dt$ for a sphere agrees with Eq.~(5.5) in
Ref.~\cite{lipowsky}. As already stated in this reference it is
difficult to obtain a general expression for $\mupret$ and
$m_{pre}^{(\tau)}$. However,
our microscopic model allows us
to obtain quantitative estimates for $\mupret$ and $m_{pre}^{(\tau)}$
for specific interaction potentials of the system.
From the first-order wetting phase diagram of a cylinder
and a sphere 
of radius $r_0=100\sigma$ (see Fig.~8) one can infer $\dmupre$ directly
as a function of the temperature. If one chooses $k_BT=0.95\epsilon$,
one has $k_BT_w^{(p)}\simeq 0.785\epsilon$, $\dmu_{pre}^{(s)}\simeq 0.01\epsilon$,  
$|m_{pre}^{(s)}|^{-1}\simeq 1.65$, and 
$\dmu_{pre}^{(c)}\simeq 0.026\epsilon$, $|m_{pre}^{(c)}|^{-1}\simeq
1.45$, respectively. 
Thus with 
$\rg\sigma^3\left(k_BT=0.95\epsilon\right)\simeq 0.076$,
$\rl\sigma^3\left(k_BT=0.95\epsilon\right)\simeq 0.551$,
as well as 
the substrate parameters $\ew=0.625\epsilon$,
$\sw=1.35\sigma$, and $\rw\sigma^3=1.0$, Eq.~(4.11) yields
\be
\frac{|\dt|}{T_w^{(p)}}\simeq 2.3\cdot 10^{-4}\quad
\mbox{(sphere, region II)}.
\ee
Equation (4.12) for a cylindrical substrate involves also the bulk
correlation length $\xi$ defined as 
$\xi^2=\frac{1}{6}\int d^3r\,r^2 {\cal G}(r)/\int d^3\, r\, {\cal G}(r)$
where ${\cal G}(|\vecr -\vecr'|)=\,<\hat\rho(\vecr)\hat\rho(\vecr')>$
$-$$<\hat\rho(\vecr)>$ $<\hat\rho(\vecr')>$ is the two-point correlation
function of the fluctuating number density $\hat\rho(\vecr)$.
Within our model this leads to
$\xi^2=\frac{1}{6}|w_2|\rho^2\kappa_T$ with the isothermal
compressibility $\kappa_T=\rho^{-2}\left(\frac{\partial^2
    f_h}{\partial\rho^2} +w_0\right)^{-1}$ and $w_i=\int d^3 r\,
r^i\tilde w(r)$. 
For the model potential of the fluid-fluid interaction given by
Eqs.~(2.3) and (2.4) one obtains $\xi(T=0.95\epsilon/k_B)\simeq
1.7\sigma$ which implies
\be
\frac{|\dt|}{T_w^{(p)}}\simeq  10^{-140}\quad
\mbox{(cylinder, region II)}.
\ee
We conclude that for $r_0=100\sigma$ the rounding of the prewetting
transition on a sphere is at the lower end of the experimentally
accessible temperature resolution whereas for a cylinder the rounding
cannot be detected. 

\vskip 1cm
\centerline{\large \bf V. Summary}
\vskip 1cm

\addtocounter{section}{1}
\setcounter{equation}{0}

We have studied wetting phenomena on cylindrical and spherical
substrates (Fig.~1) for simple fluids whose particles are governed by
dispersion forces and are exposed to long-ranged substrate
potentials. 
Our approach is based on a microscopic density functional
theory. Using a
sharp-kink approximation for the density profile (Sect.~II A), 
we have determined the 
effective interface potential $\Omega_s(l)$ (Eq.~(2.10)) for the
emerging liquidlike  film of thickness $l$. 
From a detailed analytical as well as numerical discussion of 
$\Omega_s(l)$ for a cylinder ($\tau=1$) and a sphere ($\tau=2$) of
radius $r_0$  we have obtained the following main results:
\begin{enumerate}
\item As a contribution to $\Omega_s(l)$ we have determined the
  liquid-vapor surface tension $\sigma_{l,g}^{(\tau)}(h)$ 
  of a spherical drop and a liquid thread of radius $h$ (see
  Subsec.~II B).
  For small values of $\frac{1}{h}$ it approaches its planar value
  $\siglp=\sigma_{l,g}^{(\tau)}(h=\infty)$ nonanalytically: the
  next-to-leading order term beyond the contribution $\sim\frac{1}{h}$
  is proportional to $\ln(h)/h^2$ (see Eqs.~(2.21) - (2.23) and
  Fig.~2). 
\item  The interaction contribution $\omega^{(\tau)}(l,r_0)$ to the
  effective interface potential $\Omega_s(l)$ has been determined in
  closed form (Eq.~(2.29)). It can be expressed in terms of the
  corresponding planar quantity $\omega^{(p)}(l)$ (Eq.~(2.27)) by
  scaling functions ${\cal S}_\tau(r_0/l)$ and ${\cal R}_\tau(r_0/l)$
  (see Eqs.~(2.35), (2.39), and (3.48) and Figs.~3 and 10). These
  scaling functions are universal, i.~e.~independent of interaction
  potential parameters, and reflect the geometrical properties of the
  substrate (see Eqs.~(A16), (B30), (C4), and (C5)).
\item At liquid-vapor  coexistence $\mu=\mu_0(T)$ and above the
  wetting temperature $T_w^{(\tau)}(r_0)$ (Sect.~III A) the maximum
  film  thickness $l_{max}$ is determined by 
  Eq.~(3.2)  depending on the dimensionless parameter
  $\kappa=(r_0^2\siglp/a)^{1/3}$ with the Hamaker constant $a$ (see
  Fig.~4).  
  In the limiting cases $\kappa\gg 1$ and $\kappa\ll 1$ one
  obtains for $l_{max}$ power laws as given by Eqs.~(3.9) - (3.11).
\item  For a small undersaturation $\dmu=\mu_0(T)-\mu>0$ and above the
  wetting temperature $T_w^{(\tau)}(r_0)$ the equilibrium film
  thickness 
  $l_0$ on curved substrates equals the equilibrium film
  thickness on the corresponding planar substrate 
  for the fluid taken to be off
  coexistence at an effective undersaturation $\dmu_{eff}$. 
  In the particular case of large 
  radii $r_0$, i.~e., $\kappa\to\infty$, the film thickness on a curved
  substrate equals that on the corresponding planar substrate as if
  the liquid-vapor coexistence curve was shifted by a constant value
  $\delta\mu_\infty^{(\tau)} = \tau\siglp/(r_0\drho)$ into the vapor
  phase (see Eqs.~(3.9) and (3.28) and Fig.~5). 
  In general, however, $l_0$ and $\dmu_{eff}$ are determined by
  $\kappa$ and the additional dimensionless parameter
  $\lambda=r_0\drho\dmu/\siglp$ (Eq.~(3.14)) which 
  measures the actual distance $\dmu$ from liquid-vapor coexistence
  in terms of the difference $\drho=\rl-\rg$ of the liquid and vapor
  number densities. 
  This $(\kappa,\lambda)$ parameter space separates
  into three regions (Fig.~6) in each of which $\dmu_{eff}$ and thus
  the film thickness $l_0$ reveals a characteristic power law
  behavior as function of 
  $\kappa$ and $\lambda$ (see Eqs.~(3.29) - (3.31), (3.35), and (3.36)
  and Fig.~7).
\item For suitably chosen interaction potential parameters the
  numerical analysis of the effective interface potential in 
  its full form (Eqs.~(2.10), (2.16), (2.17), and (2.29)) renders
  the phase 
  diagram for first-order wetting of curved substrates (see
  Fig.~8(a)). 
  The curvature of the substrate  raises the wetting temperature
  $T_w^{(\tau)}(r_0)$ compared to the wetting temperature $T_w^{(p)}$
  of the corresponding planar substrate. In the limit of large radii
  $r_0$  this shift vanishes as 
  $T_w^{(\tau)}(r_0)-T_w^{(p)}\sim r_0^{-3/2}$. At $T_w^{(\tau)}(r_0)$
  the jump $\Delta l_0$ in the film
  thickness is finite (Fig.~8(b)) and diverges  for $r_0\to\infty$ as
  $\Delta  l_0\sim r_0^{1/3}$. 
\item The curvature of a substrate erases a
  critical wetting transition such that there is a continuous and
  smooth increase of the film thickness (see Fig.~9). Even for very
  large radii $r_0$ the effects of curvature on the film thickness
  near this smeared out critical wetting transition are still
  significant (see Eq.~(3.59) and Fig.~9(b)). At coexistence this film
  thickness is determined by the two dimensionless parameters
  $\kappa=\left(r_0^2\siglp/a\right)^{1/3}$ and $\gamma=3b/(2ar_0)$
  formed from the expansion coefficients $a$ and $b$ of the planar
  effective interface potential  (Eq.~(2.27)). This $(\kappa,\gamma)$
  parameter space separates into a rich structure (Fig.~11) such that
  its various regions correspond to different power laws for $l_0$
  (see Eqs.~(3.59), (3.65), (3.68), (3.69), and (3.72) - (3.75)).
\item Since the surface of a sphere and a cylinder is quasi-zero- and
  quasi-onedimensional, respectively, finite size effects smear out
  the first-order phase transitions shown in Fig.~8(b) as obtained by
  mean-field theory. However, according to the estimates in Sect.~IV,
  these finite-size induced fluctuation effects
  are confined to a very narrow temperature interval
  $\delta T$ around 
  the prewetting line. For a typical spherical substrate with a radius
  $r_0=100\sigma$ where $\sigma$ denotes the diameter of the fluid
  particles one finds $\delta T/T_w^{(p)}\sim 10^{-4}$. For a
  cylindrical substrate with the same radius $\delta T$ is vanishingly
  small.

\end{enumerate}
\vskip 1cm

\centerline{\bf Acknowledgements}

We thank R.~Evans, A.~Hanke, and M.~Napi\'orkowski for many
helpful discussions.

\newpage

\begin{appendix}

\section{Explicit form of the effective interface potential for a sphere}

According to Eqs.~(2.17) and (2.29) $\sigls$ and part of
$\omega^{(s)}(l)$ are determined by
$t^{(s)}(r;a)$ which represents the interaction potential
of a fluid particle at distance $r$ 
outside a sphere of radius $a$ filled with the same fluid particles. 
In polar coordinates $t^{(s)}(r;a)$ takes on the following
form:
\be
t^{(s)}(r;a)&=&\int_{|\vecrs'|\leq a}d^3r'\tilde w(|\vecr-\vecr'|)\\
&=&2\pi\int^a_0 dr'\ r'^2 \int_0^\pi d\theta\sin\theta 
\ \tilde w\left(\sqrt{r^2+r'^2-2rr'\cos\theta}\right),\nn
\ee
where we have already used the rotational invariance of $t^{(s)}(r;a)$.
With the substitution $s^2=r^2+r'^2-2rr'\cos\theta$, Eq.~(A1) yields
\be
t^{(s)}(r;a)&=&2\pi\int_0^a dr'\frac{r'}{r} \int_{r-r'}^{r+r'}ds\ s\tilde
w(s)\nn\\
&=&\frac{2\pi}{r}\int_{r-a}^{r+a}dr'(r-r')\int_{r'}^\infty ds\ s\tilde
w(s).\label{ts2}
\ee
Equation (2.3) leads to (on physical grounds only the case
$a\geq\sigma$ is of interest)
\be
t^{(s)}(r\geq a&;&a\geq\sigma)=\nn\\
&=&\left\{\begin{array}{l@{\;}r}
t^{(s)}_>(r;a)=\frac{2\pi}{r}\int_{r-a}^{r+a}dr'(r-r')\int_{r'}^\infty ds
s\phi_{LJ}(s),& r\geq a+\sigma\label{ts3}\\
t^{(s)}_<(r;a)=\frac{2\pi}{r}\int^{r+a}_\sigma dr'(r-r')\int_{r'}^\infty ds
s\phi_{LJ}(s)&  \\
\quad\quad\quad\quad +\frac{2\pi}{r}\int_{r-a}^\sigma dr' (r-r') 
\int_\sigma^\infty ds s
\phi_{LJ}(s),&a<r<a+\sigma\label{ts4}
\end{array}\right.
\ee
One has $t^{(s)}_>(a+\sigma,a)=t^{(s)}_<(a+\sigma,a)$. In the next step we now
explicitly use Eq.~(2.4) so that
\be
t^{(s)}_>(r;a)&=&2\pi\epsilon\left[\frac{\sigma^{12}}{20}\left(\frac{1}{r(r+a)^8}
-\frac{1}{r(r-a)^8}\right)-\frac{2\sigma^{12}}{45}\left(\frac{1}{(r+a)^9}
-\frac{1}{(r-a)^9}\right)\right.\label{tsg}\nn\\
&&\left.\quad\quad-\frac{\sigma^6}{2}\left(\frac{1}{r(r+a)^2}
-\frac{1}{r(r-a)^2}\right)+\frac{\sigma^6}{3}\left(\frac{1}{(r+a)^3}
-\frac{1}{(r-a)^3}\right)\right]
\ee
and
\be
t^{(s)}_<(r;a)&=&\pi\epsilon\left(-\frac{16}{9}\sigma^3+\frac{3}{5}r\sigma^2
-\frac{3}{5}\frac{a^2\sigma^2}{r}+\frac{3}{2}\frac{\sigma^4}{r}\right)
+\frac{4\epsilon\sigma^{12}\pi}{5}\left(\frac{1}{8}\frac{1}{r(r+a)^8}\right.\nn\\
&&\quad\left.-\frac{1}{9}\frac{1}{(r+a)^9}\right)
-2\epsilon\sigma^6\pi\left(\eh\frac{1}{r(r+a)^2}
-\frac{1}{3}\frac{1}{(r+a)^3}\right).
\label{tsk}
\ee
The integral
$\int_b^\infty dr\ r^2\ t^{(s)}(r;a)$, $b\geq a$, is given as
\be
\int_b^\infty dr\ r^2\ t^{(s)}(r;a)=\left\{
\begin{array}{l@{\; ,\;}r}
\int_b^\infty dr\ r^2\ t^{(s)}_>(r;a)&b\geq a+\sigma\\
\int_b^{a+\sigma} dr\ r^2 t^{(s)}_<(r;a)+\int_{a+\sigma}^\infty dr\ r^2 
\ t^{(s)}_>(r;a)&b<a+\sigma.
\end{array}\right.
\label{intts}
\ee
From Eq.~(\ref{intts}) one can directly infer the liquid-gas surface tension
$\sigls$ (Eq.~(2.17))
\be
\sigls(h)=
\siglp\left(
1-\frac{2}{9}\frac{\ln (h/\sigma)}{(h/\sigma)^2}
-\frac{1}{(h/\sigma)^2}\left(\frac{2}{9}\ln 2+\frac{4}{27}\right)
-\frac{1}{20736}\frac{1}{(h/\sigma)^8}\right),
\ee
where $\sigma_{l,g}^{(p)}=\frac{3\pi}{4}\epsilon\sigma^4(\drho)^2$.
According to Eqs.~(2.29) and (A6) one has to distinguish the cases
$l> d_w+\sigma$ and $d_w\leq l\leq d_w+\sigma$. For $l>d_w+\sigma$
one has
\be
\lefteqn{\omega^{(s)}\left(l>d_w+\sigma\right)=}\nn\\
&=&\drho\left(\rl\int_h^\infty 
dr \left(\frac{r}{r_0}\right)^2\, t^{(s)}_>(r;r_1)
-\rw\int_h^\infty 
dr \left(\frac{r}{r_0}\right)^2\, v^{(s)}(r;r_0)\right),
\ee
whereas for $d_w\leq l\leq d_w+\sigma$ 
\be
\lefteqn{\omega^{(s)}\left(d_w\leq l\leq d_w+\sigma\right)=}\nn\\
=\drho\Bigg(
&\rl&\int_{h}^{r_1+\sigma}dr\,\left(\frac{r}{r_0}\right)^2
t_<^{(s)}(r;r_1)
+\rl\int_{r_1+\sigma}^\infty dr\,\left(\frac{r}{r_0}\right)^2 
t_>^{(s)}(r;r_1)\nn\\
-&\rw&\int_h^\infty
dr\,\left(\frac{r}{r_0}\right)^2v^{(s)}(r;r_0)\Bigg).
\ee
If $v^{(s)}(r;r_0)$ is approximated as the linear superposition of
$\phi_{LJ}^{wf}(r)$ (Eq.~(2.5)) it has the same form as
$t_>^{(s)}(r;r_0)$ with $(\epsilon,\sigma)$ replaced by
$(\ew,\sw)$. For additional simplifications see Ref.~[34(b)]. In
this case and with Eq.~(A4) we obtain 
\be
\omega^{(s)}\left(l>d_w+\sigma\right)&=&
\drho\Big[\rl\epsilon\sigma^{12}f^{(s)}_{rep}(r_1,h)
-\resi^{12}f^{(s)}_{rep}(r_0,h)\nn\\
&&\quad-\rl\epsilon\sigma^6 f_{attr}^{(s)}(r_1,h)
+\resi^6f_{attr}^{(s)}(r_0,h)\Big]
\ee
where 
\be
f^{(s)}_{rep}(a,b)=\frac{\pi}{540r_0^2}\left(
\frac{b^2+8ab+a^2}{(b+a)^8}
-\frac{b^2-8ab+a^2}{(b-a)^8}\right)
\ee
and
\be
f_{attr}^{(s)}(a,b)=\frac{\pi}{3r_0^2}\left(
2ab\frac{b^2+a^2}{(b^2-a^2)^2}
-\ln\frac{b+a}{b-a}\right).
\ee
Expansion in powers of $\frac{d_w}{r_0}$ yields
\be 
\omega^{(s)}(l)=\omega^{(s)}_{rep}(l)+\omega^{(s)}_{attr}(l)
\ee
where
\be
\omega^{(s)}_{rep}(l)&=&\frac{\pi}{540 r_0^2}\drho\left(\rl
\epsilon\sigma^{12}-\resi^{12}\right)
\left(\frac{h^2+8hr_0+r_0^2}{(h+r_0)^8}-\frac{h^2-8hr_0+r_0^2}{(h-r_0)^8}\right)\nn\\
&+&\frac{\pi}{90}\drho\rl\epsilon\sigma^{12}
\left(\frac{9h-r_0}{(h-r_0)^9}-\frac{9h+r_0}{(h+r_0)^9}\right)\frac{d_w}{r_0}\nn\\
&+&{\cal O}\left(\left(\frac{d_w}{r_0}\right)^2\right)
\ee
and
\be
\omega^{(s)}_{attr}(l)&=&-\frac{\pi}{3r_0^2}\drho\left(\rl
\epsilon\sigma^6-\resi^6\right)
\left(2hr_0\frac{h^2+r_0^2}{(h^2-r_0^2)^2}-\ln\frac{h+r_0}{h-r_0}\right)\nn\\
&-&\frac{d_w}{r_0}\frac{16}{3}\pi\rl\drho\frac{r_0h^3}
{(h^2-r_0^2)^3}+{\cal O}\left(\left(\frac{d_w}{r_0}\right)^2\right).
\ee
In the limit $l\gg\sigma$ the functions ${\cal
S}_s(r_0/l)=\omega_{attr}^{(s)}(l)/\omega_{attr}^{(p)}(l)$
and $s_s(x)=\left(\left(1+x^{-1}\right)/
({\cal S}_s(x)+\frac{x}{2}{\cal S}'_s(x))\right)^{1/3}$ are given by
\be
{\cal S}_s(x)=\frac{1}{x^2}\left(2x(x+1)^2\frac{2x^2+2x+1}{(1+2x)^2}
-\ln(2x+1)\right)
\ee
and 
\be 
s_s(x)=\frac{1}{2}\frac{1+2x}{(x^2(1+x))^{1/3}}\quad ,
\ee
respectively.

\section{ The effective interface potential of a cylindrical substrate}

For a cylinder the calculation of the $l$-dependent terms $\siglc$ and
$\omega^{(c)}(l)$ is somewhat more complicated than in
the case of a sphere. 
They have the form
\be
\siglc=-\eh(\drho)^2f^{(c)}(h,h;[\tilde w])
\ee
and
\be
\omega^{(c)}(l)=\drho\left(
\rl f^{(c)}(r_1,h;[\tilde w])-\rw g^{(c)}(r_0,h;[v])\right),
\ee
with
\be
f^{(c)}(a,b;[\tilde w])=\frac{1}{2\pi Mr_0}
\int\limits_{{\cal C}(b,\infty)}d^3r
\int\limits_{{\cal C}(0,a)}d^3r'\tilde w(|\vecr-\vecr'|)
\ee
and
\be
g^{(c)}(a,b;[v])=\frac{1}{2\pi Mr_0}
\int\limits_{{\cal C}(b,\infty)}d^3r
\int\limits_{{\cal C}(0,a)}d^3r' v(|\vecr-\vecr'|)
\ee
where ${\cal C}(r_<,r_>)$ denotes the region between two cylinders of
radii $r_<$ and $r_>$. Eq.~(B3) can be rewritten in the form
\be
f^{(c)}(a,b;[\tilde w]):=I(a,\infty)-I(a,b) 
\ee
where
\be
I(a,b)=\frac{1}{2\pi Mr_0}
\int\limits_{{\cal C}(0,b)} d^3r
\int\limits_{{\cal C}(0,a)} d^3r'
\tilde w(|\vecr-\vecr'|).
\ee
With $\vecr=(\vep,z)$ where the $z$-axis is the cylinder axis and
$\vep$ orthogonal to it one has
\be
I(a,b)&=&\frac{1}{\pi r_0}\int_0^\infty dz\int\limits_{0\leq|\veps|\leq b} \dvep
\int\limits_{0\leq|\veps'|\leq a}\dveps\tilde w
\left(\sqrt{(\vep-\vep')^2+z^2}\right)\nn\\
&=:&\frac{1}{\pi r_0}\int_0^\infty dz J(z)\label{iab}
\ee
with
\be
J(z)=\int\limits_{0\leq|\veps|\leq b}\dvep
\int\limits_{0\leq|\veps'|\leq a}\dveps
\tilde w\left(\sqrt{(\vep-\vep')^2+z^2}\right).
\ee
The coordinate transformation
\be
\begin{array}{l@{\quad}l}
\ver=\eh(\vep +\vep'),&\tvec=\vep-\vep'\\
\vep=\ver+\eh\tvec,&\vep'=\ver-\eh\tvec,\label{trafo}
\end{array}
\ee
yields
\be
J(z)=2\pi\int_0^{a+b}d\tr\,\tr\,\tilde w\left(\sqrt{\tr^2+z^2}\right)
\int_D d^2R.\label{jz}
\ee
Equation (\ref{trafo}) provides the following 
limits of integration for $\ver$:
\be
0\leq\left(\ver+\eh\tvec\right)^2\leq b^2,\quad
0\leq\left(\ver-\eh\tvec\right)^2\leq a^2.\label{circles}
\ee
This corresponds to two circles,
one with radius $b$ centered at
$\ver=-\eh\tvec$, the other with radius $a$ centered at
$\ver=+\eh\tvec$. Thus
$\int_D d^2R=:D(\tr,a,b)$ is the area of intersection of these
two circles which depends on $\tvec$ for
$b-a\leq|\tvec|\leq b+a$ and equals $\pi a^2$ for
$0\leq|\tvec|\leq b-a$:
\be
D(\tr,a,b)=\left\{\begin{array}{l@{\;\;}l}
g_{a,b}(\tr),
&b-a\leq|\tvec|\leq b+a\\
\pi a^2,&0\leq|\tvec|<b-a\\
0,&\mbox{otherwise}
\end{array}\right.\label{rint}
\ee
where
$g_{a,b}(\tr):=\frac{\pi}{2}n^2-\eh\sqrt{-\tr^4+2n^2\tr^2-m^4}
+a^2\arcsin\frac{m^2-\tr^2}{2\tr a}-b^2\arcsin\frac{m^2+\tr^2}{2\tr b}$
with $m^2=b^2-a^2$ and $n^2=b^2+a^2$. One has $g_{a,\infty}=\pi a^2$
because the smaller circle lies always within the larger one and
\be
D(\tr,a,a)=\left\{\begin{array}{l@{\;\;}l}
g_{a,a}(\tr),& 0\leq|\tr|\leq 2a\\
0,&|\tr|>2a.
\end{array}\right.
\ee
with $g_{a,a}(\tr)=\pi a^2-\frac{\tr}{2}\sqrt{4a^2-\tr^2}
-2a^2\arcsin\frac{\tr}{2a}$.
Inserting Eq.~(\ref{rint}) into Eq.~(\ref{jz}) yields
\be
J(z)=2\pi^2a^2\int_{|z|}^{\sqrt{(b-a)^2+z^2}}ds &s&\tilde w(s)\nn\\
+2\pi\int_{\sqrt{(b-a)^2+z^2}}^{\sqrt{(b+a)^2+z^2}}ds &s&\tilde w(s) 
g_{a,b}\left(\sqrt{s^2-z^2}\right)
\ee
so that
\be
I(a,b)=\frac{2\pi a^2}{r_0}\int_0^\infty &dz&
\int_{|z|}^{\sqrt{(b-a)^2+z^2}} ds s\tilde w(s)\nn\\
+\quad\frac{2}{r_0}
\int_0^\infty &dz&
\int_{\sqrt{(b-a)^2+z^2}}^{\sqrt{(b+a)^2+z^2}} ds s\tilde w(s)
g_{a,b}\left(\sqrt{s^2-z^2}\right)
\ee
and especially
\be
I(a,\infty)=\frac{2\pi a^2}{r_0}\int_0^\infty dz
\int_{|z|}^\infty ds s \tilde w(s).
\ee
Thus independent of the explicit form of $\tilde w$ 
$f^{(c)}(a,b;[\tilde w])$ is given by
\be
f^{(c)}(a,b;[\tilde w])=\frac{2\pi a^2}{r_0}
\int_0^\infty &dz&\int_{\sqrt{(b-a)^2+z^2}}^\infty ds s\tilde w(s)\nn\\
-\quad\frac{2}{r_0}\int_0^\infty &dz&
\int_{\sqrt{(b-a)^2+z^2}}^{\sqrt{(b+a)^2+z^2}}
ds s\tilde w(s) g_{a,b}\left(\sqrt{s^2-z^2}\right).
\ee
Equation (2.3) requires the consideration of two cases:
\be
f^{(c)}(b>a+\sigma)=\frac{2}{r_0}\Bigg(\pi a^2\int_0^\infty
&dz&\int^\infty_{\sqrt{(b-a)^2+z^2}}
ds s w(s)\nn\\ 
-\int_0^\infty &dz&\int_{\sqrt{(b-a)^2+z^2}}^{\sqrt{(b+a)^2+z^2}}
ds s w(s)g_{a,b}
\left(\sqrt{s^2-z^2}\right)\Bigg)
\ee 
and 
\be
\lefteqn{f^{(c)}(a\leq b\leq a+\sigma)=}\nn\\
&=&\frac{2\pi
a^2}{r_0}\left(\int_0^{\sqrt{\sigma^2-(b-a)^2}}dz\int_\sigma^\infty ds 
+\int_{\sqrt{\sigma^2-(b-a)^2}}^\infty
dz\int_{\sqrt{(b-a)^2+z^2}}^\infty ds\right) 
s w(s)\nn\\
&-&\frac{2}{r_0}\left(\int_0^{\sqrt{\sigma^2-(b-a)^2}}dz
\int_\sigma^{\sqrt{(b+a)^2+z^2}} ds
+\int_{\sqrt{\sigma^2-(b-a)^2}}^\infty
dz\int_{\sqrt{(b-a)^2+z^2}}^{\sqrt{(b+a)^2+z^2}} ds\right)\times\nn\\
&&\quad\quad\quad\quad\quad\quad\quad\quad\quad\quad\quad\quad\quad\quad\quad
\times\, s  w(s)g_{a,b}\left(\sqrt{s^2-z^2}\right).\label{fc}
\ee
For $a=b$ one obtains
\be
f^{(c)}(a=b)&=&\frac{2\pi a^2}{r_0}\left(\int_0^\sigma
dz\int_\sigma^\infty ds +\int_\sigma^\infty dz\int_z^\infty ds
\right) s w(s)\nn\\
&-&\frac{2}{r_0}\left(\int_0^\sigma dz\int_\sigma^{\sqrt{4a^2+z^2}} ds
+\int_\sigma^\infty dz\int_z^{\sqrt{4a^2+z^2}} ds\right)\times\nn\\
&&\quad\quad\quad\quad\quad\quad\quad\quad\quad\quad
\times s w(s)g_{a,a}\left(\sqrt{s^2-z^2}\right).
\ee
If we now adopt the Lennard-Jones form for $w(r)$ (Eq.~(2.4))
$f^{(c)}$ separates into a contribution $f^{(c)}_{rep}$ from the
repulsive part proportional to $r^{-12}$
and into a contribution $f^{(c)}_{attr}$ due
to the attractive part $\sim r^{-6}$ of this potential so that 
\be
f^{(c)}=\epsilon\sigma^{12}f^{(c)}_{rep}
-\epsilon\sigma^6f^{(c)}_{attr}
\ee
and
\be
f^{(c)}_{rep}=\frac{\pi}{20r_0}\int_0^\infty \frac{dz}{z^{10}}
\left(\frac{\sum_{i=0}^9a_{2i}(m,n)z^{2i}}{\left(z^4+2n^2z^2+m^4\right)^{9/2}}
-8m^2\right)
\ee
with the coefficients 
$a_0=8m^{20},
 a_2=72m^{16}n^2,
 a_4=36m^{16}+252m^{12}n^4,
 a_6=252m^{12}n^2+420m^8n^6,
 a_8=63m^{12}+630m^8n^4+315m^4n^8,
 a_{10}=70n^{10}+620m^4n^6+318m^8n^2,
 a_{12}=155n^8+55m^8+462m^4n^4,
 a_{14}=132n^6+156m^4n^2,
 a_{16}=52n^4+20m^4,
 a_{18}=8n^2$ 
where $m^2=b^2-a^2$ and $n^2=b^2+a^2$.
By using a symbol-manipulation program one can show that the latter
expression is identical with
\be
f^{(c)}_{rep}=\frac{7\pi^2}{64r_0}\frac{a^2}{b^9}
\,{}_2F_1\left(\frac{11}{2},\frac{9}{2};2;\left(\frac{a}{b}\right)^2\right).
\ee
The attractive contribution has a similar form:
\be
f^{(c)}_{attr}&=&\frac{\pi}{2r_0}\int_0^\infty\frac{dz}{z^4}
\left(\frac{n^2z^6+z^4\left(2n^4+m^4\right)+3m^4n^2z^2+m^8}
{\left(z^4+2n^2z^2+m^4\right)^{3/2}}-m^2\right)\nn\\
&=&\frac{\pi^2}{2r_0}\frac{a^2}{b^3}
{}_2F_1\left(\frac{5}{2},\frac{3}{2};2;\frac{a^2}{b^2}\right).
\ee
For $a=r_1$ and $b=h$ one obtains $f^{(c)}(r_1,h;[\tilde w])$.

In order to provide an explicit formula for the effective interface
potential $\omega^{(c)}(l)$ one finally has to specify the contribution
$g^{(c)}$ from the substrate potential (see Eqs.~(B2)) and (B4). If
the substrate potential is approximated as the linear superposition of
Lennard-Jones interaction potentials between pairs of substrate atoms
and fluid atoms (see Eq.~(2.5)), $g^{(c)}$ has the same functional as
$f^{(c)}$ with $\epsilon$ and $\sigma$ replaced by $\ew$ and $\sw$,
respectively: 
\be
g^{(c)}(r_0,h)=f^{(c)}\left(a=r_0,b=h;\epsilon\to\ew,\sigma\to\sw\right).
\ee
Equations (B23) and (B24) are only valid for $h>r_0+d_w+\sigma$, because due to
the Heaviside function entering Eq.~(2.3) $f^{(c)}$ changes its form
for $d_w<l<d_w+\sigma$. Since in the substrate potential there is no
such Heaviside function, $g^{(c)}$ retains its form for {\em all}
values of $h$. Keeping this in mind one has 
\be
\omega^{(c)}(l>d_w+\sigma)=\drho\left(\rl f^{(c)}\left(r_1,h\right)-\rw
f^{(c)}\left(r_0,h;\epsilon\to\ew,\sigma\to\sw\right) \right).
\ee
In the limit $\frac{d_w}{r_0}\to 0$ Eq.~(B26) yields explicitly
\be
\omega^{(c)}(l)=\omega^{(c)}_{rep}(l)+\omega^{(c)}_{attr}(l),
\quad d_w/r_0\ll 1,
\ee
where
\be
\lefteqn{\omega^{(c)}_{rep}(l)=\frac{7\pi^2}{64}
\drho\left(\rl\epsilon\sigma^{12}-\resi^{12}\right)
\frac{r_0}{h^9}\,{}_2F_1\left(\frac{11}{2},\frac{9}{2};2;\left(\frac{r_0}{h}\right)^2\right)}\nn\\
& &+\frac{7\pi^2}{64}\drho\rl\epsilon\sigma^{12}\frac{r_0}{h^9}
\left(2\,{}_2F_1\left(\frac{11}{2},\frac{9}{2};2;\left(\frac{r_0}{h}\right)^2\right)
+\frac{99}{4}\frac{r_0^2}{h^2}
\,{}_2F_1\left(\frac{13}{2},\frac{11}{2};3;\left(\frac{r_0}{h}\right)^2\right)\right)\frac{d_w}{r_0}\nn\\
& &+{\cal O}\left(\left(\frac{d_w}{r_0}\right)^2\right)
\ee
and
\be
\lefteqn{\omega^{(c)}_{attr}(l)=-\frac{\pi^2}{2}
\drho\left(\rl\epsilon\sigma^6-\resi^6\right)
\frac{r_0}{h^3}\,{}_2F_1\left(\frac{5}{2},\frac{3}{2};2;\left(\frac{r_0}{h}\right)^2\right)}\nn\\
& &-\frac{\pi^2}{2}\drho\rl\epsilon\sigma^6\frac{r_0}{h^3}
\left(2\,{}_2F_1\left(\frac{5}{2},\frac{3}{2};2;\left(\frac{r_0}{h}\right)^2\right)
+\frac{15}{4}\frac{r_0^2}{h^2}
\,{}_2F_1\left(\frac{7}{2},\frac{5}{2};3;\left(\frac{r_0}{h}\right)^2\right)\right)\frac{d_w}{r_0}\nn\\
& &+{\cal O}\left(\left(\frac{d_w}{r_0}\right)\right).
\ee
The functions ${\cal S}_c(x)$ and $s_c(x)$ (Eqs.~(2.36) and (3.4)) are thus
given by
\be
{\cal S}_c(x)=\frac{3\pi}{2}\frac{x}{(x+1)^2}
\,{}_2F_1\left(\frac{5}{2},\frac{3}{2};2;\left(\frac{x}{x+1}\right)^2\right)
\ee
and
\be
s_c(x)=\frac{\left(\frac{4\left(1+x\right)^4}{9\pi x}\right)^{1/3}}
{\left({}_2F_1\left(\frac{5}{2},\frac{3}{2};2;\left(\frac{x}{x+1}\right)^2\right)
+\frac{5x^2}{4\left(x+1\right)^2}{}_2F_1\left(\frac{7}{2},\frac{5}{2};3;
\left(\frac{x}{x+1}\right)^2\right)\right)^{1/3}},
\ee
respectively.

The liquid-vapor surface tension follows from Eqs.~(B1) and (B20). A
lengthy calculation yields the following result based on Eqs.~(2.3)
and (2.4):
\be
\siglc=-(\drho)^2\epsilon\sigma^4 K\left(\frac{h}{\sigma}\right)
\ee
where 
\be
K(x)=\sum_{i=1}^{4} K_i(x)
\ee
with
\be
K_1(x)=
&&-\frac{2}{5}(\drho)^2\left[
\pi\sqrt{4x^2+1}\bigg(\frac{35}{2048x}
-\frac{125}{49152x^3}+\frac{337}{786432x^5}-\frac{209}{2097152x^7}
\right.\nn\\
&&\quad+\frac{1}{1024x^3(4x^2+1)^3}+\frac{1}{12288x^5(4x^2+1)^2}
+\frac{3}{16384x^7(4x^2+1)}\bigg)\nn\\
&&\left.\quad-\pi\frac{175}{4194304}\frac{\ln\left(2x+\sqrt{4x^2+1}\right)}
{x^8}\right],
\ee
\be
K_2(x)&=&
-\frac{4}{5}x\int\limits_0^1dz\Bigg\{
\frac{35x^6+30x^4z^2+9x^2z^4+z^6}{z^9(4x^2+z^2)^{7/2}}\times\nn\\
&\times&\left(\frac{\pi}{2}-\arcsin\left(
\sqrt{(1-z^2)\left(1+\frac{z^2}{4x^2}\right)}\right)\right)
+\arcsin\left(\frac{\sqrt{1-z^2}}{2x}\right)\nn\\
&+&\frac{1}{4x^2}
\sqrt{(1-z^2)(4x^2+z^2-1)}\bigg(1-\frac{1}{4z^2}
-\frac{14x^2+3z^2}{12z^4(4x^2+z^2)}\nn\\
&-&\frac{70x^4+32x^2z^2+3z^4}{12z^6(4x^2+z^2)^2}
-\frac{420x^6+290x^4z^2+62x^2z^4+3z^6}{12z^8(4x^2+z^2)^3}
\bigg)\Bigg\},
\ee
\be
K_3(x)=
\frac{\pi}{8x}\sqrt{4x^2+1}-\frac{\pi}{16x^2}
\ln\frac{\sqrt{4x^2+1}+2x}{\sqrt{4x^2+1}-2x},
\ee
and
\be
K_4(x)&=&
2x\int\limits_0^1 dz\left\{\arcsin\left(\frac{\sqrt{1-z^2}}{2x}\right)
-\frac{1}{4x^2}\frac{(1-z^2)^{3/2}\sqrt{4x^2+z^2-1}}{z^2}\right.\nn\\
&+&\left.\frac{1}{2z^3\sqrt{4x^2+z^2}}
\left(\frac{\pi}{2}-\arcsin\frac{(2x^2+z^2)(1-z^2)-2x^2z^2}{2x^2}
\right)\right\}.
\ee
In the asymptotic limit $h\to\infty$ this expression reduces to 
\be
\lefteqn{\siglc(h\gg\sigma)=}\nn\\
\sigma_{l,g}^{(p)}\Bigg(1&-&\frac{1}{6}\frac{\ln (h/\sigma)}{(h/\sigma)^2}
-\frac{1}{(h/\sigma)^2}\left(\frac{5}{3}\ln
2+\frac{1}{144}\right)\nn\\
&-&\frac{5}{1536}\frac{1}{(h/\sigma)^4}
+{\cal O}\left(\frac{1}{(h/\sigma)^6}\right)\Bigg).
\ee

\section{Effective interface potential for coated spheres and cylinders}

We consider spherical and cylindrical substrates which are composed of
two different species for $0\leq r\leq r_0-\delta$ and
$r_0-\delta<r\leq r_0$. They interact with the fluid particles
according to 
pair potentials $\phi_{wf}^{(0)}(r)$ and $\phi_{wf}^{(1)}(r)$,
respectively. For reasons of simplicity we assume that the number
density of the substrate particles $\rw$ is constant. If one
approximates the substrate potential $V(\vecr;r_0)=\rw v(\vecr;r_0)$
by the homogeneous (i.~e.~disregarding the discrete arrangement of the
substrate atoms)
linear superposition of these pair potentials one
has $v(\vecr;r_0)=v(r;r_0)$ with
\be
v(r;r_0)&=&\int_{{\cal S}(r_0-\delta)} d^3r'
\phi_{wf}^{(0)}(|\vecr-\vecr'|)
+\int_{{\cal S}(r_0)\backslash{\cal S}(r_0-\delta)} d^3r'
\phi_{wf}^{(1)}(|\vecr-\vecr'|)\nn\\
&=&\int_{{\cal S}(r_0)} d^3r' \phi_{wf}^{(1)}(|\vecr-\vecr'|)
+\int_{{\cal S}(r_0-\delta)} d^3r'\left(\phi_{wf}^{(0)}(|\vecr-\vecr'|)
-\phi_{wf}^{(1)}(|\vecr-\vecr'|)\right)\nn\\
&=:&v_1(r;r_0)+v_{0,1}(r;r_0-\delta)\label{secondorder}
\ee
where ${\cal S}(r)$ denotes a substrate of radius $r$.
If the functions $\phi_{wf}^{(0)}(r)$ and $\phi_{wf}^{(1)}(r)$
are approximated by Eq.~(2.5) with
interaction potential parameters $\{\ew^{(0)},\sw^{(0)}\}$  and
$\{\ew^{(1)},\sw^{(1)}\}$, respectively, the computation of
$v_1(r;r_0)$ and $v_{0,1}(r;r_0-\delta)$ can be carried out along the
lines of the 
Appendices A and B. On this basis we obtain
\be
\lefteqn{\omega^{(c,s)}(l)=}\nn\\
&=&\left(
\rw\left(\ew^{(1)}(\sw^{(1)})^{12}-\ew^{(0)}(\sw^{(0)})^{12}\right)
f^{(c,s)}_{rep}(r_0-\delta,h)
-\rw\ew^{(1)}(\sw^{(1)})^{12}f^{(c,s)}_{rep}(r_0,h)\right.\nn\\
&&-\rw\left(\ew^{(1)}(\sw^{(1)})^6-\ew^{(0)}(\sw^{(0)})^6\right)
f^{(c,s)}_{attr}(r_0-\delta,h)
+\rw\ew^{(1)}(\sw^{(1)})^6f^{(c,s)}_{attr}(r_0,h)\nn\\
&&\left.+\rl\epsilon\sigma^{12}f_{rep}^{(c,s)}(r_1,h)
-\rl\epsilon\sigma^6f_{attr}^{(c,s)}(r_1,h)\right).
\ee
The expansions of the effective interface potentials 
$\omega^{(\tau)}(l,r_0)$ into powers of $r_0/l$ under
the condition that both $\delta\ll r_0,l$ and $d_w\ll r_0,l$ can be
calculated straightforwardly. 
As a generalization of Eq.~(2.27) one obtains 
\be
\omega^{(\tau)}(l,r_0)=\frac{a}{l^2}
{\cal S}_\tau\left(\frac{r_0}{l}\right)
+\frac{b}{l^3}{\cal R}_\tau\left(\frac{r_0}{l}\right)
+{\cal O}\left(\frac{1}{l^4}\right)
\ee
where ${\cal S}_\tau(x)$ and ${\cal R}_\tau(x)$
are given by Eqs.~(2.36) and (3.51), respectively,
and with the constants $a$ and $b$ given by
Eqs.~(3.44), (3.47), and (3.48); they are identical to those of the
corresponding planar substrate.
For the model potentials given by Eqs.~(2.4) and (2.5) one has
\be
{\cal R}_c(x)=\frac{3\pi}{2}\frac{1}{(1+x)^3}
{}_2F_1\left(\frac{5}{2},\frac{3}{2};2;\left(\frac{x}{x+1}\right)^2\right)
\ee
for a cylinder and
\be
{\cal R}_s(x)=\frac{8(1+x)^3}{(1+2x)^3}
\ee
for a sphere, respectively.
From these  results one can infer the 
explicit form of the functions $q_\tau(x,\gamma)$ (Eqs.~(3.55) and (3.56)):
\be
q_c(x)=\frac{2}{3x}
\left[
{}_2F_1\left(\frac{5}{2},\frac{3}{2},2,\left(\frac{x}{x+1}\right)^2\right)
+\frac{5}{4}\frac{x^2}{(x+1)^2}\right.
&{}_2F_1&\left.
\left(\frac{7}{2},\frac{5}{2},3,\left(\frac{x}{x+1}\right)^2\right)
\right]^{-1}\times\nn\\
\times\left[
{}_2F_1\left(\frac{5}{2},\frac{3}{2},2,\left(\frac{x}{x+1}\right)^2\right)
+\frac{35}{8}\frac{x^2}{(x+1)^2}\right.
&{}_2F_1&\left(\frac{7}{2},\frac{5}{2},3,\left(\frac{x}{x+1}\right)^2\right)
+\nn\\
+\frac{175}{48}\frac{x^4}{(x+1)^4}
&{}_2F_1&\left.\left(\frac{9}{2},\frac{7}{2},4\left(\frac{x}{x+1}\right)^2\right)
\right]\ee
for a cylinder and
\be
q_s(x)=\frac{1}{x}\frac{1+2x+2x^2}{1+2x}
\ee
for a sphere, respectively.

\end{appendix}

\vfill
\newpage

\centerline{\large \bf Figure captions}
\vskip 1 cm

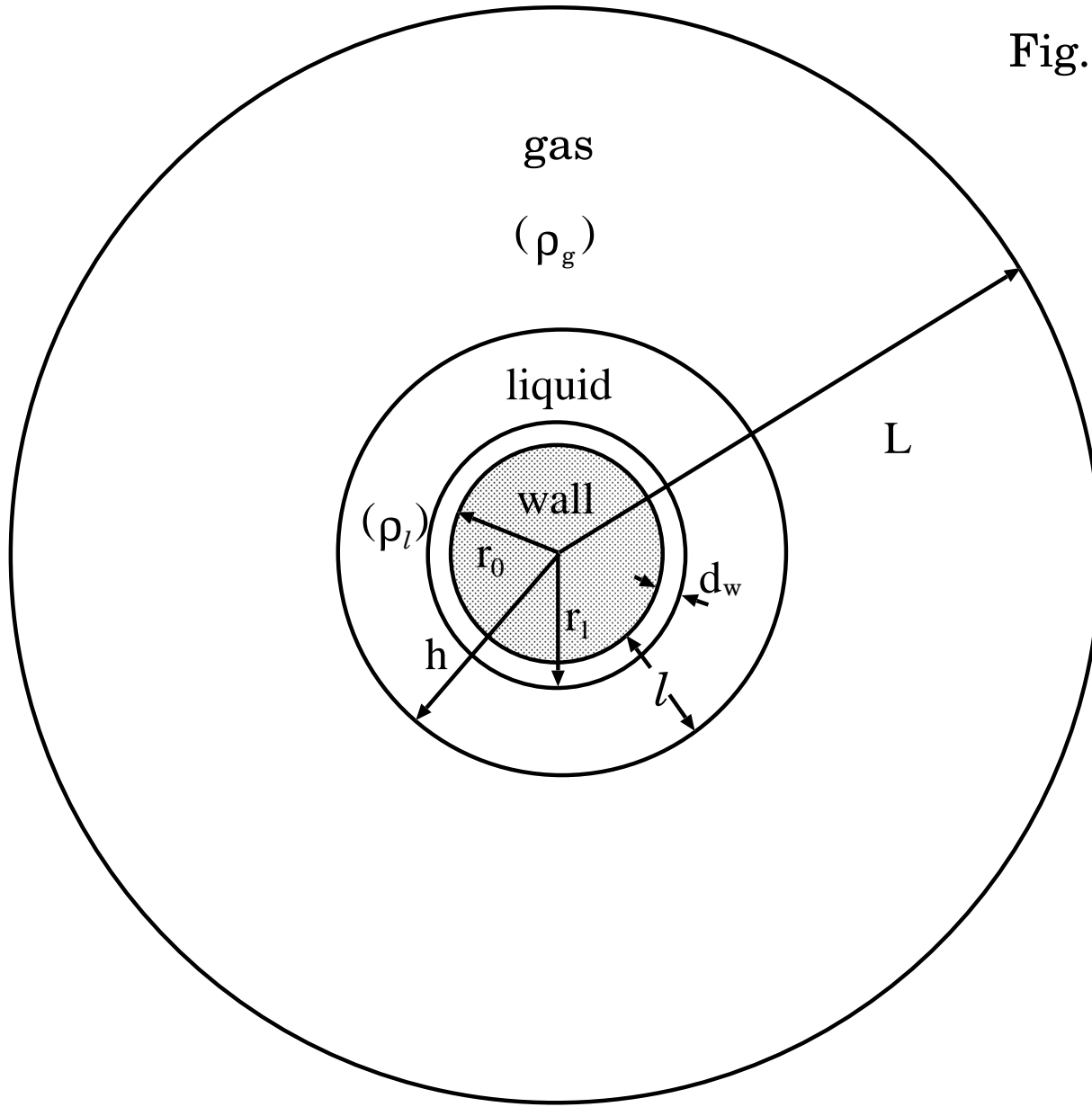
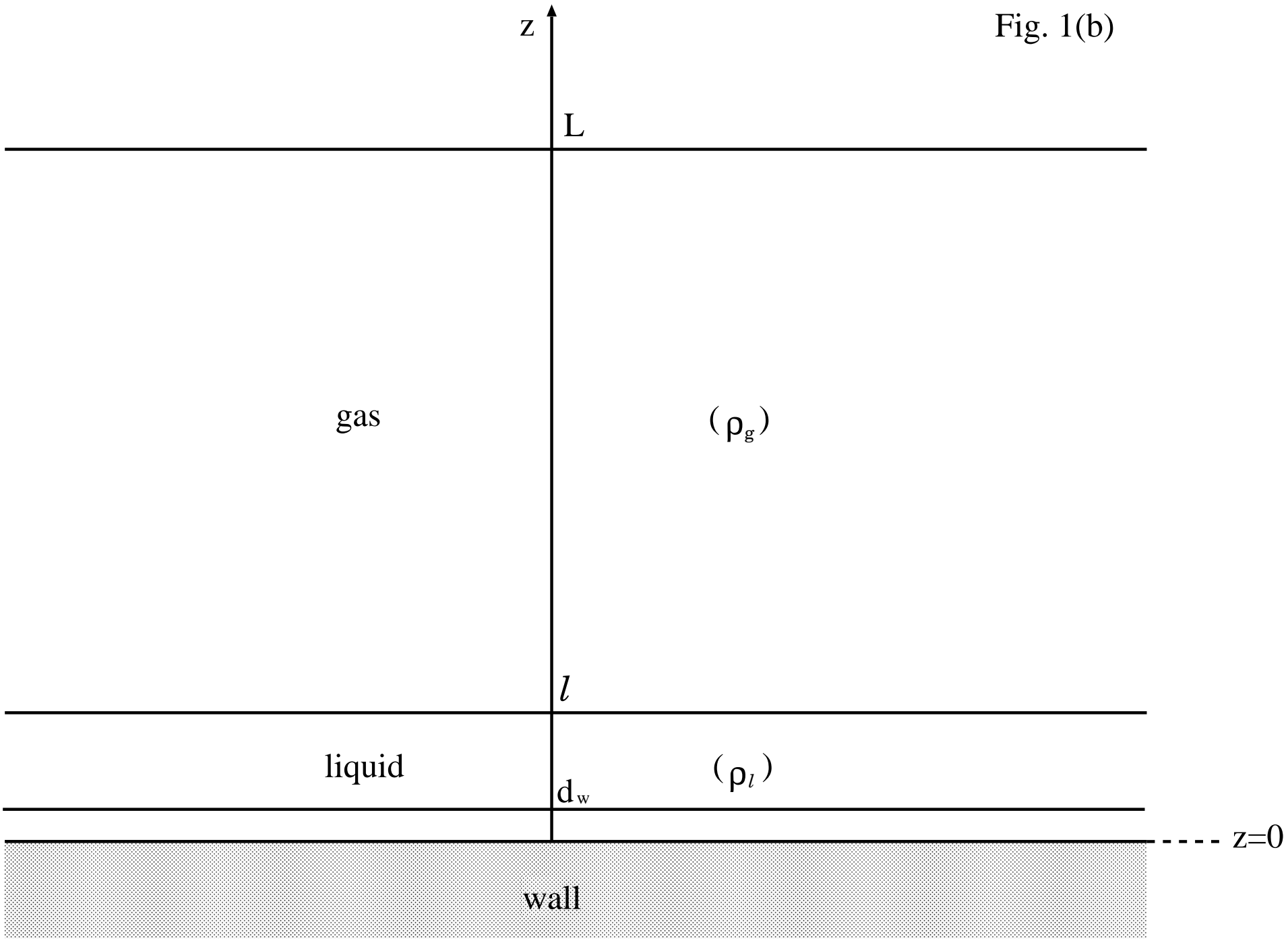
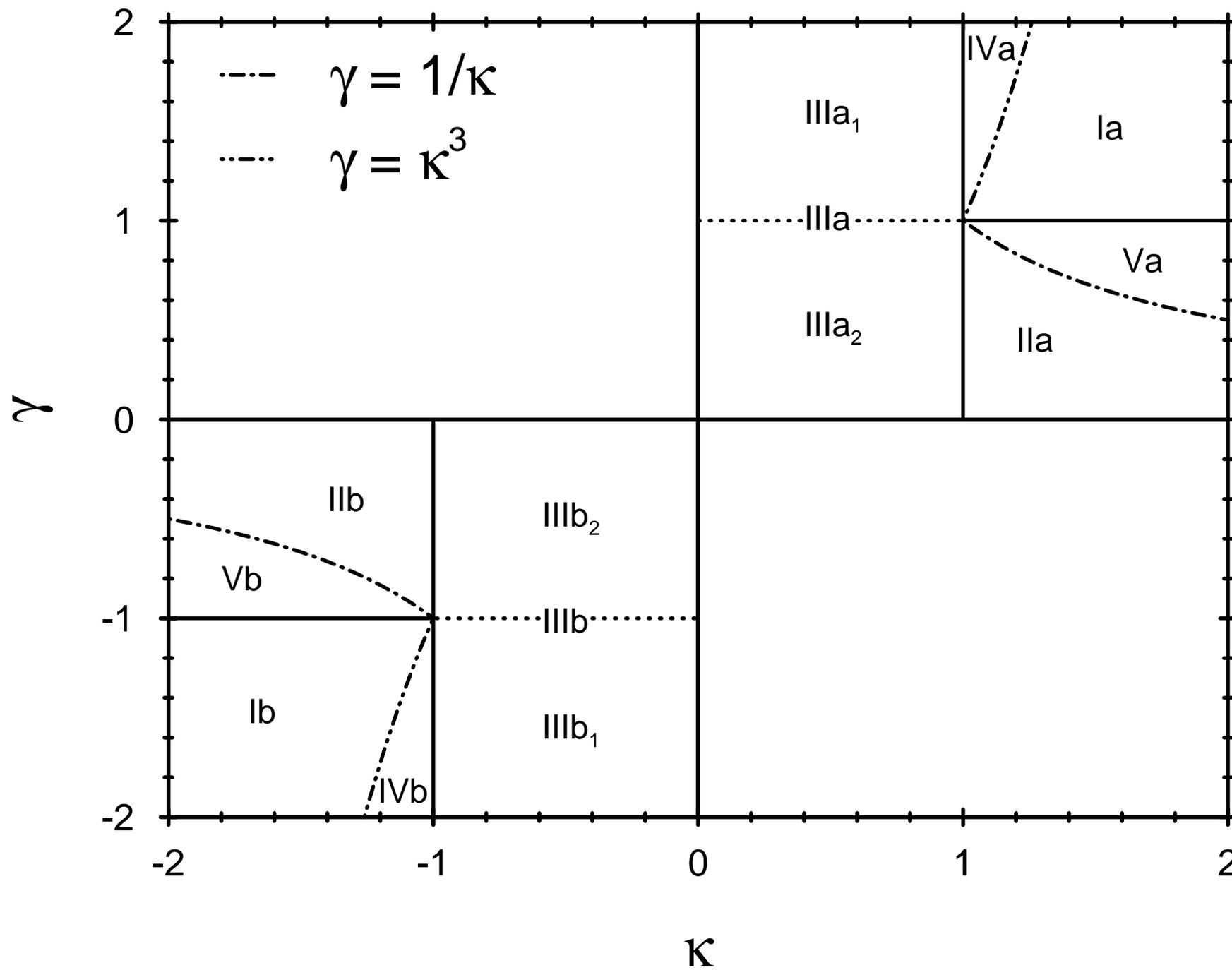
\noindent {\bf FIG.~1.}  (a) Cross section of the density 
      configuration around a sphere or
      a cylinder of radius $r_0$ with a number density $\rw$.
      The gas phase with number density $\rho_g$ is truncated for
      $r>L$.  The liquidlike film with number density $\rho_l$
      has a radius of curvature given by $h$ and a thickness
      $l=h-r_0$. The density distribution vanishes within 
      an excluded volume $r_0<r<r_1=r_0+d_w$.
      (b) Corresponding density configuration close to a planar wall.

\noindent {\bf FIG.~2.}  Liquid-vapor surface tension 
      of a liquid drop (s) and thread (c)
      as compared with the planar surface tension (p) as function of
      their radius $h$ in units of the diameter $\sigma$ of the 
      fluid particles
      based on Eqs.~(2.16) and (2.17) for the model
      potential given in Eq.~(2.4) with the parameters
      $\sigma=1.0$ and $\epsilon=1.0$.
      Within the sharp-kink approximation the ratios 
      $\sigma^{(\tau)}_{l,g}(h)/\siglp$ are constant as function
      of the temperature and behave asymptotically for 
      $\sigma/h\to 0$ as given in Eqs.~(2.22) and (2.23).
      For $\sigma/h < 1.89\cdot 10^{-7}$ one has $\siglc(h)<\sigls(h)$;
      otherwise $\sigls(h)$ is smaller than $\siglc(h)$.
      Beyond the sharp-kink approximation the ratio
      $\sigma^{(s)}_{l,g}(h)/\siglp$ for a sphere is expected 
      to approach 1 linearly as function of $\sigma/h\to 0$ with a slope 
      $-2\delta/\sigma$ given by the Tolman length $\delta$. 
      (See also Ref.~\cite{ergaenzung}.)

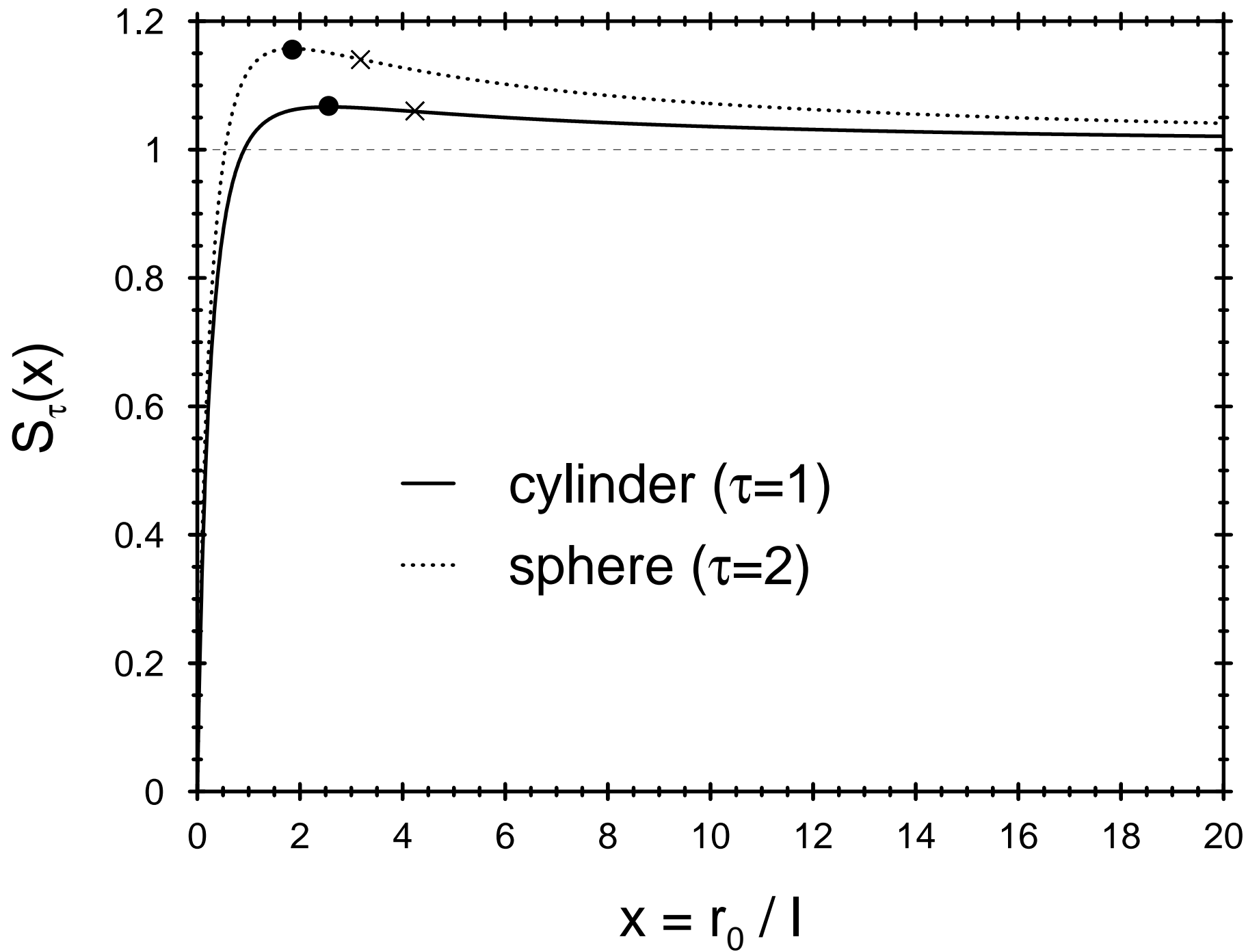
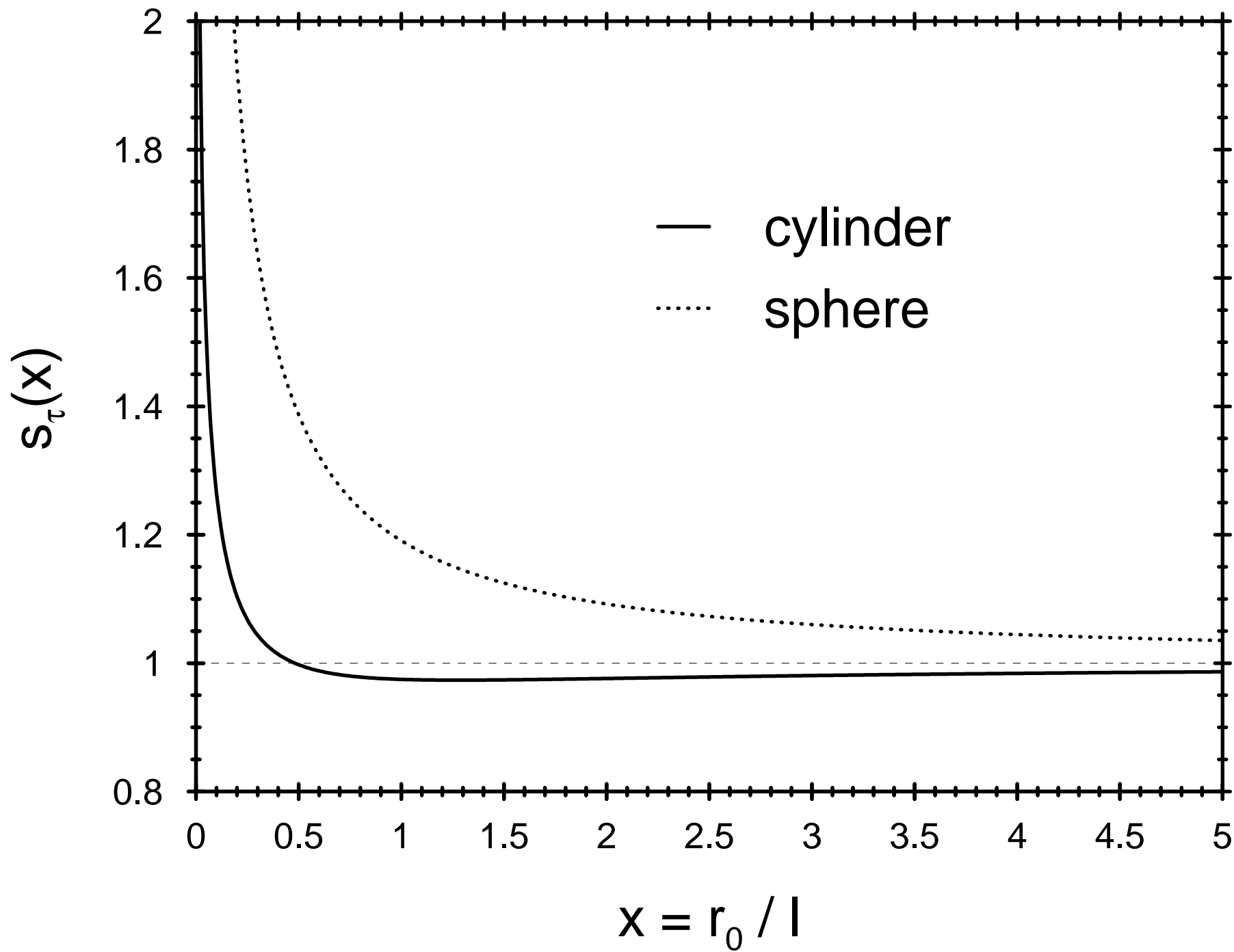
\noindent {\bf FIG.~3.} (a) Scaling function 
      ${\cal S}_\tau(x)\equiv{\cal S}_{attr}^{(\tau)}(x)$ 
      of the effective 
      interface potential (see Eqs.~(2.29), (2.30), (2.32),
      (2.34), (2.36), and (2.39)) 
      as function of $x=r_0/l$. 
      ${\cal S}_\tau(x)$ vanishes linearly for $x\to 0$ (Eq.~(2.41))
      and approaches 1 for $x\to\infty$ proportional to $x^{-1}$ (Eq.~(2.40)).
      The maxima ($\bullet$) and the inflection points ($\times$) are
      located at
      $x_{max}^{(c)}\simeq 2.56$, $x_{max}^{(s)}\simeq 1.85$, and 
      $x_{t}^{(c)}\simeq 4.24$, $x_{t}^{(s)}\simeq 3.18$, respectively.
      (b) Corresponding scaling functions $s_\tau(x)$ (see
      Eq.~(3.41)). According to this figure Eq.~(3.3) has indeed one
      unique solution.

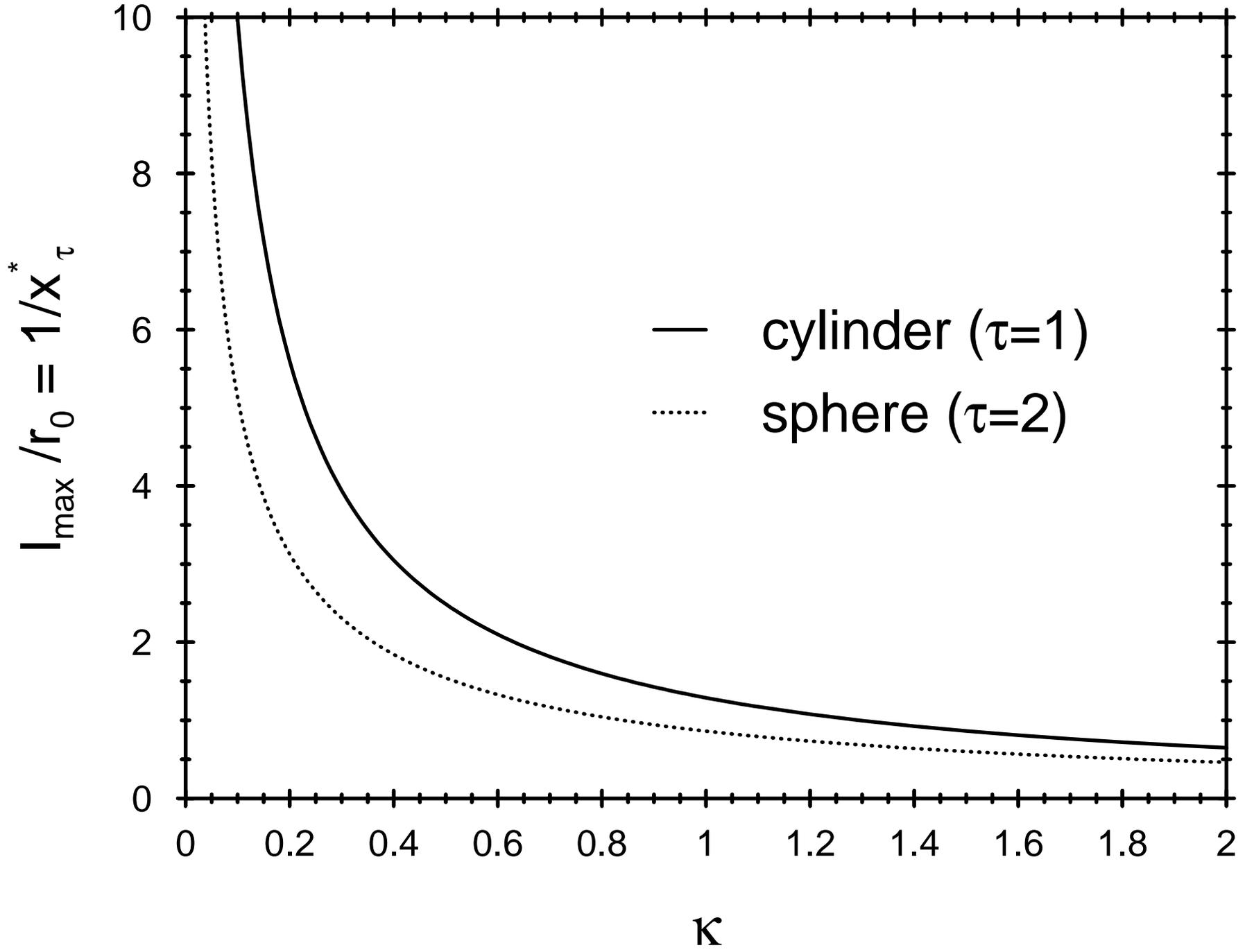
\noindent {\bf FIG.~4.} Ratios $l_{max}/r_0=1/x_\tau^*$ as a function of 
      $\kappa=(r_0^2\siglp/a)^{1/3}$ obtained by solving Eq.~(3.3) 
      numerically for the same interaction potential
      parameters as those used in Fig.~2. According to Eqs.~(3.6)
      and (3.7) both ratios vanish for $\kappa\to\infty$ proportional 
      to $\kappa^{-1}$ whereas for $\kappa\to 0$ they diverge
      proportional to $\kappa^{-3/4}$ and $\kappa^{-3/5}$ for the
      cylinder and the sphere, respectively.

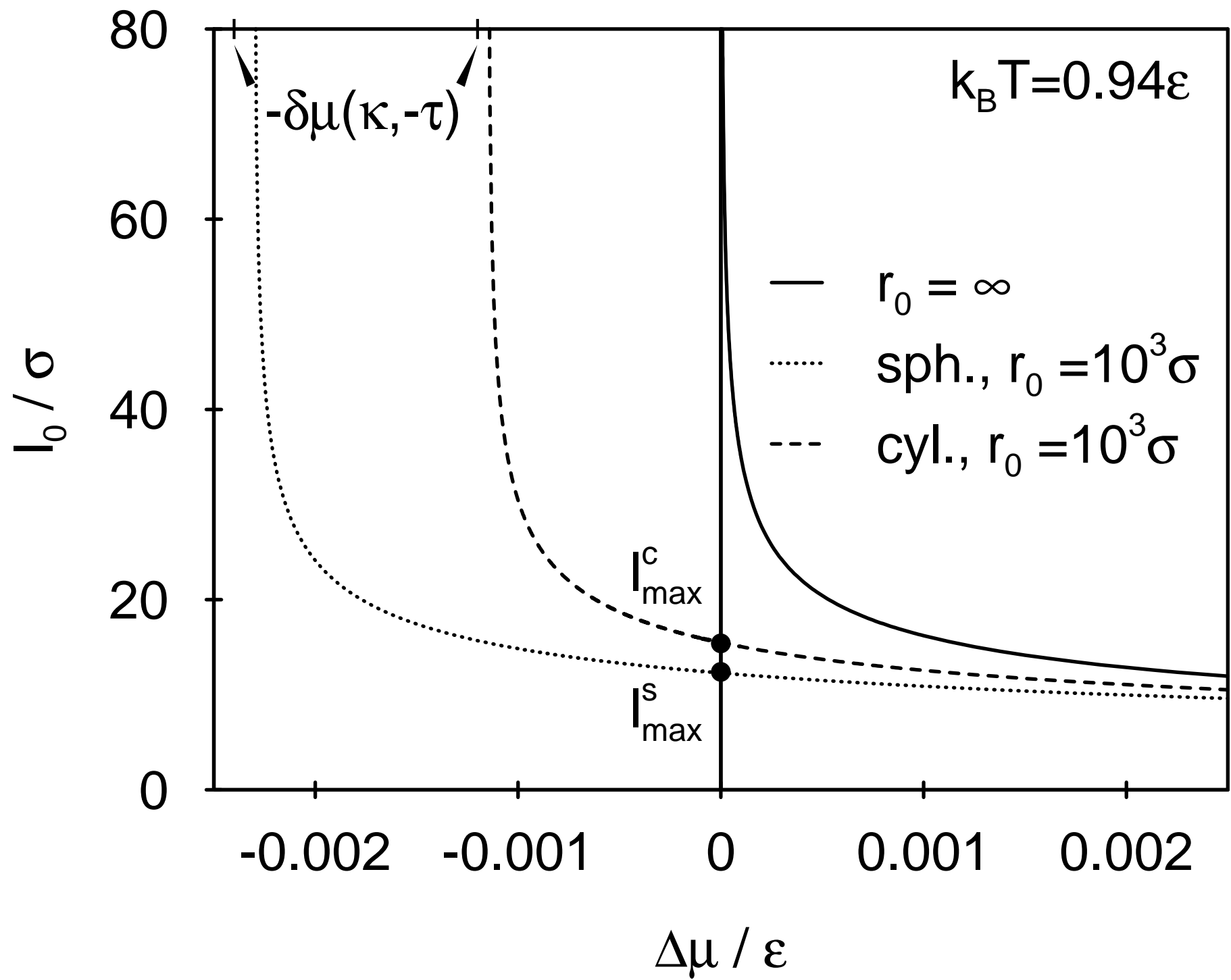
\noindent {\bf FIG.~5.} Complete wetting film thickness along an
      isotherm $k_BT/\epsilon=0.94$ for a 
      sphere and cylinder,
      respectively, of radius $1000\sigma$, compared with that 
      of a planar wall ($r_0=\infty$). 
      The curves correspond to Eq.~(3.14) with 
      the interaction parameters 
      chosen to be the same as in Fig.~2. In this case $\kappa\simeq 75.5$
      so that Eqs.~(3.21), (3.25), and (3.26) apply. Therefore
      the complete wetting film thickness on a curved substrate
      resembles that on the corresponding 
      planar substrate as if liquid-vapor coexistence would occur 
      at 
      $\Delta\mu(\kappa,\lambda)=-\delta\mu(\kappa,\lambda=-\tau)<0$ 
      (see Eqs.~(3.17) and (3.27)) instead of at $\dmu=0$.
      ($\dmu_{eff}=0$ implies $\lambda=-\tau$, see
      Eqs.~(3.24)-(3.26).) 

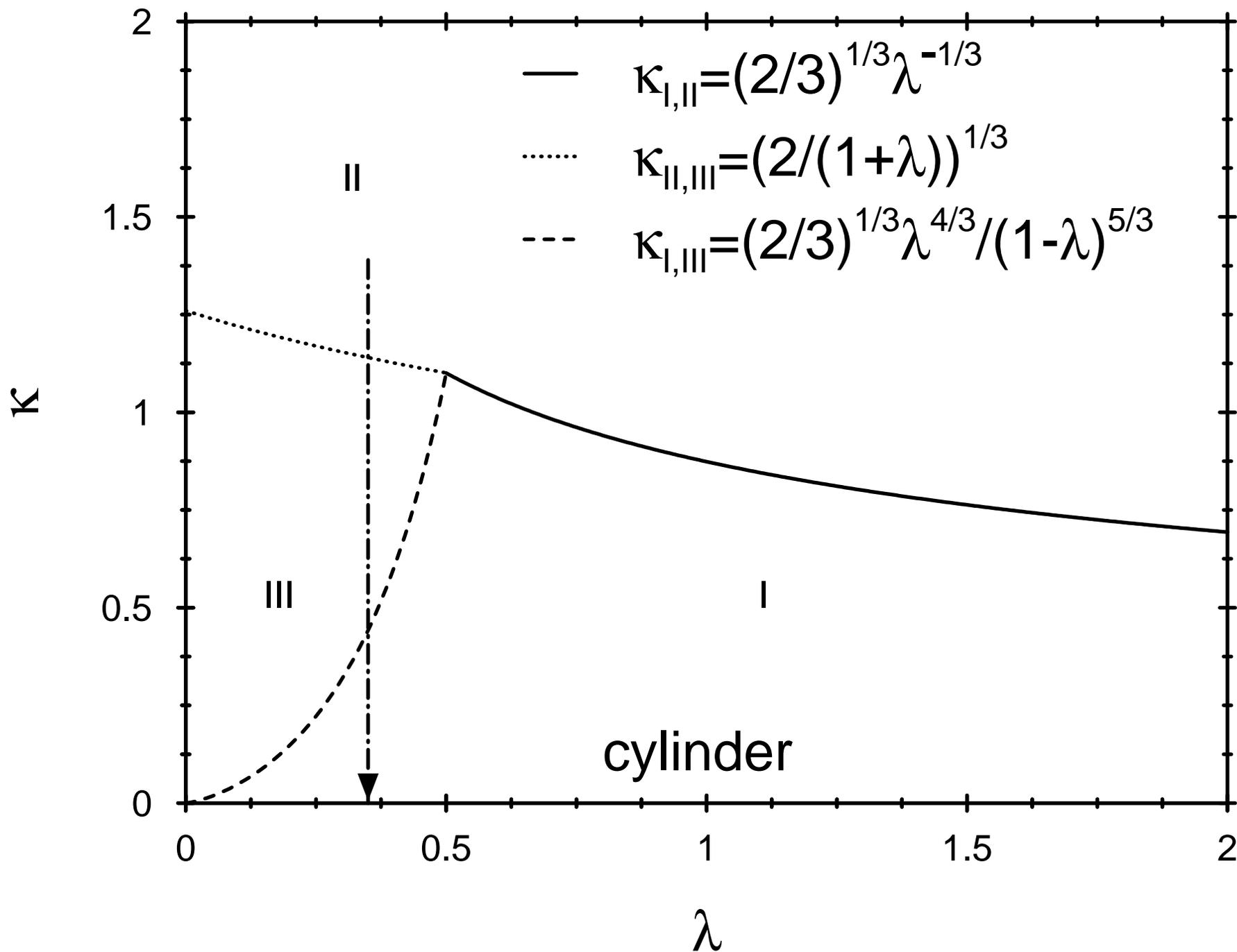
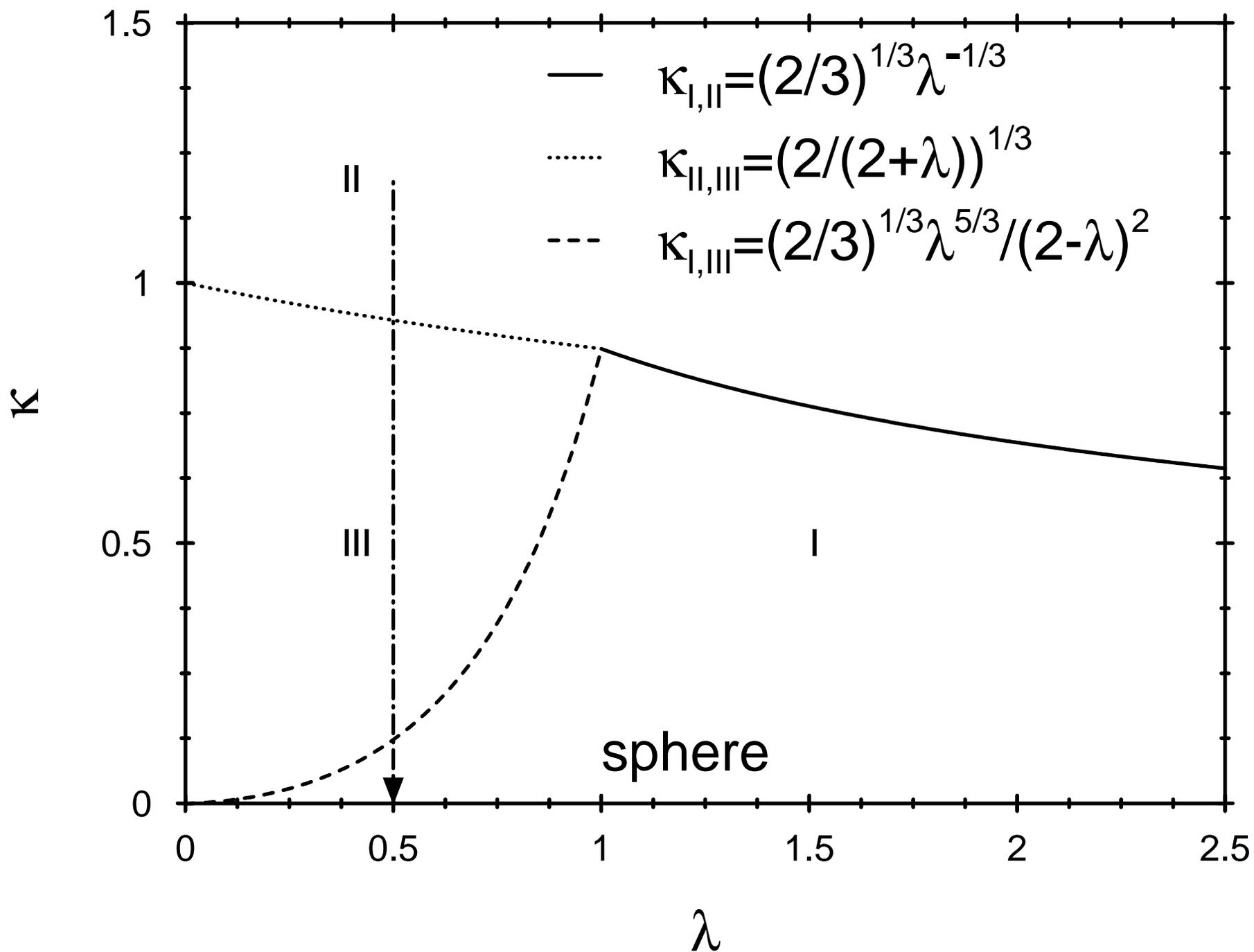
\noindent {\bf FIG.~6.} According to Eqs.~(3.25) and (3.26) the
      thickness of a complete wetting film on a curved substrate
      is given by the film thickness on the corresponding planar
      substrate for an effective undersaturation 
      $\Delta\mu_{eff}(\dmu;r_0,\tau)$. In terms of the relevant 
      dimensionless parameters $\kappa=(\siglp r_0^2/a)^{1/3}$
      and $\lambda=\drho r_0\dmu/\siglp$ one finds three distinct
      regimes I, II, and III within which
      $\lambda_{eff}(\kappa,\lambda)$ exhibits the characteristic
      power laws given by Eqs.~(3.29), (3.30), (3.31), (3.35),
      and (3.36), respectively.
      (a) shows the crossover lines $\kappa_{I,II}$,
      $\kappa_{II,III}$, and $\kappa_{I,III}$ between these regimes 
      for a cylinder, (b) the corresponding ones for a sphere.
      Their amplitudes are chosen such that they meet in a single
      point. The variation of $\lambda_{eff}(\kappa,\lambda)$
      along the dashed-dotted paths is shown in Fig.~7

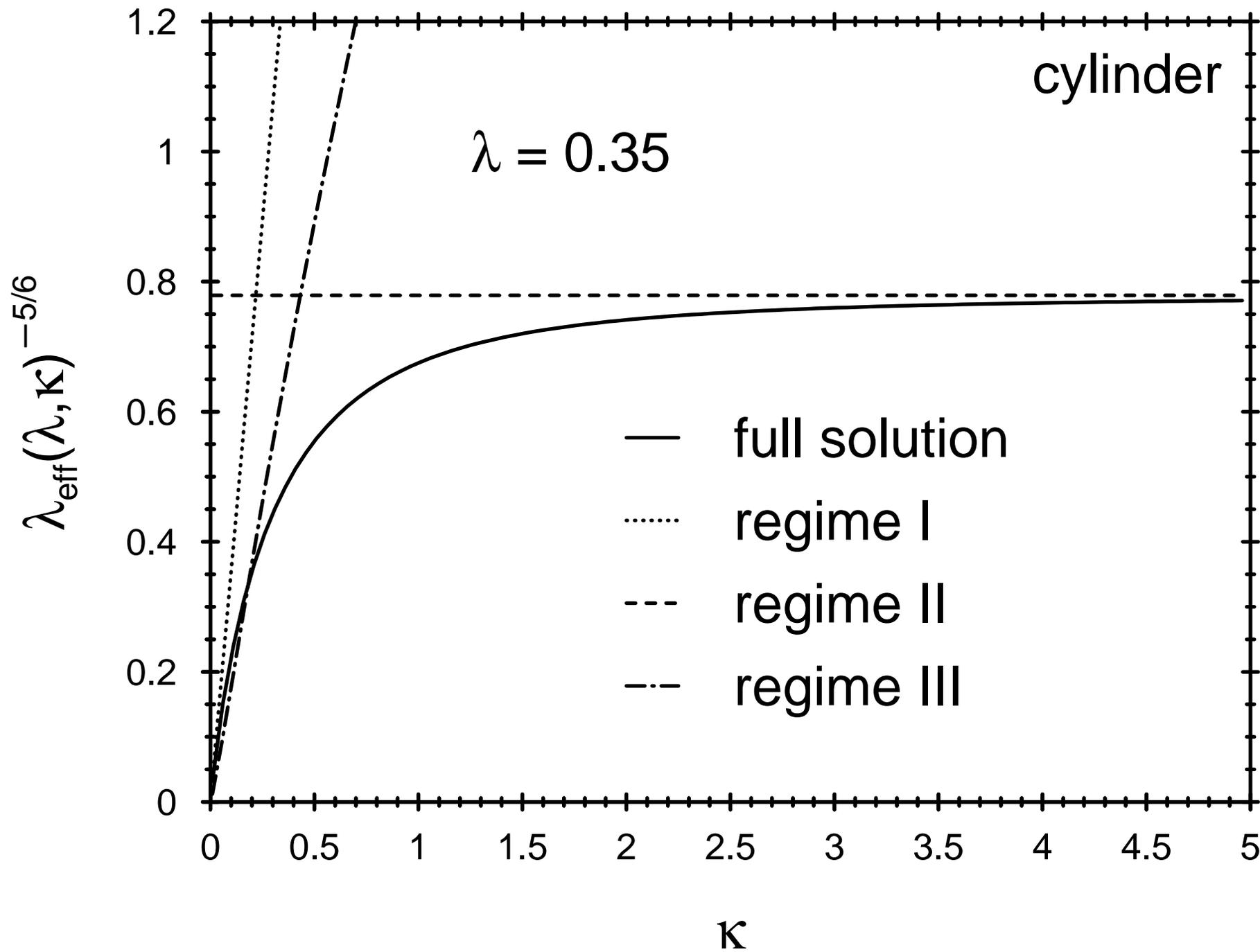
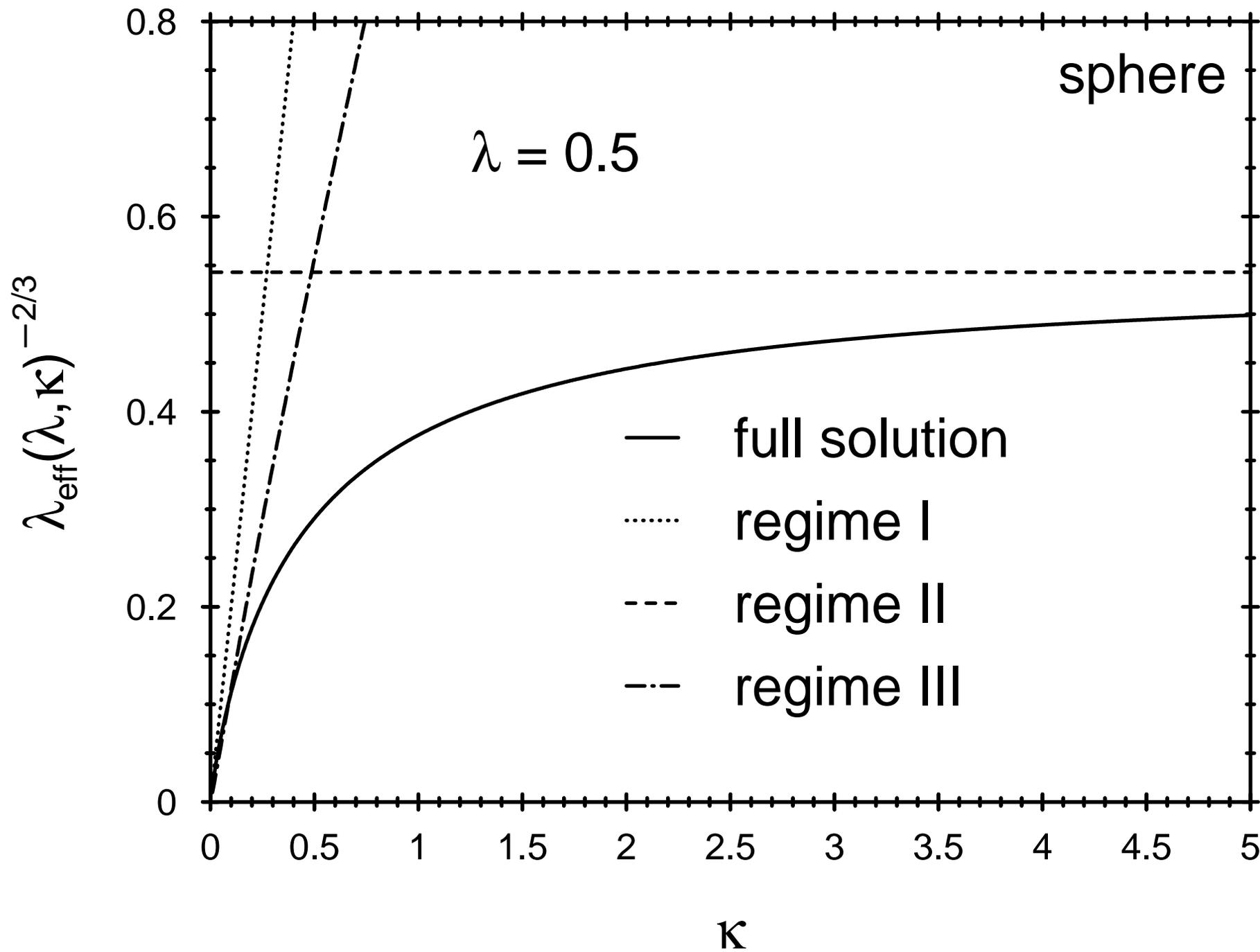
\noindent {\bf FIG.~7.} (a) shows the effective undersaturation 
      $\lambda_{eff}(\kappa,\lambda)$ for a cylinder 
      along the dashed-dotted path in Fig.~6(a), i.~e., for 
      $\lambda=0.35$, and (b) for a sphere 
      along the dashed-dotted path in Fig.~6(b), i.~e.,
      for $\lambda=0.5$, as function of $\kappa$.
      The full curves correspond to the full solution determined
      by Eq.~(3.26). These solutions display the crossover behaviors
      which occur according to the corresponding vertical paths in
      Fig.~6. For large values of $\kappa$ one is in regime II where
      $\lambda_{eff}^{(\tau)}\simeq\tau+\lambda$ is constant as
      function of $\kappa$ (dashed curves). Upon decreasing $\kappa$
      one enters regime III where Eqs.~(3.35) and (3.36)
      (dashed-dotted curves) hold. Ultimately for $\kappa\to 0$
      one reaches regime I where Eqs.~(3.29) and (3.30) (dotted curves)
      are valid. $\lambda_{eff}$ is raised to the powers $-5/6$ and
      $-2/3$, respectively, so that the dotted curves are straight lines.

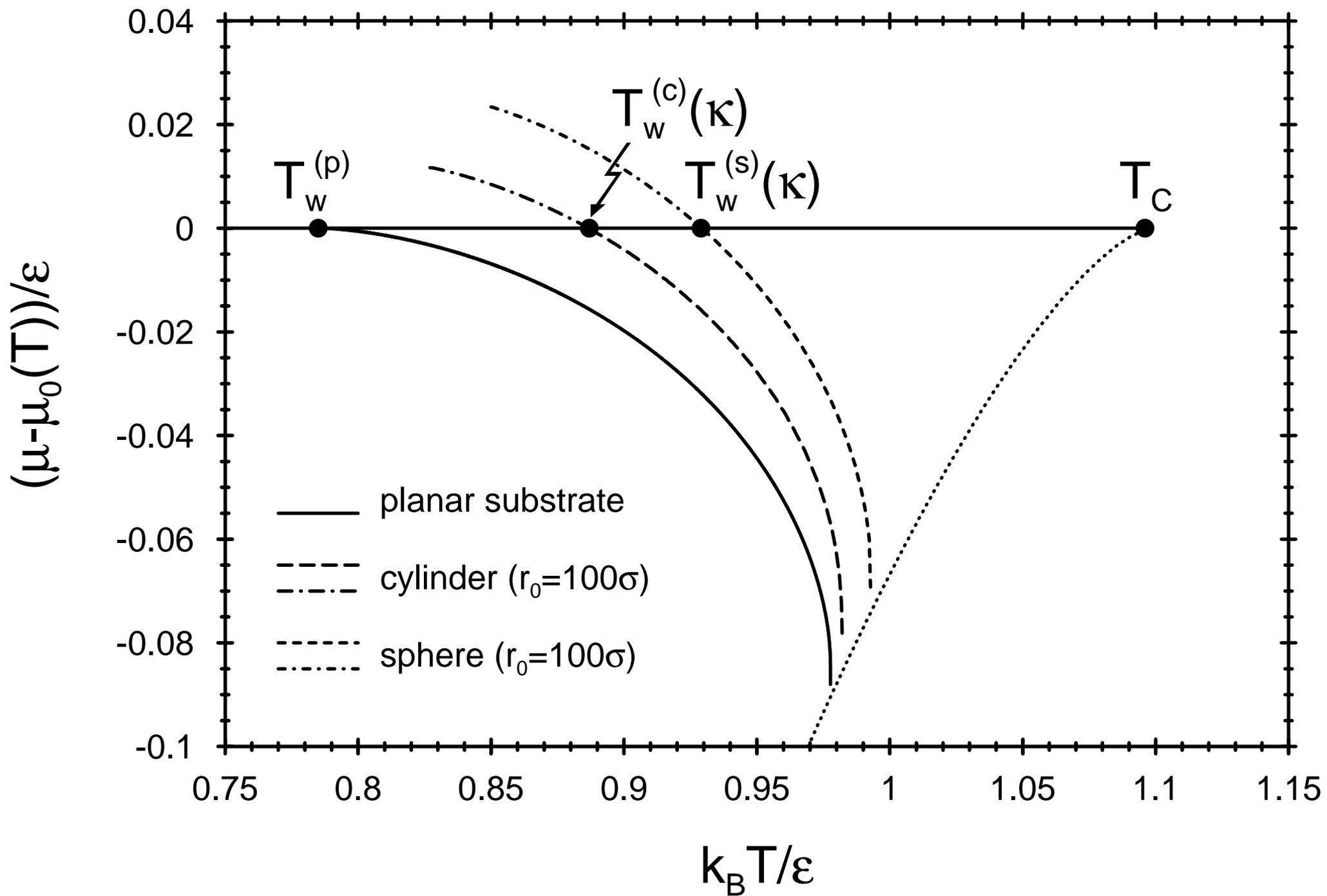
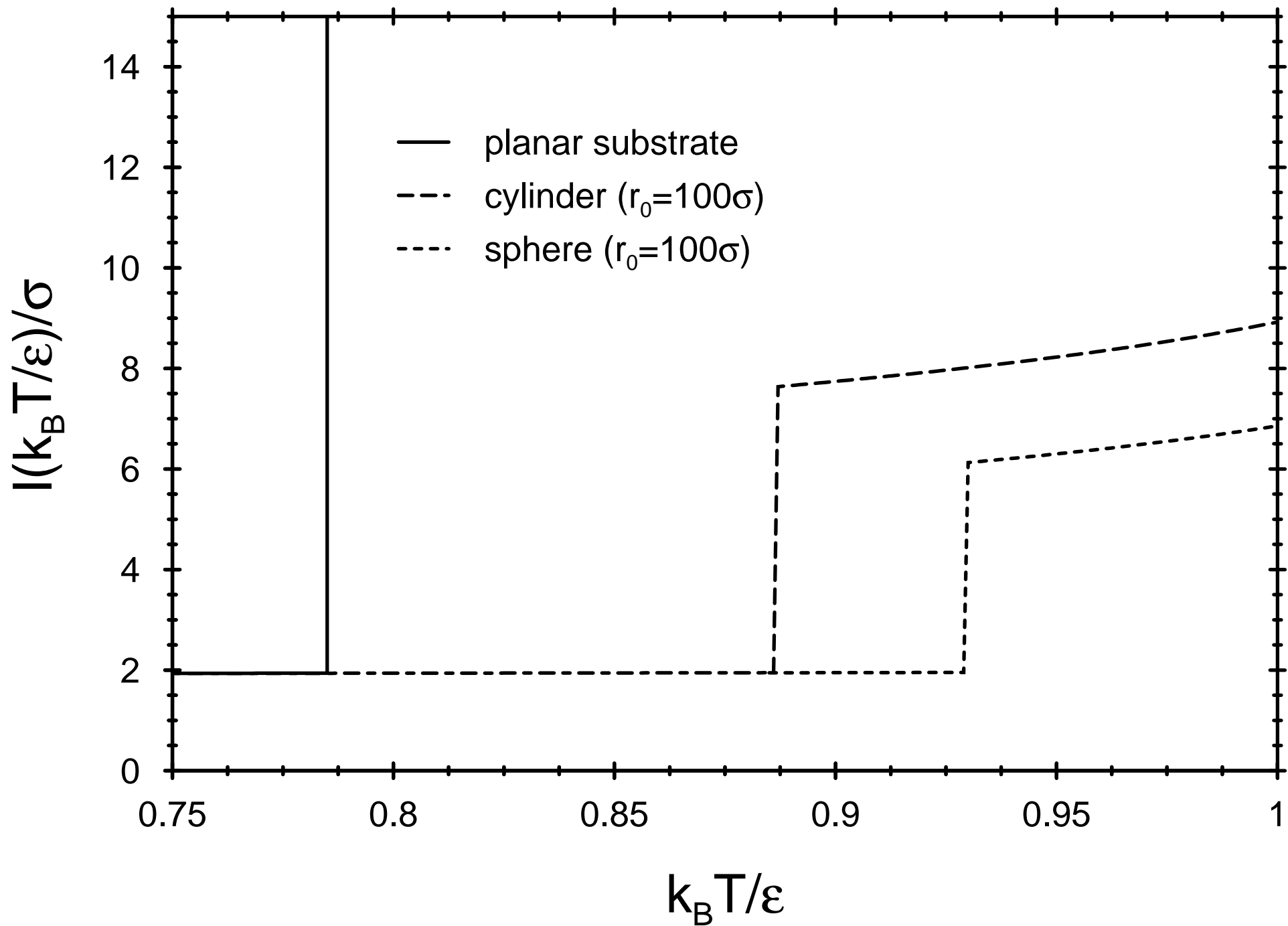
\noindent {\bf FIG.~8.} (a) Phase diagram of a first-order wetting
      transition on a 
      cylindrical and a spherical substrate of radii
      $r_0/\sigma=100$ obtained by a numerical minimization of 
      Eq.~(2.10) with the substrate potential parameters 
      $\sigma_{wf}/\sigma=1.35$, $\epsilon_{wf}/\epsilon=0.625$,
      $d_w/\sigma=1.175$,
      and $\rho_w\sigma^3=1.0$. The prewetting line of the 
      corresponding planar substrate (solid curve) joins the
      liquid-vapor coexistence curve $\mu-\mu_0(T)=0$ tangentially at 
      the first-order transition temperature $T_w^{(p)}$.
      The wetting behavior on a curved substrate is similar to
      that of a planar substrate whose coexistence curve is shifted
      upwards. Consequently for a curved substrate at coexistence 
      there is a thin-thick transition at $T_w^{(\tau)}(\kappa)$
      where the remaining part of the prewetting line intersects 
      the coexistence curve {\em linearly}.
      Note, however, that along the prewetting line of the 
      curved substrates $\kappa$ varies as function of the temperature
      and is thus not a constant; for the given 
      parameter values $\kappa$ ranges between 13.8 and 17.1 for
      $0.78\le k_B T/\epsilon\le 1.0$. Therefore the prewetting lines
      belonging to the curved surfaces are not parallel to the
      corresponding prewetting line of the flat substrate.
      The thin-thick transition can be traced into the oversaturated
      vapor region (dashed-dotted lines) provided it is possible to
      maintain the thermodynamically instable vapor phase for these
      values of $\mu$. At the upper end points
      of these curves the minimum in $\Omega_s(l)$ corresponding to the
      thicker film ceases to exist.
      Each prewetting line ends at its lower end
      in a critical point. However, in the present model and
      approximation the 
      prewetting lines of both the planar and the curved substrates
      can only be followed up to 
      the dotted curve which denotes the loci where in the bulk 
      the metastable liquid phase ceased to exist. Therefore beyond
      the dotted curve the present sharp-kink approximation (Eq.~(2.6))
      for the density  distribution close to the wall is no longer
      applicable. In that region of the phase diagram for the present
      choice of interaction potentials the thicknesses of the wetting
      films are so small that these films cannot be described
      adequately by a liquidlike layer with a number density $\rho_l$
      which corresponds to that of the metastable bulk liquid phase.
      (b) Temperature dependence of the equilibrium film thickness
      $l_0$ at coexistence for the systems described by the phase
      diagram in (a). It illustrates the shift of the wetting
      transition temperature and the finite discontinuity induced by
      the curvature of the substrate.

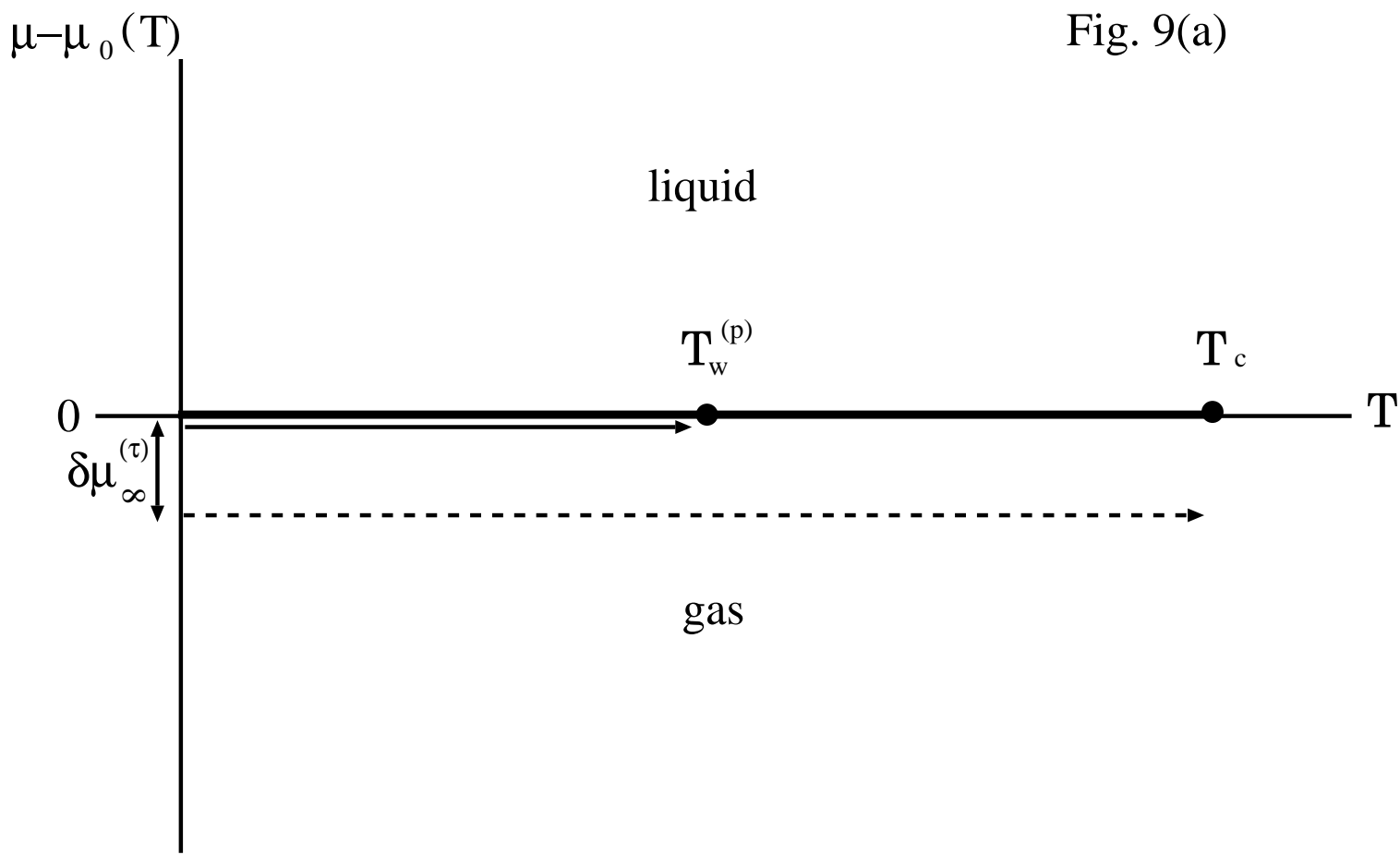
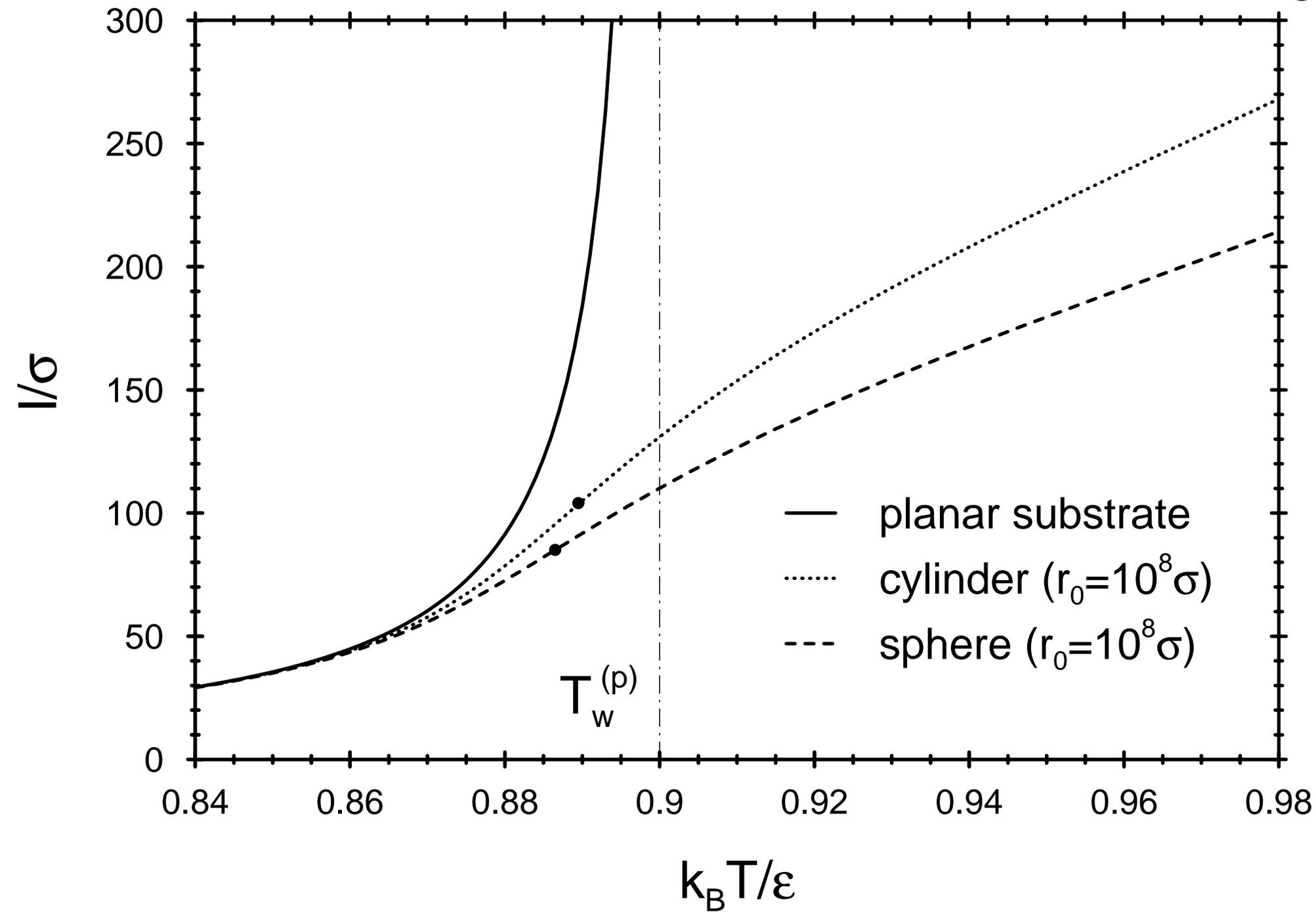
\noindent {\bf FIG.~9.} (a) Schematic phase diagram for a planar
      substrate exhibiting a second-order wetting transition at
      $T_w^{(p)}$. At $\mu=\mu_0(T)$ and for $T\leq T_c$ the bulk
      liquid and gas phases 
      coexist; only temperatures above the triple point are
      considered. If one
      records the film thickness $l/\sigma$ as function of the
      temperature $k_BT/\epsilon$ along a path slightly below coexistence
      (thin line) and at a fixed undersaturation
      $\delta\mu_\infty^{(\tau)}>0$  (dotted line) for a given set of
      interaction parameters (see below) one obtains the curves shown
      in (b). For the film thickness on a 
      cylinder or a sphere the path along the coexistence curve across
      $T_w^{(p)}$ is equivalent to the dotted path for the
      corresponding 
      planar substrate because of the effective upward shift of the
      bulk phase diagram due to the curvature (see main text). 
      (b) Film thickness $l/\sigma$ as function of the temperature
      $k_BT/\epsilon$ for a planar substrate and for a cylinder and a
      sphere of radius $r_0/\sigma=10^8$, respectively,
      on a thermodynamic path along the coexistence line, $\mu=\mu_0(T)$,
      crossing the second order wetting transition of the planar
      substrate. The interaction parameters have been chosen such that
      for the planar substrate there is a wetting transition at
      $k_BT_w^{(p)}/\epsilon=0.9$. This requires (Eqs.~(3.41) - (3.46))
      $\ewnull/\epsilon=(\sigma/\swnull)^6\rl(T_w^{(p)})/\rw\simeq
      0.6002$ and $\swnull/\sigma=1.0$ so that
      $a\left(T=T_w^{(p)}\right)=0$ (Eq.~(3.41)) is fulfilled. 
      In order to satisfy the
      remaining conditions for a critical wetting transition (Eq.~(3.41))
      one may choose, e.~g., $\eweins/\epsilon=1.4$, 
      $\sweins/\sigma=1.0$, and  $\delta/\sigma=1.5$. 
      The inflection points ($\bullet$) in the film thickness 
      $l^{(\tau)}(T)$ on the curved substrates occur {\em below} 
      the wetting temperature $T_w^{(p)}$ of the planar substrate.
      The levelling off of the curves for the cylinder and the sphere
      is not yet complete even at $T_c$. It is remarkable that for
      $\left(T_w^{(p)}-T\right)/T_w^{(p)}\quad\lsim\quad 10^{-2}$ even
      for radii 
      of the order of $1\mbox{cm}$ the effect of the curvature on the
      wetting film thickness as compared with a truly planar geometry
      is still visible. We have checked that the above choice of
      interaction potential parameters is not atypical in the sense
      that they cause an unusually small temperature gradient of the
      Hamaker constant at $T_w$ as compared with the experimental
      observation of critical wetting \cite{bonn}. Therefore we conclude
      that this extraordinarily strong sensitivity of critical wetting
      to curvature is a generic feature.

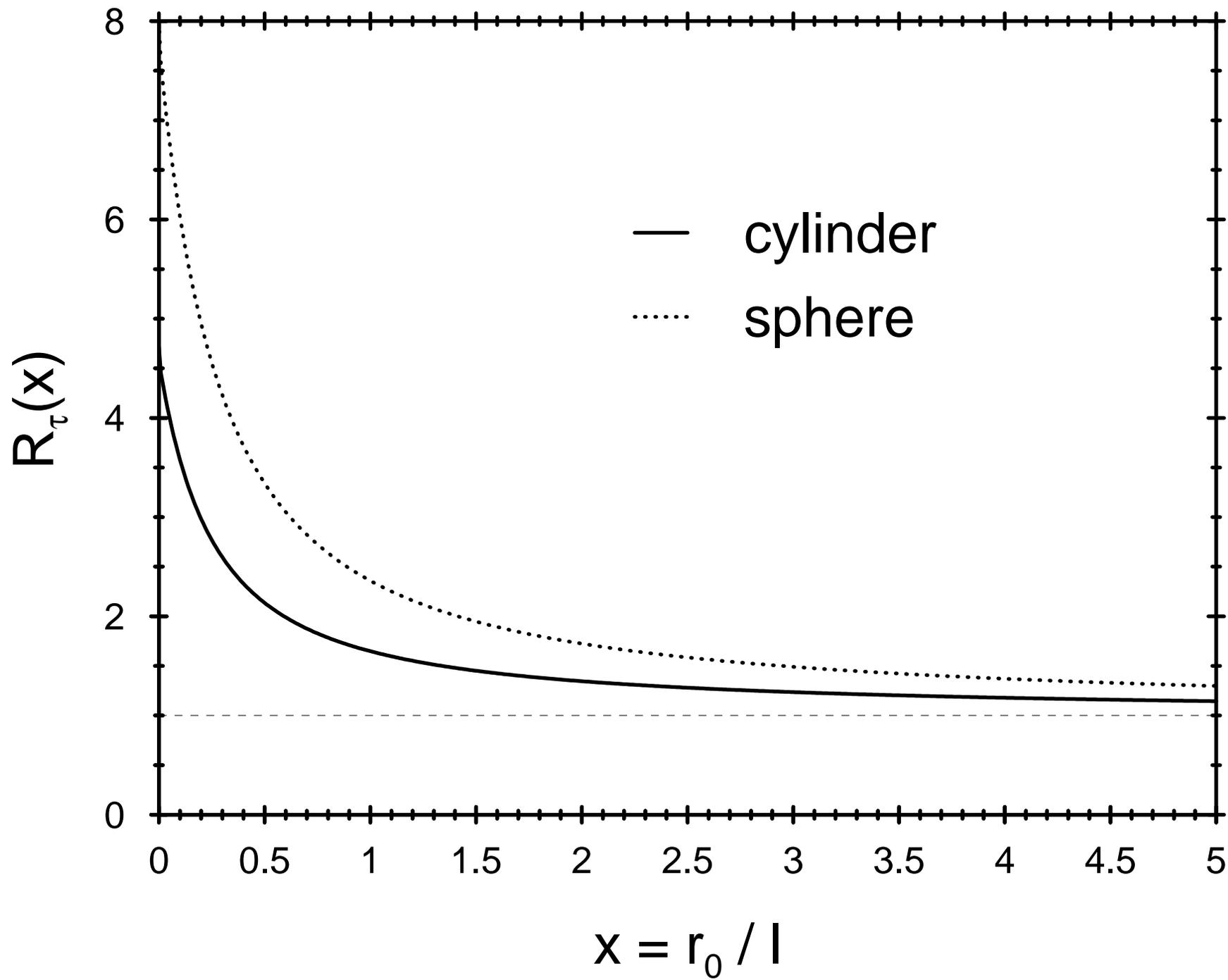
\noindent {\bf FIG.~10.} Scaling function ${\cal R}_\tau(x)$ given by
      Eqs.~(3.49), (C4), and (C5), as function of $x=r_0/l$. ${\cal
      R}_\tau(x)$ rises linearly to a finite value for $x\to 0$ (${\cal
      R}_c(x=0)=\frac{3\pi}{2}$ and ${\cal R}_s(x=0)=8$, respectively;
      Eqs.~(3.51) and (3.52)) and approaches 1 for $x\to\infty$
      proportional to $x^{-1}$ (Eq.~(3.50)).

\noindent {\bf FIG.~11} In the case of critical wetting at coexistence
      the film thickness on the curved substrates is determined by
      Eq.~(3.54) with the dimensionless parameters
      $\kappa=\left(r_0^2\siglp/a\right)^{1/3}$ and
      $\gamma=3b/\left(2ar_0\right)$. Since $x=r_0/l$ is
      positive, it is sufficient to consider the regions ($\kappa\geq
      0$, $\gamma\geq 0$) and
      ($\kappa<0$, $\gamma<0$) which correspond to $T>T_w^{(p)}$
      and $T<T_w^{(p)}$, respectively. The
      limits $T=T_w^{(p)}\pm 0$ correspond to
      $\gamma\to\pm\infty$ whereas on the line $\gamma=0$ one has,
      apart from possible multicritical wetting transitions,
      first-order wetting which we have studied in Subsect.~IIIC. By
      means of the limits $x\to 0$ and $x\to\infty$ of the functions
      $q_\tau(x)$ and $s_\tau(x)$ entering into Eq.~(3.55) one can
      evaluate 
      Eq.~(3.54) for the regions I, II, and III of the ($\kappa$,
      $\gamma$) parameter space. In the regions Ia,b one has
      $|\kappa|\gg1$ 
      and $|\gamma|\gg1$ so that $\gamma\ll\kappa^3$, i.~e., $T\to
      T_w^{(p)}\pm 0$ and $r_0\gg\left(3b/(2\siglp)\right)^{1/3}$,
      which corresponds to film thicknesses given by Eqs.~(3.57) -
      (3.59). For $|\kappa|\gg1$ and $|\gamma|\ll1$ with
      $\kappa\gamma\ll1$  (large $r_0$ and $b/a\to 0$) the film 
      thicknesses are determined by Eqs.~(3.60)-(3.62) for positive 
      $\gamma$ (region IIa) and by Eqs.~(3.64)-(3.65) for negative 
      $\gamma$ (region IIb). In the case $|\kappa|\ll1$ (regions
      IIIa,b), i.~e. $r_0\ll\left(|a|/\siglp\right)^{1/2}$, Eqs.~(3.66) -
      (3.69) are valid.  For $|\gamma|\to\infty$ (regions IIIa$_1$,b$_1$),
      i.~e. $T\to T_w^{(p)}\pm 0$, one can expand Eqs.~(3.66) and
      (3.67) into powers of
      $\kappa^3/\gamma=2r_0^3\siglp/(3b)$ which leads to
      Eqs.~(3.70) - (3.73). The opposite limit, $\gamma\to 0$ 
      makes sense only for $\gamma\geq 0$ (region IIIa$_2$) where 
      the film thicknesses are determined by
      Eqs.~(3.74) and (3.75).
      In the remaining regions IVa,b and Va,b there is no suitable
      expansion parameter so that there 
      Eq.~(3.54) can only be solved numerically.
 
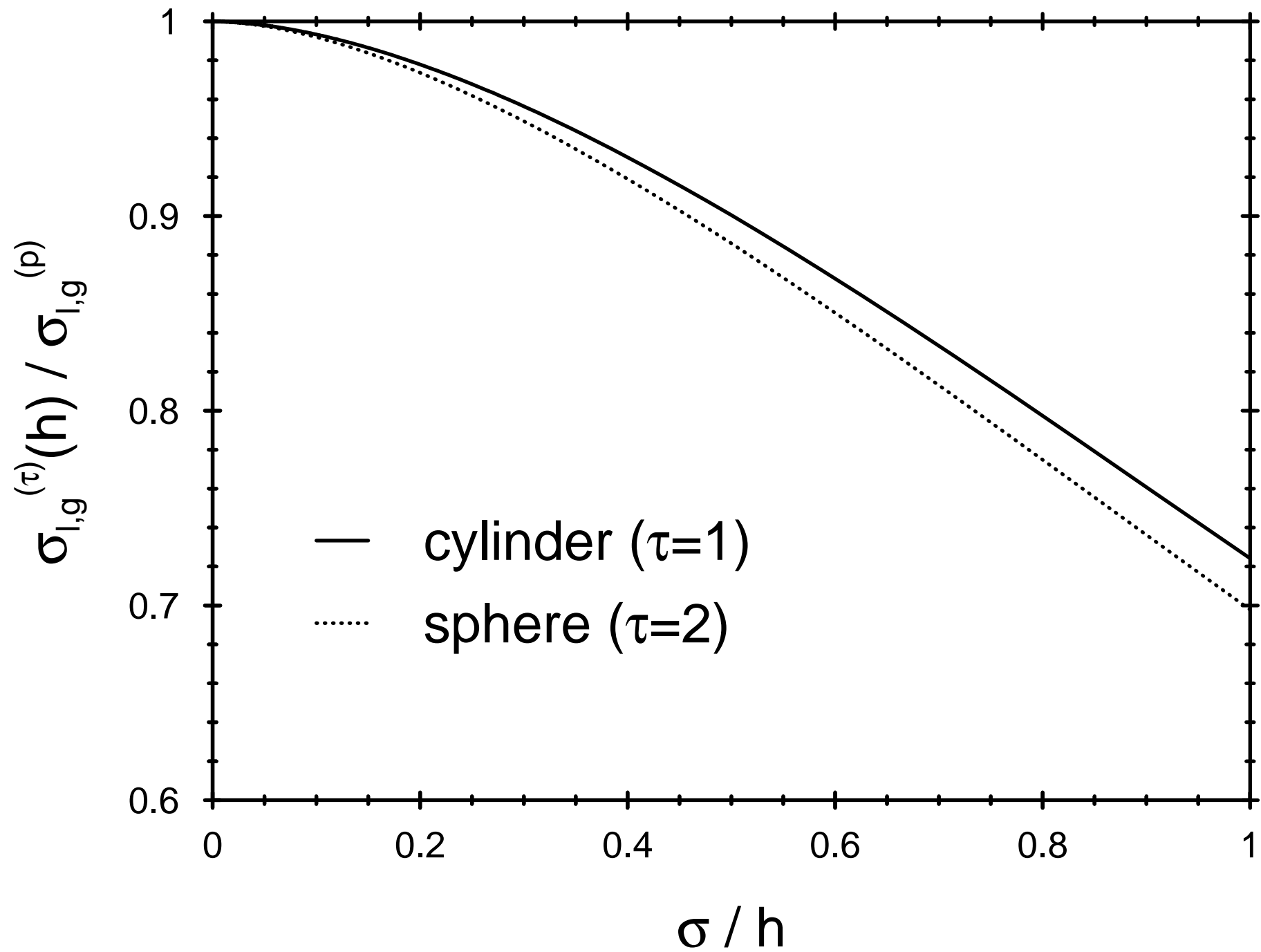
\end{document}